\documentclass[12pt, a4paper]{article}
\pdfoutput=1
\usepackage{jheppub}
\usepackage{latexsym, graphicx}
\usepackage{amsmath, amssymb}
\usepackage{color, hyperref}

\title{Relativistic single-electron wavepacket in quantum electromagnetic fields II: 
Quantum radiation emitted by a uniformly accelerated electron}
\author[a]{Shih-Yuin Lin}
\author[b]{and Bei-Lok Hu}

\affiliation[a]{Department of Physics, National Changhua University of Education, Changhua 500207, Taiwan}
\affiliation[b]{Maryland Center for Fundamental Physics and Joint Quantum Institute \\
University of Maryland, College Park, Maryland 20742-4111, USA}

\emailAdd{sylin@cc.ncue.edu.tw} 
\emailAdd{blhu@umd.edu}

\abstract{We compute the quantum radiation emitted by wavepackets of relativistic single electrons, both at rest and undergoing uniform acceleration in the Minkowski vacuum of the electromagnetic field. We find that the cubic terms in the original nonlinear action of electrodynamics should be considered in obtaining the quantum radiation to the leading order. We show that the quantum radiation from a single-electron wavepacket at rest vanishes exactly. For a uniformly accelerated electron, the quantum radiated power has secular growth in the long-time regime. We demonstrate that this secular growth has a classical interpretation, and argue that the resummed quantum radiation at late times would not diverge. Regarding experimental proposals for the detection of the Unruh effect from the quantum radiation in the `blind spots' of classical radiation we ascertain that quantum corrections in the two blind spots are fully contributed by the transverse-deviation correlators, where the dominant contributions are irrelevant to the Unruh effect in electron microscopes.}


\begin{document}

\maketitle

\newpage

\section{Introduction}
\label{intro}

Building on Paper I \cite{LH24}, this is the second of a planned series of papers seeking to describe the real time evolution of relativistic particles represented by wavepackets interacting with quantum fields in a nonequilibrium dynamical setting (e.g., ref. \cite{CalHuBook}). We aim in particular at this particle-field system's quantum features such as thermal-like fluctuations 
(in the context of the Unruh effect \cite{Unr76}), emitted quantum radiation with back-reaction, quantum dissipation, as well as the quantum informational aspects such as decoherence, correlation, and entanglement. We choose to work on the familiar subject of a single-electron wavepacket interacting with quantum electromagnetic (EM) fields so that our theory can find a good anchor in the well-studied and proven quantum electrodynamics (QED), and that our approaches and new results can be compared with and understood against this benchmark theory.   The other equally important goal of this series is to use the further developed theory to examine vintage experiments performed (such as with transmission electron microscopes (TEM) \cite{TE89}) or novel detection schemes proposed (such as emitted quantum radiation \cite{CT99}). We aim at identifying features bearing on foundational issues of quantum mechanics and quantum field theory, as in the Unruh effect, which has generated many theory papers on Unruh-DeWitt detectors \cite{DeW79} and attracted many interesting experimental proposals for its detection with accelerated particles \cite{BL83}-\cite{SSH08}. Furthermore, by way of analog gravity models and considerations \cite{WU11, St16}, the present studies can provide a useful conduit for the investigation of fundamentally interesting effects in quantum field theory in curved spacetime (QFTCS)\cite{BD82}, such as cosmological particle creation \cite{Par69} and the Hawking effect \cite{Ha75}.   
 
With these motivations and our stated goals, we presented in Paper I a linearized effective theory for a Gaussian wavepacket description of a spinless, charged relativistic particle interacting with  quantum electromagnetic fields \cite{LH24}. Using techniques in QFTCS, we studied the interplay between single electrons and quantum fields in free space, at an energy scale well below the Schwinger limit \cite{Sc51}. In particular, we calculated the evolution of the purity of the Gaussian wavepacket of a single electron, and compared it with experimental data of electron interference in TEM. We found that our result {is broadly consistent with} the blurring of the interference pattern observed by Tonomura et al. \cite{TE89}, if we choose the initial state and 
parameter values according to the specification {of 
that} experiment.

In this paper, we turn our attention to the quantum radiation emitted by accelerated single electrons.  Our study encompasses issues somewhat known before, such as the cutoffs in the coincidence limit.  We want to remove the veil of mystery which some theories seem to carry, and show that the cutoffs just need to be chosen according to the experiments. We find that the secular growth in the corrections to the radiation is not of quantum origin, but has a classical interpretation. There are also some unexpected but important results, namely, the nonlinear (cubic) terms must be included in the action. Otherwise, our quantum theory will not be compatible with the well-established theory of classical radiation, even to the leading order approximation. These are the main theoretical findings of this work.  From a practical angle, our study addresses directly some experimental proposals to detect the Unruh effect in the quantum radiation emitted from (not experienced by) accelerated electrons (e.g. \cite{CT99}).

Quantum corrections to classical radiated power, or simply the ``quantum radiations", emitted by Unruh-DeWitt detectors in uniform linear acceleration \cite{RSG91, HR00, FOC06, LH06, IYZ11, GON25, WP26}, linear oscillatory motion \cite{Lin17}, and by electrons in linear acceleration \cite{SSH06, SSH08, OYZ16}, circular motion \cite{Sc54, Lin03c}, and in the non-relativistic limit \cite{HW09}, have been studied via various methods and approximations. Here we apply our effective theory introduced in ref. \cite{LH24} to consider a single-electron wavepacket at rest or in uniform linear acceleration. Notice that many studies of uniformly accelerated detectors are set in the Rindler space, 
because it offers a welcoming mathematical simplicity. However, such convenience cannot be easily carried over to a more realistic physical description. For example, 
in our present model of a wavepacket, the tail of our Gaussian wavepacket defined in the Minkowski coordinates apparently goes beyond the event horizon of the uniformly accelerated particle's mean trajectory {(similar to figure \ref{FigWLSet}), and so the Rindler frame would not help \footnote{This is by nature similar to the added effort one needs to put in for understanding the real time nonequilibrium dynamics of a system compared to an imaginary-time formulation of finite temperature description of systems in equilibrium, where attractive conditions like the Kubo-Martin-Schwinger condition and elegant theorems like the Bisognano-Wickmann theorem can be invoked.}: 
To keep the Rindler-time slices smooth, the direction of time in the opposite Rindler wedge where the tail of the wavepacket is supported will be opposite to the direction of the Minkowski time. Moreover, the null infinity of the Rindler frame is different from the null infinity of the Minkowski frame, so that the radiations defined in these two different coordinate systems may not agree \cite{Hi01, Hi02}.
Thus in this paper, we will always be working in the Minkowski frame, which is considered as our laboratory frame, though the proper time of the particle's mean trajectory can serve as a convenient working parameter in our calculation. The structural similarity between our effective theory and the Unruh-DeWitt detector theory still enables a natural incorporation of the Unruh effect into our result, as shown in section 4.2 of ref. \cite{LH24}.}

In the conceptual framework of open quantum systems, our attention in this paper is focused on the EM fields, which are thus treated as the system while the single-electron wavepacket as the environment. Our main task will be to calculate the expectation value of the stress-energy tensor of the EM fields in the presence of accelerated single electrons, and then determine the radiated power of the electrons in the far zone.  This is elaborated in section \ref{SecRadFormula}. 
It turns out that the cubic terms of the original nonlinear action must be considered in order to make a correct correspondence to the classical result given in appendix \ref{SecClRadUAC}.
We will see in section \ref{SecQRadRest} that the quantum radiation emitted by a single-electron wavepacket at rest vanishes exactly while the wavepacket is spreading. In section \ref{SecQRadUAC} we further show that the quantum radiation by a uniformly accelerated electron in a typical electron microscope is much smaller than its classical radiation, and dominated by those contributions \textit{irrelevant} to the Unruh effect. Thus, attempts to observe consequences of the Unruh effect in the quantum radiation of linearly accelerated electrons are practically untenable in a linear accelerator with electric field of the same order as in TEM. In section \ref{SecSumOut} we summarize our findings in this paper and give some final remarks and indications for further theoretical developments. Some selected explicit expressions for the formulas and coefficients we used in our calculation are collected in appendix \ref{ApxCoeff}. Finally, as a supplement to Paper I, the explicit leading-order result for the two-point correlator of the transverse deviations of the uniformly accelerated electron considered in ref. \cite{LH24} is given in appendix \ref{ApxzTzTF}.


\section{Radiated power of single electrons}
\label{SecRadFormula}

Consider {a charged, spinless relativistic particle (our `single electron') of mass $m$, charge $q$, in position $z_{}^i(t)$  at time $t$ and locally interacting with EM fields $A^\mu_{\bf x}(t) \equiv A^\mu_{}(t,{\bf x})$, described by the action
\begin{eqnarray}
  S &=& -mc^2 \int dt
    \sqrt{1-\frac{d z_{}^i}{dt} \frac{dz^{}_i}{dt} }+
  \int \frac{dt}{\mu^{}_0} \sum_{\bf x} \left[-\frac{1}{4}  F^{\bf x}_{\mu\nu} F_{\bf x}^{\mu\nu} -\frac{\bar{\alpha}}{2}\left( {\partial^{}_\mu} 
  A_{\bf x}^\mu\right)^2\right] \nonumber\\
 &&+ q \int dt \sum_{\bf x} \left[ c A^{\bf x}_0 +\frac{dz_{}^i}{dt} A^{\bf x}_i\right]\delta^3\big[{\bf x}-{\bf z}(t)\big]
\label{Stot}
\end{eqnarray}
as given by eqs. (2.1)-(2.3) in ref. \cite{LH24}, or (I.2.1)-(I.2.3) for short, with the choices of the Minkowski-time gauge for the particle, the Lorentz gauge for the EM fields, and the spacetime signature $(-+++)$. 
Here, $F^{}_{\mu\nu} \equiv \partial^{}_\mu A^{}_\nu- \partial^{}_\nu A^{}_\mu$ with $A^\mu=(\Phi/c, {\bf A})$ where $\Phi$ and ${\bf A}$ are the `scalar' and vector potentials, respectively, in the non-relativistic expressions of electromagnetism,  $\bar{\alpha}$ is an arbitrary constant, {and} $\mu^{}_0 = 4\pi \times 10^{-7} \, {\rm N}\cdot{\rm A}^{-2}$ is the vacuum permeability. As in Paper I, we denote 
$\bar{z}_{}^i(t)$ and $\bar{A}_{\bf x}^\mu(t)$ as the classical solutions, and $\tilde{z}_{}^i$ and $\tilde{A}_{\bf x}^\mu$ as the particle (electron) worldline deviations and field fluctuations, respectively, assumed to be perturbations. Inserting $z_{}^i = \bar{z}_{}^i + \tilde{z}_{}^i$ and $A_{\bf x}^\mu = \bar{A}_{\bf x}^\mu +\tilde{A}_{\bf x}^\mu$ into (\ref{Stot}), then expanding $S$ about the classical solutions into a series of the $\tilde{z}_{}^i$ and  $\tilde{A}^\mu_{\bf x}$, we would obtain approximated results by considering the leading orders of the series.  Up to the quadratic action (I.2.5), a linear effective theory has been studied in Paper I, where only the deviations and fluctuations are quantized ($\tilde{z}_{}^i \to \hat{z}_{}^i$ and $\tilde{A}^\mu_{\bf x} \to \hat{A}^\mu_{\bf x}$), and the associated quantum states are characterized by the correlators of $\hat{z}^i$ and $\hat{A}_{\bf x}^\mu$. Our effective theory is similar to the Unruh-DeWitt detector theory \cite{Unr76, DeW79, LH06, LH07, HH19}, particularly the local interaction around the classical worldline of the particle $\sim \delta^3 \big[ {\bf x}-\bar{\bf z}(t)\big]$.} 

\begin{figure}[ht]
\center{\includegraphics[width=7.6cm]{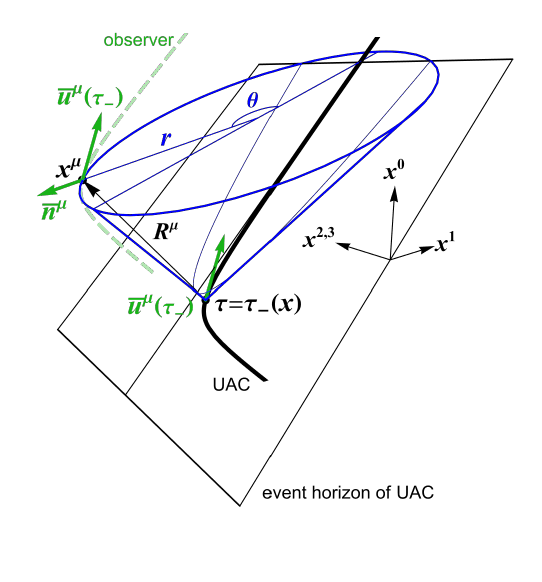}}
\caption{{Spacetime diagram of the setting for (\ref{dPdOmegaCl}) in \cite{Ro65} yielding the relativistic Larmor formula (\ref{ClLarmorF}). A uniformly accelerated charge (UAC) moves along the worldline $\bar{z}^\mu (\tau)$ marked in black, and a retardedly comoving observer's worldline is marked by a green dashed line. At the event $x^\mu$, the observer receives the radiation emitted by the UAC at $\tau=\tau^{}_-(x)$ alone the future lightcone of $\bar{z}_{}^\mu(\tau^{}_-)$ (blue), when the four-velocity of the UAC is $\bar{u}^\mu (\tau^{}_-)$ (green). Then in the instantaneous moving frame of the UAC at $\tau= \tau^{}_-$, $\bar{n}^\mu (\tau^{}_-)$ (green), $r$, and $\theta$ are determined by $R^\mu$ (black) in (\ref{Rmudef}), according to which the observer determines the radial direction and then the angular radiated power (\ref{dPdOmegaCl}).}}
\label{FIGsetting}
\end{figure}

To see the angular radiated power emitted by the electron with respect to its proper time in our model, we calculate
\begin{equation}
  \frac{d{P}}{d\Omega^{}_{\rm II}} \equiv -{\rm Re}\lim_{r\to \infty} r^2 \,
  \bar{u}_{}^\mu\big(\tau^{}_-(x)\big) 
  \big< \hat{T}^{}_{\mu\nu}(x) 
  \big>^{}_{ren}    \bar{n}_{}^{\nu}\big(\tau^{}_-(x)\big) \label{dPdOmega}
\end{equation} 
generalized from  {Rohrlich's classical formula \cite{Ro65}.} 
Here, $x$ is a shorthand for $x^\mu=(ct, {\bf x})$, the spacetime coordinates of the event of observation,
$\big< \hat{T}^{}_{\mu\nu}(x)\big>^{}_{ren}$ 
is the renormalized expectation value of the stress-energy tensor of the EM fields observed at $x$ {with respect to the quantum state of the electron worldline deviations and EM-field fluctuations}, $\bar{u}_{}^\mu (\tau)= d\bar{z}_{}^\mu/d\tau$ is the classical four-velocity of the electron at its proper time $\tau$ ($\bar{u}^{}_\mu \bar{u}_{}^\mu =-c^2$), the retarded time $\tau_-(x)$ is the solution to $\sigma\big(x,\bar{z}(\tau_-)\big)=0$ (similarly, $\bar{z}$ is a shorthand for $\bar{z}^\mu$) {and $x_{}^0 > \bar{z}_{}^0 (\tau^{}_-)$} with the Synge's world function $\sigma(x,x')\equiv -(x^{}_\mu-x'_\mu)(x_{}^\mu-x'{}^\mu)/2$, and the spatial distance $r(x)$ and the normalized spacelike four vector $\bar{n}_{}^\mu(\tau)$ are defined by the null vector ({see figure \ref{FIGsetting}; cf.} figure 7 in ref. \cite{LH06}, and appendix A in ref. \cite{OLHM12})
\begin{equation}
  R^\mu \equiv x^\mu - \bar{z}_{}^\mu\big(\tau^{}_-(x)\big)\equiv r(x)
\left[\bar{n}_{}^\mu\big(\tau^{}_-(x)\big)+ \frac{\bar{u}_{}^\mu\big(\tau^{}_-(x)\big)}{c}\right]
  \label{Rmudef}
\end{equation}
with the conditions $R_\mu R^\mu=0$, $\bar{u}_{}^\mu (\tau)\bar{n}_\mu^{}(\tau)=0$, and $\bar{n}_{}^\mu(\tau) \bar{n}_\mu^{}(\tau)=1$ \footnote{In refs. \cite{Ro65, LH06}, the four velocity and the associated normalized spacelike four-vector of the point particle or detector are denoted as $v_{}^\mu$ and $u_{}^\mu$, respectively. Here we denote them as $\bar{u}_{}^\mu$ and $\bar{n}_{}^\mu$ instead, since we have been denoting $\bar{v}_{}^\mu = d\bar{z}_{}^\mu/dt$ as the velocity vector with respect to $t$ and $\bar{u}_{}^\mu$ as the four velocity in ref. \cite{LH24}.}.

Note that the observers here are retardedly comoving with the electron, rather than at rest in the laboratory frame. $\bar{u}^\mu (\tau^{}_-(x))$ and $\bar{n}^\nu(\tau^{}_-(x))$ in (\ref{dPdOmega}) refer to the motion of the observer [according to the classical trajectory of electron at $\tau_-(x)$], whose uncertainty is assumed to be much lower than the single electron's. Thus we do not consider here the correlators with them such as $\langle \hat{u}_{}^\mu \hat{n}^\nu\rangle \bar{T}_{\mu\nu}$, $\langle \hat{u}_{}^\mu \hat{F}^{}_{\mu\rho}\rangle \bar{F}^\rho{}_\nu \bar{n}^\nu$, etc.
{Introducing such retarded comoving observers in (\ref{dPdOmega}) helps to obtain a classical relativistic generalization of the Larmor formula (\ref{ClLarmorF}), which is a constant of time, for a point-like uniformly accelerated charge (UAC), whose radiation reaction is zero \cite{Ro65, Bo80}.}

The classical stress-energy tensor of EM fields in the Lorentz gauge reads
\begin{equation}
  \bar{T}^{}_{\mu\nu}
  =-\frac{1}{\mu_0}\left[ \bar{F}_{\mu\rho}
     \bar{F}^\rho{}_\nu 
  + \frac{1}{4}\eta_{\mu\nu}\bar{F}^{}_{\rho\sigma}
  \bar{F}^{\rho\sigma}
  \right],
\end{equation}
where $\bar{F}_{}^{0j} = -\partial_0 \bar{A}^j - \partial_j \bar{A}^0 = E^j/c$ is the electric field and $\bar{F}^{}_{ji} = \partial_j \bar{A}_i - \partial_i \bar{A}_j = \epsilon^{}_{jik}B_{}^k$ is the magnetic field \cite{Gr17}. 
Then the classical part of the angular radiated power (\ref{dPdOmega}) is
\begin{equation}
  \frac{d\bar{P}}{d\Omega^{}_{\rm II}} =-\lim_{r\to \infty} r^2 \, 
  \bar{u}_{}^\mu\big(\tau^{}_-(x)\big)\bar{T}^{}_{\mu\nu}(x)\bar{n}_{}^{\nu}\big(\tau^{}_-(x)\big) 
  = \lim_{r\to \infty} \frac{r^2}{\mu_0} \bar{u}_{}^\mu\left(\tau^{}_-\right)\,\bar{F}_{\mu\rho}\bar{F}^\rho{}_\nu 
  \bar{n}_{}^{\nu}\left(\tau^{}_-\right),
  \label{dPdOmegaCl}
\end{equation}
since $\eta_{\mu\nu}\bar{u}^\mu(\tau_-)\bar{n}^\nu(\tau_-) =0$. 
Suppose a classical charged particle at its proper time $\tau^{}_-$ is instantaneously at rest at the origin of the observer's reference frame. Then one has
\begin{equation}
  -\bar{u}_{}^\mu(\tau^{}_-) \bar{T}_{\mu\nu}(x) \bar{n}^{\nu}(\tau^{}_-) = -c\,\bar{T}_{0i}\, \hat{r}_{}^i = 
  \frac{c}{\mu^{}_0} \bar{F}_{0j}\bar{F}^j{}_i \frac{x_{}^i}{|{\bf x}|} 
  = \frac{1}{\mu^{}_0}\big( \vec{E}\times \vec{B}\, \big)\cdot\hat{r} \label{ClTmn}
\end{equation}
 ($\hat{r}^i \equiv x_{}^i/|{\bf x}|$),
which is indeed the radial component of the Poynting vector observed at spacetime point $x$.

For a uniformly, linearly accelerated charge at proper acceleration $a=\sqrt{\bar{a}^{}_\rho\bar{a}_{}^\rho}$, inserting (\ref{UAC4vec})-(\ref{UAC4acc}) into (\ref{CluFFnGerenal}) and then (\ref{dPdOmegaCl}),
one obtains
\begin{equation}
  \frac{d\bar{P}}{d\Omega^{}_{\rm II}} = \frac{\mu^{}_0 q^2}{16 \pi^2 c} a^2 \sin^2\theta
  = \frac{3 s\bar{m}c^2}{8\pi}\alpha^2\sin^2\theta \label{ClassicaldRdOmega}
\end{equation}
with 
$\alpha\equiv a/c$ and 
\begin{equation}
s \equiv \frac{\mu^{}_0 q^2}{6\pi c\bar{m}} \approx 6.3\times 10^{-24} \,{\rm s} 
\end{equation}
($\bar{m}$ is the renormalized electron mass) as defined in (I.3.2) and (I.2.19), and so the radiated power with respect to the charge's proper time is
\begin{equation}
  \bar{P} =\int_0^{2\pi}d\varphi\int_0^\pi d\theta\sin\theta
  \frac{d\bar{P}}{d\Omega^{}_{\rm II}} = \frac{\mu^{}_0 q^2}{4 \pi c} \, \frac{2}{3}a^2
\label{ClLarmorF}
\end{equation}
which is a relativistic generalization of the Larmor formula in classical electrodynamics \cite{Ro65}.


In our effective theory, the symmetric stress-energy tensor operator $\hat{T}^{}_{\mu\nu}$ of the EM fields would be slightly modified from (\ref{ClTmn}).
{Generalizing the action 
(I.2.1)-(I.2.3) to curved spacetime ($\eta_{\mu\nu}\to g_{\mu\nu}$, $\partial_\mu \to \nabla_\mu$, and $d^4 x \to d^4 x \sqrt{-g}$) and performing the variation with respect to $g^{\mu\nu}(x)$,
letting $x_{}^\mu \not= z_{}^\mu(\tau)$ for all $\tau$, then going back to flat space and quantizing the fluctuations ($\tilde{A}_{ }^\mu \to \hat{A}_{}^\mu$), we obtain \cite{BD82}
\begin{eqnarray}
  \hat{T}_{\mu\nu}(x) &\equiv& \left. -\frac{2}{\sqrt{-g}}\frac{\delta S}{\delta g^{\mu\nu}(x)} 
  \right|_{g_{\mu\nu} \to \eta_{\mu\nu}} \nonumber\\
  &=& -\frac{1}{\mu_0}\bigg[ F_{\mu\rho} F^\rho{}_{\nu}  
  + \bar{\alpha}A_\mu \partial^{}_\nu\partial^{}_\rho A^\rho 
  + \bar{\alpha}A_\nu \partial^{}_\mu\partial^{}_\rho A^\rho \nonumber\\
  &&\hspace{.5cm}\left. + 
  \eta_{\mu\nu}\left( \frac{1}{4}F^{}_{\rho\sigma} F^{\rho\sigma} - 
  \frac{\bar{\alpha}}{2} \partial^{}_\rho A_{}^\rho\partial^{}_\sigma A_{}^\sigma -\bar{\alpha} 
  A_{}^\rho \partial^{}_\rho\partial^{}_\sigma A_{}^\sigma\right)\right] \label{QTmnop}
\end{eqnarray}
off the worldline of the relativistic charged particle, with $A_{}^\mu(x) =\bar{A}^\mu_{ \bf x}(t) + \hat{A}^\mu_{\bf x}(t)$ and $F^{}_{\mu\nu}(x) = \partial^{}_\mu A^{\bf x}_\nu(t) - \partial^{}_\nu A^{\bf x}_\mu(t) \equiv \bar{F}_{\mu\nu}^{\bf x}(t)+ \hat{F}_{\mu \nu}^{\bf x}(t)$. Here $\hat{T}_{\mu\nu}$ explicitly depends on the factor $\bar{\alpha}$ associated with the gauge-fixing term in (I.2.3).}

Inserting (\ref{QTmnop}) into (\ref{dPdOmega}), 
one can show that the gauge-fixing terms contribute no angular radiated power in the far zone,
and (\ref{dPdOmega}) reduces to 
\begin{eqnarray}
  \frac{d{P}}{d\Omega^{}_{\rm II}} &=& {\rm Re}
  \lim_{r\to \infty} \lim_{x'\to x} \frac{r^2}{\mu^{}_0} \, \bar{u}_{}^\mu(x)
  \langle \big[\bar{F}^{}_{\mu\rho}(x)+\hat{F}^{}_{\mu\rho}(x) \big]
  \big[\bar{F}_{}^{\rho\nu}(x')+\hat{F}_{}^{\rho\nu}(x')\big]\rangle^{}_{ren} \bar{n}^{}_\nu(x')
  \nonumber\\
  &=& \frac{d\bar{P}}{d\Omega^{}_{\rm II}} + \frac{d{P}^{}_{qm}}{d\Omega^{}_{\rm II}}
  + \frac{d{P}^{}_{mq}}{d\Omega^{}_{\rm II}}
  + \frac{d{P}^{}_{dd}}{d\Omega^{}_{\rm II}} \label{RenAngRadPower}
\end{eqnarray} 
with the `quadrupole-monopole' corrections
\begin{eqnarray}
\frac{d{P}^{}_{qm}}{d\Omega^{}_{\rm II}} &\equiv & {\rm Re}
  \lim_{r\to \infty} \lim_{x'\to x} \frac{r^2}{\mu^{}_0} \, \bar{u}_{}^\mu(x)
  \langle \hat{F}^{}_{\mu\rho}(x)\rangle^{}_{ren} \bar{F}_{}^{\rho\nu}(x') \bar{n}^{}_\nu(x'),
  \label{RenAngRadPowerQC}\\
\frac{d{P}^{}_{mq}}{d\Omega^{}_{\rm II}} &\equiv & {\rm Re}
  \lim_{r\to \infty} \lim_{x'\to x} \frac{r^2}{\mu^{}_0} \, \bar{u}_{}^\mu(x)
  \bar{F}^{}_{\mu\rho}(x)\langle \hat{F}_{}^{\rho\nu}(x')\rangle^{}_{ren} \bar{n}^{}_\nu(x'),
  \label{RenAngRadPowerCQ}
\end{eqnarray} 
and the `dipole-dipole' correction
\begin{equation}
  \frac{d{P}^{}_{dd}}{d\Omega^{}_{\rm II}} \equiv {\rm Re}
  \lim_{r\to \infty} \lim_{x'\to x} \frac{r^2}{\mu^{}_0} \, \bar{u}_{}^\mu(x)
  \langle \hat{F}^{}_{\mu\rho}(x)\hat{F}_{}^{\rho\nu}(x')\rangle^{}_{ren} \bar{n}^{}_\nu(x').
\label{RenAngRadPowerDD}
\end{equation} 
The above `monopole', `dipole', and `quadrupole' refer to the quantum multipole fields. The nomenclature will become clear later. 

{In this series of papers, we are working in the Heisenberg picture where the time evolution of the operators are calculated and their expectation values are taken with respect to the initial state of the combined system.}

In appendix \ref{SecClRadUAC}, we show that the classical correspondence of $d{P}_{qm}/d\Omega^{}_{\rm II} + d{P}_{mq}/d \Omega^{}_{\rm II}$ for a distribution of uniformly accelerated charges would be comparable in magnitude with $d{P}_{dd}/d \Omega^{}_{\rm II}$ in the long-time regime. Within the linear approximation of ref. \cite{LH24}, however, {the operators of dynamical variables will evolve to linear combinations of the operators defined at the initial moment $t^{}_0$, e.g., in (I.2.39),
\begin{eqnarray}
 \hat{z}_{}^i(t) &=& \sum_{j=1}^3 \left[ {\cal Z}^i_{z_{}^j}(t) \hat{z}_{}^j +{\cal Z}^i_{p^{}_j}(t) \hat{p}^{}_j \right] + \sum_{\bf k} \sum_{\lambda =0}^3 \left[ {\cal Z}^i_{(\lambda){\bf k}}(t) \hat{b}_{\bf k}^{(\lambda)}+{\cal Z}^{i*}_{(\lambda){\bf k}}(t) \hat{b}_{\bf k}^{ (\lambda)\dagger}\right],
 \label{zjExpand}
\end{eqnarray}
where ${\cal Z}^i_\Omega(t)$, $\Omega = z_{}^j, p^{}_j, (\lambda){\bf k}$, are mode functions with ${\cal Z}^i_{z_{}^j}(t^{}_0)=1$ while ${\cal Z}^i_{p^{}_j}(t^{}_0) = {\cal Z}^i_{(\lambda){\bf k}}(t^{}_0)=0$ \footnote{{Note that in ref. \cite{LH24} we quantize the combined system in the Minkowski coordinates, though it is more convenient to parametrize the classical worldline $\bar{z}_{}^\mu (\tau)$ of the relativistic particle in proper time $\tau$ than the Minkowski time $t$. Below when we write ${\cal Z}^i_\Omega(\tau)$, we mean ${\cal Z}^i_\Omega\big(\tau (t) \big)$, where $\tau(t)$ refers to the proper time when the corresponding classical worldline $\bar{z}^\mu(\tau)$ intersecting the Minkowski $t$-slice.}}. If we choose an initial state such that the expectation values of all the quantized dynamical variables defined at the initial moment are vanishing, namely, $\langle \hat{z}_{}^j\rangle = \langle \hat{p}^{}_j\rangle = \langle \hat{b}_{\bf k}^{ (\lambda)} \rangle =0$ such that $\langle A^{\mu}_{\bf x}(t^{}_0)\rangle = \bar{A}^{\mu}_{\bf x}(t^{}_0 )$, $\langle \pi^{\mu}_{\bf x}(t^{}_0)\rangle = \bar{\pi}^{\mu}_{\bf x}(t^{}_0)$, $\langle z_{}^i (t^{}_0) \rangle = \bar{z}_{}^i(t^{}_0)$ and $\langle \hat{p}_{}^i(t^{}_0) \rangle = \bar{p}_{}^i (t^{}_0)$, 
then we will have $\langle \hat{z}_{}^j(t) \rangle = \langle \hat{p}_{}^j(t) \rangle =\langle \hat{A}^{ \mu}_{\bf x}(t) \rangle = \langle \hat{\pi}^{ \mu}_{\bf x}(t)\rangle =0$ for all $t \ge t^{}_0$ because of (I.2.39-40), (I.2.42-43), and (I.2.54). This implies that} $\langle \hat{F}^{}_{\mu \nu}(x) \rangle = \partial^{}_\mu \langle \hat{A}^{\nu}_{ \bf x}(t) \rangle - \partial^{}_\nu\langle \hat{A}^{\mu}_{\bf x}(t) \rangle$ will always be zero, and the important components $d{P}_{qm}/ d\Omega^{}_{\rm II}$ in (\ref{RenAngRadPowerQC}) and $d{P}_{mq}/ d\Omega^{}_{\rm II}$ in (\ref {RenAngRadPowerCQ}) will be completely missed out in the result. To obtain the full leading-order quantum correction to radiated power, we have to go beyond the linear approximation with the quadratic action $S^{}_2$ in \cite{LH24}, as follows.

\subsection{Beyond the quadratic action}
\label{SecS3S2S1}

Expanding (I.2.1)-(I.2.3) beyond the quadratic action $S_2$ in (I.2.5), the next order is the cubic part of $S$, namely,
\begin{eqnarray}
  S_3 &=&\int dt L_3 \nonumber\\
  &=& \int dt\left\{ \frac{m \bar{\gamma}^3}{6 c^2}\bar{V}^{}_{ijk} 
\dot{\tilde{z}}_{}^i\dot{\tilde{z}}_{}^j \dot{\tilde{z}}_{}^k
    + q \sum_{\bf x} 
    \left[ \dot{\tilde{z}}_{}^i \tilde{z}_{}^j \partial^{}_j \tilde{A}_i^{\bf x} + 
    \frac{1}{2}\bar{v}_{}^\mu \tilde{z}_{}^i\tilde{z}_{}^j \partial^{}_i \partial^{}_j 
        \tilde{A}_\mu^{\bf x} + \right.\right. \nonumber\\ && \left. \left.   
    \frac{1}{2}\dot{\tilde{z}}_{}^k \tilde{z}_{}^i\tilde{z}_{}^j \partial^{}_i \partial^{}_j 
        \bar{A}_k^{\bf x} +
    \frac{1}{6}\bar{v}_{}^\mu \tilde{z}_{}^i\tilde{z}_{}^j 
    \tilde{z}_{}^{{k}} 
        \partial^{}_i \partial^{}_j \partial^{}_k \bar{A}_\mu^{\bf x} 
  \right] \delta^3\big( {\bf x}-\bar{\bf z}(t) \big) \right\}, \label{S3def}
\end{eqnarray}  
where $\dot{\tilde{z}}_{}^j(t) = \partial^{}_t \tilde{z}_{}^j(t)$ here, and
\begin{equation}
  \bar{V}_{ijk} \equiv \eta^{}_{ij}\bar{v}^{}_k +\eta^{}_{ki}\bar{v}^{}_j +
\eta^{}_{jk}\bar{v}^{}_i+3\frac{\bar{\gamma}^2}{c^2}\bar{v}^{}_i \bar{v}^{}_j \bar{v}^{}_k
\end{equation}
is a function of $t$ and symmetric. Let $S'=S^{}_1+S^{}_2+S^{}_3$. Then the conjugate momentum of particle worldline deviations becomes
\begin{eqnarray}
  && \tilde{p}^{}_i(t) = \frac{\delta S'}{\delta \dot{\tilde{z}}^i(t)} = \nonumber\\
  && m\left[ \bar{\gamma}\bar{M}_{ij}\dot{\tilde{z}}_{}^j+ \frac{\bar{\gamma}^3}{2c^2}\bar{V}_{ijk}
    \dot{\tilde{z}}_{}^j\dot{\tilde{z}}_{}^k \right]+
  q \left[ \tilde{A}_i^{\bar{\bf z}}+ \tilde{z}_{}^j\partial^{}_j\left( \bar{A}_i^{\bar{\bf z}}
  + \tilde{A}_i^{\bar{\bf z}}\right) +\frac{1}{2} 
  {\tilde{z}_{}^j\tilde{z}_{}^k\partial^{}_j \partial^{}_k}
  \bar{A}_i^{\bar{\bf z}} \right], \label{PiS3}
\end{eqnarray}
whereas the conjugate momentum of the fields $\tilde{\pi}^\mu_{\bf x}$ are the same as (I.2.26) and (I.2.27), and the form of $\hat{T}^{}_{\mu\nu}$ off the electron's worldline is the same as (\ref{QTmnop}). The Hamiltonian up to the cubic part of $S$ defined on the $t$-slice now reads
\begin{eqnarray}
  && \tilde{H}(t) = \tilde{p}_i\dot{\tilde{z}}^i + 
  \sum_{\bf x} \tilde{\pi}_{\bf x}^\mu \partial^{}_t \tilde{A}_\mu^{\bf x} -(L_2 +L_3) =\nonumber\\
  &&\sum_{\bf x} \left[ \frac{\mu^{}_0 c^2}{2}\left( \tilde{\pi}_i^{\bf x}\tilde{\pi}^i_{\bf x}
   + \frac{1}{\bar{\alpha}} \tilde{\pi}_0^{\bf x}\tilde{\pi}^0_{\bf x} \right)
   + c\tilde{\pi}_{}^i\partial^{}_i \tilde{A}_0^{\bf x} + c\tilde{\pi}^0\partial^{}_i
   \tilde{A}^i_{\bf x}+ \frac{1}{4\mu^{}_0} \tilde{F}_{ij}^{\bf x} \tilde{F}^{ij}_{\bf x}\right]  
   \nonumber\\ &&  -q\bar{v}_{}^\mu \left[ \tilde{z}_{}^i\partial^{}_i\tilde{A}_\mu^{\bar{\bf z}}
    + \frac{1}{2}\tilde{z}_{}^i\tilde{z}_{}^j\partial^{}_i\partial^{}_j \left(
    \bar{A}_\mu^{\bar{\bf z}}+\tilde{A}_\mu^{\bar{\bf z}}\right) +\frac{1}{6}\tilde{z}_{}^i
    \tilde{z}_{}^j\tilde{z}_{}^k\partial^{}_i\partial^{}_j\partial^{}_k 
    \tilde{A}_\mu^{\bar{\bf z}} \right]\nonumber\\ && + 
    \frac{\bar{M}^{ij}}{2m\bar{\gamma}}\left(\left\{ \tilde{p}^{}_i 
      -q\left[\tilde{A}_i^{{\bar{\bf z}}}+
    \tilde{z}_{}^k \partial^{}_k\left( \bar{A}_i^{\bar{\bf z}}+\tilde{A}_i^{\bar{\bf z}}\right)
    + \frac{1}{2} \tilde{z}_{}^k\tilde{z}_{}^{k'} \partial^{}_k \partial^{}_{k'}
    \bar{A}_i^{\bar{\bf z}} \right]\right\}\times\right. \nonumber\\ && \hspace{1.3cm}
    \left. \left\{ \tilde{p}^{}_j -q\left[\tilde{A}_j^{{\bar{\bf z}}}+
    \tilde{z}_{}^l \partial^{}_l\left( \bar{A}_j^{\bar{\bf z}}+\tilde{A}_j^{\bar{\bf z}}\right)
    + \frac{1}{2} \tilde{z}_{}^l\tilde{z}_{}^{l'} \partial^{}_l \partial^{}_{l'}
    \bar{A}_j^{\bar{\bf z}} \right]\right\} -\frac{m^2\bar{\gamma}^6}{4c^4}
     \bar{V}^{}_{ik k'}\bar{V}^{}_{jll'} \dot{\tilde{z}}_{}^k\dot{\tilde{z}}_{}^{k'}
     \dot{\tilde{z}}_{}^l\tilde{\hat{z}}_{}^{l'} \right)\nonumber\\ &&  
    -\frac{m\bar{\gamma}^3}{6c^2}\bar{V}_{ijk}
      \frac{\bar{M}_{}^{ia}}{m\bar{\gamma}}\left\{\tilde{p}^{}_a -
    q\left[\tilde{A}_a^{{\bar{\bf z}}}+
    \tilde{z}_{}^l \partial^{}_l\left( \bar{A}_a^{\bar{\bf z}}+\tilde{A}_a^{\bar{\bf z}}\right)
    + \frac{1}{2} \tilde{z}_{}^l\tilde{z}_{}^{l'} \partial^{}_l \partial^{}_{l'}
    \bar{A}_{{a}}^{\bar{\bf z}} 
    \right] - \frac{m\bar{\gamma}^3}{2c^2}\bar{V}^{}_{all'}
    \dot{\tilde{z}}_{}^l \dot{\tilde{z}}_{}^{l'} \right\} 
    \times \nonumber\\ && \hspace{1cm}
    \frac{\bar{M}_{}^{jb}}{m\bar{\gamma}}\left\{\tilde{p}^{}_b -
    q\left[\tilde{A}_b^{{\bar{\bf z}}}+
    \tilde{z}_{}^n \partial^{}_n\left( \bar{A}_b^{\bar{\bf z}}+\tilde{A}_b^{\bar{\bf z}}\right)
    + \frac{1}{2} \tilde{z}_{}^n\tilde{z}_{}^{n'} \partial^{}_n \partial^{}_{n'}
    \bar{A}_b^{\bar{\bf z}} \right] - \frac{m\bar{\gamma}^3}{2c^2}\bar{V}^{}_{bnn'}
    \dot{\tilde{z}}_{}^n \dot{\tilde{z}}_{}^{n'} \right\} 
    \times \nonumber\\ && \hspace{1cm}
    \frac{\bar{M}_{}^{kc}}{m\bar{\gamma}}\left\{\tilde{p}^{}_c -
    q\left[\tilde{A}_c^{{\bar{\bf z}}}+
    \tilde{z}_{}^p \partial^{}_p\left( \bar{A}_c^{\bar{\bf z}}+\tilde{A}_c^{\bar{\bf z}}\right)
    + \frac{1}{2} \tilde{z}_{}^p\tilde{z}_{}^{p'} \partial^{}_p \partial^{}_{p'}
    \bar{A}_c^{\bar{\bf z}} \right] - \frac{m\bar{\gamma}^3}{2c^2}\bar{V}^{}_{cpp'}
    \dot{\tilde{z}}_{}^p \dot{\tilde{z}}_{}^{p'} \right\} \nonumber\\ && = \cdots          
\end{eqnarray}
which will become an infinite series of $\tilde{p}^{}_i$, $\tilde{z}_{}^i$, and $\tilde{A}_i^{\bar{\bf z}}$, if we keep substituting $\dot{\tilde{z}}_{}^{{l}} 
=\frac{\bar{M}^{{li}}}
{m\bar{\gamma}}\left[ \tilde{p}^{}_i -\cdots -\frac{m\bar{\gamma}^3}{2c^2}\bar{V}^{}_{ijk}\dot{\tilde{z}}_{}^j\dot{\tilde{z}}_{}^{k}\right]$ iteratively according to (\ref{PiS3}).

Quantize the theory by promoting $\tilde{\cal O}$ to operators $\hat{\cal O}$ and introducing the equal-$t$ commutation relations (I.2.31) and (I.2.32), then the Heisenberg equation $\partial^{}_t \hat{\pi}^i_{\bf x} = \frac{i}{\hbar}[\hat{H}, \hat{\pi}^i_{\bf x} ]$ gives
\begin{equation} 
  \partial^{}_\mu \hat{F}^{\mu i}_{\bf x} + \bar{\alpha}\partial_{}^{{i}}
  \partial^{}_\mu\hat{A}^\mu_{\bf x} = 
  - \mu^{}_0 q \left[ \dot{\hat{z}}_{}^i - \left( \bar{v}_{}^i+
  \dot{\hat{z}}_{}^i \right)\hat{z}_{}^j\partial^{}_j + \frac{1}{2}\bar{v}^i 
  \hat{z}_{}^j\hat{z}_{}^k \partial^{}_j\partial^{}_k \right]\delta^3({\bf x}-\bar{\bf z})
  \label{AiELeq}
\end{equation}
by (I.2.26), and $\partial^{}_t \hat{\pi}^0_{\bf x} = \frac{i}{\hbar}[\hat{H}, \hat{\pi}^0_{\bf x} ]$ gives
\begin{equation}
  \left( \bar{\alpha}\partial^{}_0\partial_{}^0 +\partial^{}_i\partial_{}^i \right)
  \hat{A}^0_{\bf x} = - \mu^{}_0 q c \left[ -\hat{z}_{}^i \partial^{}_i + \frac{1}{2} 
  \hat{z}_{}^j\hat{z}_{}^k \partial^{}_j\partial^{}_k \right]\delta^3({\bf x}-\bar{\bf z})
  \label{A0ELeq}
\end{equation}
by (I.2.27). In the Lorentz-Feynman gauge $\bar{\alpha}=1$, we have
\begin{equation}
  \Box \hat{A}^\mu_{\bf x}(t) = -\mu^{}_0 q 
  \left[ \dot{\hat{z}}_{}^\mu -\left( \bar{v}_{}^\mu + \dot{\hat{z}}_{}^\mu  
    \right) \hat{z}_{}^j\partial^{}_j + \frac{\bar{v}_{}^\mu}{2}\hat{z}_{}^i\hat{z}_{}^j
    \partial^{}_i\partial^{}_j \right] \delta^3\big({\bf x}-\bar{\bf z}(t)\big)
   \label{BoxAmu}
\end{equation}
from (\ref{AiELeq}) and (\ref{A0ELeq}) with $\hat{u}^0 \equiv 0$. Taking the expectation values of both sides of the above equation {with respect to the initial state}, 
we obtain
\begin{equation}
  \Box \langle\hat{A}^\mu_{\bf x}(t)\rangle = -\mu^{}_0 q 
  \left[ \langle\dot{\hat{z}}_{}^\mu\rangle - \left( \bar{v}_{}^\mu \langle\hat{z}_{}^j\rangle+ 
    \langle\dot{\hat{z}}_{}^\mu  \hat{z}_{}^j\rangle\right) \partial^{}_j + 
    \frac{\bar{v}_{}^\mu}{2}\langle\hat{z}_{}^i\hat{z}_{}^j\rangle
    \partial^{}_i\partial^{}_j \right] \delta^3\big({\bf x}-\bar{\bf z}(t)\big),
   \label{BoxExpectAmu}
\end{equation}
which is of the same nature as in the Ehrenfest theorem.

Denoting the operators of the dynamical variables in the linear approximation in ref. \cite{LH24} as $\hat{\cal O}_{{}_{\underline{\texttt{0}}}}$. From the linear part of (\ref{BoxAmu}) [quantized (I.2.29) and (I.2.30) in the Feynman gauge $\bar{\alpha}=1$], we have
\begin{equation}
  \hat{A}_{{}_{\underline{\texttt{0}}}\bf x}^\mu(t) = 
  \hat{A}^{[0]\mu}_{{}_{\underline{\texttt{0}}}\bf x}(t) + 
  \hat{A}^{[1]\mu}_{{}_{\underline{\texttt{0}}}\bf x}(t),
\end{equation}
with the homogeneous solution (free-field operator) $\hat{A}^{[0]\mu}_{{}_{\underline{\texttt{0}}}\bf x}(t)$ and the inhomogeneous solutions 
\begin{eqnarray}
  \hat{A}^{[1]0}_{{}_{\underline{\texttt{0}}}\bf x}(t) &=& -\frac{\mu^{}_0 q c}{4\pi}\partial^{}_j 
    \left[ \frac{c \hat{z}^j_{{}_{\underline{\texttt{0}}}}\big(\tau_-(x)\big)}{R(x)}\right], 
     \label{A01Om} \\
  \hat{A}^{[1]i}_{{}_{\underline{\texttt{0}}}\bf x}(t) &=&  \frac{\mu^{}_0 q c}{4\pi}\left\{
    \frac{\partial_{\tau_-}\hat{z}^i_{{}_{\underline{\texttt{0}}}}\big(\tau_-(x)\big)}
    {\bar{\gamma}^{}_-(x) R(x)}-
    \partial^{}_j\left[\frac{\bar{v}_{-}^i(x)
    \hat{z}^{{j}}_{{}_{\underline{\texttt{0}}}} 
    \big(\tau_-(x)\big)}{R(x)} \right]\right\},  \label{Ai1Om}
\end{eqnarray}
where $\tau^{}_-(x)$ is the retarded proper time determined by the observation point $x$ \cite{LH06}, $\bar{v}^j(t) = d\bar{z}^j(t)/dt$ is the three-velocity of the classical electron with respect to $t$ and $\bar{v}^j_-(x) \equiv \bar{v}^j\big( \tau_-(x) \big)$ for short, $\bar{\gamma}^{}_- \equiv 1/\sqrt{1-|\bar{\bf v}(\tau^{}_-)/c|^2}$ is the Lorentz factor of the electron at $\tau=\tau^{}_-$, and 
\begin{equation}
    R(x) \equiv \left| \frac{d\sigma(x,x')}{d t'}
    \right|_{\sigma(x,x')=0, \,\, x'^\mu=\bar{z}^\mu(\tau^{}_-(x)).} \label{Rdef}
\end{equation}
Here, $\sigma(x, x')=-(x^{}_\mu-x'_\mu)(x^\mu-x'_{}{}^\mu)/2$ is Synge's world function,
which yields 
\begin{equation}
  R(x)=\Big| \bar{v}^{}_\mu(t^{}_-)
           \left( x_{}^\mu -\bar{z}_{}^\mu(t^{}_-) \right) \Big|
      = \left| \frac{\bar{u}^{}_\mu(\tau_-)}{\bar{\gamma}^{}_-} R^\mu \right| 
      = \frac{c r(x)}{\bar{\gamma}^{}_-(x)} \label{Rxcrgamma}
\end{equation} 
with $R^\mu$ defined in (\ref{Rmudef}), $t^{}_- \equiv \bar{z}_{}^0(\tau^{}_-)/c$, and $\bar{u}_{}^\mu(\tau^{}_-) = \partial^{}_{\tau^{}_-} \bar{z}_{}^\mu(\tau^{}_-) = \bar{\gamma}^{}_- \bar{v}_{- }^\mu$. Thus (\ref{A01Om}) and (\ref{Ai1Om}) can be 
combined and written as
\begin{equation} 
\hat{A}^{[1]\mu}_{{}_{\underline{\texttt{0}}}\bf x}(t) =  
\frac{\mu^{}_0 q}{4\pi}\left\{\frac{\dot{\hat{z}}^\mu_{{}_{\underline{\texttt{0}}}}(\tau_-)}{r}
-\partial^{}_j \left[ \frac{\bar{u}_{}^\mu(\tau_-)}{r}
\hat{z}^j_{{}_{\underline{\texttt{0}}}}(\tau_-)\right]\right\} \label{Amu1Om}
\end{equation}
with $\dot{\hat{z}}^i_{{}_{ \underline{\texttt{0}}}}(\tau^{}_-) \equiv \partial^{}_{ \tau^{}_-}\hat{z}^i_{{}_{\underline{\texttt{0}}}}(\tau^{}_-)$ and $\hat{z}^0_{{}_{\underline{ \texttt{0}}}}(\tau) \equiv 0$ for all $\tau$. Here $\hat{A}^{[1]\mu}_{{}_{\underline{ \texttt{0}}}\bf x}(t)$ is analogous to a dipole field as the classical position of the electron $\bar{z}^j$ is the origin of the spatial deviations $\hat{z}^j_{{}_{ \underline{\texttt{0}}}}$ \cite{Gr17,Ja99,Fr05,Sc98}. Writing
\begin{eqnarray}
\hat{F}^{\mu\nu}_{{}_{\underline{\texttt{0}}}\bf x}(t) = 
\hat{F}^{[0]\mu\nu}_{{}_{\underline{\texttt{0}}}\bf x}(t)+
\hat{F}^{[1]\mu\nu}_{{}_{\underline{\texttt{0}}}\bf x}(t), 
  \label{F0F1}
\end{eqnarray}
where $\hat{F}^{[B]\mu\nu}_{ {}_{\underline{\texttt{0}}}\bf x}\equiv\partial_{}^\mu \hat{A}^{[B]\nu}_{{}_{\underline{\texttt{0}}}\bf x}- \partial_{}^\nu \hat{A}^{[B]\mu}_{{}_{\underline{\texttt{0}}}\bf x}$, $B=0,1$, we have
\begin{equation}
\hat{F}_{{}_{\underline{\texttt{0}}}\mu\nu}^{[1]\bf x} = \frac{\mu^{}_0 q}{4\pi}\left\{
\partial^{}_\mu\left[\frac{\dot{\hat{z}}^{}_{{}_{\underline{\texttt{0}}}\nu}(\tau_-)}{r}\right]- 
  \partial^{}_\mu\partial^{}_j \left[\frac{\bar{u}^{}_\nu(\tau_-)}{r} \hat{z}^j_{{}_{ \underline{\texttt{0}}}}(\tau_-)\right] -(\mu \leftrightarrow \nu) \right\}
\label{F1mndi}
\end{equation}
from (\ref{Amu1Om}).

\subsection{Quadrupole-monopole corrections}
\label{subsecRmq}

As we mentioned, if the expectation values $\langle \hat{A}^{\mu}_{{}_{ \underline{ \texttt{0}}} \bf x}(t) \rangle$, $\langle \hat{\pi}^{\mu}_{{}_{\underline{ \texttt{0}}}\bf x}(t)\rangle$, $\langle \hat{z}^i_{{}_{\underline{ \texttt{0}}}}(t) \rangle$, and $\langle \hat{p}^i_{{}_{\underline{\texttt{0}}}}(t) \rangle$ in linear approximation are vanishing at the initial moment $t=t^{}_0$, they will be vanishing for all $t \ge t^{}_0$. However, when the nonlinear terms in (\ref{BoxExpectAmu}) are taken into account, $\langle\hat{A}^\mu_{\bf x}(t)\rangle \sim \langle\hat{z}^i(t)\hat{z}^j(t)\rangle$ become nonzero in general after the coupling is switched on. Write $\hat{\cal O} = \sum_{{\underline{\texttt{n}}}=0}^\infty \hbar_{}^n \hat{\cal O}^{}_{{}_{\underline{\texttt{n}}}}$ with $\langle \hat{\cal O}^{}_{{}_{\underline{\texttt{0}}}}\hat{\cal O}^{}_{{}_{\underline{\texttt{0}}}}\rangle \sim O(\hbar)$ in mind.
To the first order approximation, we may take the operators on the right hand side of (\ref{BoxAmu}) as $\hat{\cal O}_{{}_{\underline{\texttt{0}}}}$, and assume $\langle \hat{A}^{\mu}_{{}_{\underline{\texttt{0}}}{0}\bf x}(t^{}_0) \rangle =\langle \hat{\pi}^{\mu}_{{}_{\underline{\texttt{0}}}\bf x}(t^{}_0) \rangle = \langle \hat{z}^i_{{}_{\underline{\texttt{0}}}}(t^{}_0) \rangle = \langle \hat{p}^i_{{}_{\underline{\texttt{0}}}}(t^{}_0) \rangle = 0$ as we did in ref. \cite{LH24}. Then (\ref{BoxExpectAmu}) reduces to
\begin{equation}  
\Box \langle\hat{A}^\mu_{{}_{\underline{\texttt{1}}}{\bf x}}(t)\rangle = -\mu^{}_0 q \left[
-\langle \dot{\hat{z}}_{{}_{ \underline{\texttt{0}}}}^\mu
\hat{z}_{{}_{\underline{\texttt{0}}}}^j\rangle 
    \partial^{}_j + \frac{\bar{v}_{}^\mu}{2}\langle\hat{z}_{{}_{\underline{\texttt{0}}}}^i
\hat{z}_{{}_{\underline{\texttt{0}}}}^j\rangle \partial^{}_i\partial^{}_j 
   \right]\delta^3\big({\bf x}-\bar{\bf z}(t)\big). \label{BoxExpectAmu1st}
\end{equation}
Solving the above equation, we find
\begin{equation}
  \langle\hat{A}^\mu_{\bf x}(t)\rangle \approx
\langle\hat{A}^\mu_{{}_{\underline{\texttt{1}}}\bf x}(t)\rangle = 
\langle\hat{A}^{[0]\mu}_{{}_{\underline{\texttt{1}}}\bf x}(t)\rangle + \langle\hat{A}^{[1]\mu}_{{}_{ \underline{\texttt{1}}}\bf x}(t)\rangle, 
\end{equation}
where $\langle\hat{A}^{[0]\mu }_{{}_{\underline{\texttt{1}}}\bf x}(t)\rangle$ is the homogeneous solution to (\ref{BoxExpectAmu1st}), and
\begin{equation} 
\langle\hat{A}^{[1]\mu}_{{}_{\underline{\texttt{1}}}\bf x}(t)\rangle =
  \frac{\mu^{}_0 q}{4\pi}\left\{ 
    -\partial^{}_i \left[ \frac{1}{r}\langle\dot{\hat{z}}^\mu_{{}_{\underline{\texttt{0}}}}(\tau_-)
\hat{z}^i_{{}_{\underline{\texttt{0}}}}(\tau_-) \rangle \right]
+ \partial^{}_i\partial^{}_j\left[ \frac{\bar{u}^\mu(\tau_-)}{2r} 
\langle \hat{z}_{{}_{\underline{ \texttt{0}}}}^i(\tau_-)
\hat{z}_{{}_{\underline{\texttt{0}}}}^j(\tau_-)\rangle \right]\right\},
\end{equation} 
is the inhomogeneous solution which could be interpreted as a quadrupole field \cite{Gr17,Ja99,Fr05,Sc98}. 
One can see that the nonlinear terms make the expectation values of $\hat{A}^\mu$ (and $\hat{z}^i$) gradually deviating from the classical field configuration (and trajectories), though the initial values of the {expected deviations} are zero. 

Choose $\langle\hat{A}^{[0]\mu }_{{}_{\underline{\texttt{1}}}\bf x}(t)\rangle=0$ to match the initial condition $\langle\hat{A}^{\mu}_{\bf x}(t^{}_0)\rangle = 0$. Then the first-order corrected expectation values of the EM fields read
\begin{eqnarray}
&&\langle\hat{F}_{\mu\nu}^{\bf x}\rangle \approx 
\langle\hat{F}_{{}_{\underline{\texttt{1}}}\mu\nu}^{\bf x}\rangle =
  \partial^{}_\mu \langle\hat{A}^\nu_{{}_{\underline{\texttt{1}}}\bf x}(t)\rangle -
  \partial^{}_\nu \langle\hat{A}^\mu_{{}_{\underline{\texttt{1}}}\bf x}(t)\rangle =
  -\frac{\mu^{}_0 q}{4\pi}\times \nonumber\\ &&
  \left\{ \partial^{}_\mu\partial^{}_i 
    \left[ \frac{1}{r}\langle\dot{\hat{z}}_{{}_{\underline{\texttt{0}}}\nu}(\tau_-)
\hat{z}^i_{{}_{\underline{\texttt{0}}}}(\tau_-) \rangle \right]
  - \partial^{}_\mu \partial^{}_i\partial^{}_j\left[ \frac{\bar{u}_\nu(\tau_-)}{2r} 
    \langle \hat{z}_{{}_{\underline{\texttt{0}}}}^i(\tau_-)
\hat{z}_{{}_{\underline{\texttt{0}}}}^j(\tau_-)\rangle \right]\right\} - 
    (\mu\leftrightarrow\nu),
\end{eqnarray}
or
\begin{eqnarray}
  \langle\hat{F}_{\mu\nu}^{\bf x}\rangle &\approx& -\frac{\mu^{}_0 q}{4\pi} 
  \left\{ {\sf D}^{}_{\mu i}
\langle\dot{\hat{z}}_{{}_{\underline{\texttt{0}}}\nu}(\tau_-)
\hat{z}^i_{{}_{\underline{\texttt{0}}}}(\tau_-) \rangle 
  -{\sf D}^{}_{\mu\nu ij} \langle \hat{z}_{{}_{\underline{\texttt{0}}}}^i(\tau_-)
\hat{z}_{{}_{\underline{\texttt{0}}}}^j(\tau_-)\rangle\right\} -(\mu\leftrightarrow\nu),
\end{eqnarray}
where 
\begin{eqnarray}
  &&{\sf D}^{}_{\mu i} = \frac{\tau^{}_{-,\mu}\tau^{}_{-,i}}{r}\,\partial^2_{\tau^{}_-} + 
    \left(\frac{\tau^{}_{-,i\mu}}{r}
-\frac{r^{}_{,\mu}\tau^{}_{-,i}+ r^{}_{,i}\tau^{}_{-,\mu}}{r^2}\right)\partial^{}_{\tau^{}_-} + \frac{2r^{}_\mu r^{}_{,i}-r r^{}_{,\mu i}}{r^3},\\
&&{\sf D}^{}_{\mu\nu ij} = \frac{\bar{u}^{}_\nu(\tau^{}_-)}{2r}
  \tau^{}_{-,\mu}{{\tau^{}_{-,i}}}\tau^{}_{-,j}\partial^3_{\tau^{}_-} 
+ \cdots    
+ \left[\frac{\bar{u}^{}_{\nu}(\tau^{}_-)}{2r}\right]^{}_{,\mu ij} .
\end{eqnarray}
One can keep working out the derivatives of $\bar{u}^{}_\nu(\tau^{}_-)/r$, then use the formulas in appendix \ref{ApxCoeff} to convert the derivatives of $\tau^{}_-$ and $r$ to functions of $\bar{n}_\mu$, $\bar{u}^{}_\mu$, and their $\tau$-derivatives. 
Inserting the result into (\ref{RenAngRadPowerQC}) and (\ref{RenAngRadPowerCQ}), then combining (\ref{ClFmunuN}) and (\ref{ClUFmunu}), we obtain 
\begin{eqnarray}
&&\frac{d{P}^{}_{qm}}{d\Omega^{}_{\rm II}} =
    \frac{d{P}^{}_{mq}}{d\Omega^{}_{\rm II}} \approx 
\lim_{r\to\infty}\frac{\mu^{}_0 q^2}{16\pi^2} 
    \big( \bar{a}_\nu^- - \bar{a}^{}_{\bar{n}}\bar{n}_\nu^- \big)\frac{R^-_j}{r} \times
    \nonumber\\ && \left\{ \left[ \frac{1}{c_{}^2}\partial_{\tau^{}_-}^2+
    \frac{3\bar{a}^-_{\bar{n}}}{c^3}\partial^{}_{\tau^{}_-} +
     \frac{\dot{\bar{a}}^-_\rho\bar{n}^\rho_-}{c^3}+
       \frac{3(\bar{a}^-_{\bar{n}})^2-\bar{a}_\rho^- \bar{a}^\rho_-}{c^4} \right]
\langle \dot{\hat{z}}^{\nu-}_{{}_{\underline{\texttt{0}}}}, \hat{z}^{j-}_{{}_{\underline{\texttt{0}}}}\rangle\right. \nonumber\\
  && +\frac{R^-_i}{2r}\left[ \frac{3\bar{a}^\nu_-}{c^3}\partial_{\tau^{}_-}^2 + 
     3\left( \frac{\dot{\bar{a}}^\nu_-}{c^3}+ 
       \frac{4\bar{a}_{\bar{n}}^-\bar{a}^\nu_-}{c^4}\right)\partial^{}_{\tau^{}_-}
       + \frac{\ddot{\bar{a}}^\nu_-}{c^3} + \frac{6\bar{a}_{\bar{n}}\dot{\bar{a}}_-^\nu}{c^4}
  \right. \nonumber\\ && \hspace{2cm} \left.\left. +
\left(\frac{4\dot{\bar{a}}_\rho^- \bar{n}_-^\rho}{c^4} +
\frac{15(\bar{a}_{\bar{n}}^-)^2-4\bar{a}_\rho^-\bar{a}^\rho_-}{c^5} \right)\bar{a}_-^\nu
     \right] \langle \hat{z}^{i-}_{{}_{\underline{\texttt{0}}}}, \hat{z}^{j-}_{{}_{\underline{\texttt{0}}}}\rangle \right\}, 
\label{dPmqdOmegaResult}
\end{eqnarray}
where $\langle A(\tau), B(\tau)\rangle \equiv \lim_{\tau'\to\tau}\langle \{A(\tau), B(\tau')\}\rangle = \lim_{\tau'\to \tau} \frac{1}{2}\big< A(\tau)B(\tau')+ B(\tau')A(\tau) \big>$, $\bar{a}_{}^\mu(\tau_-) \equiv \partial^{}_{\tau^{}_-} \bar{u}_{}^\nu(\tau^{}_-)$ is the classical four-acceleration of the electron,
$C_{}^-$ or $C^{}_-$ denotes $C^{}(\tau_-)$ for $C=\hat{z}^{i}_{{}_{\underline{\texttt{0}}}}, \bar{n}^{}_\mu, \bar{u}^{}_\mu, \bar{a}^{}_\mu$, 
$R_j^-/r = (\bar{n}_j^- + \bar{u}_j^-/c)$ from (\ref{Rmudef}), $\bar{a}^{}_{\bar{n}}\equiv \bar{a}_\mu^- \bar{n}^\mu_-$,
and we have used $\big( \bar{a}_\nu^- - \bar{a}^{}_{\bar{n}}\bar{n}_\nu^- \big)\bar{u}_-^\nu = \big( \bar{a}_\nu^- - \bar{a}^{}_{\bar{n}}\bar{n}_\nu^- \big)\bar{n}_-^\nu =0$.

{As in ref. \cite{LH24}, we assume the initial state of the combined system at $t=t_0$ to be $\rho^{\,{}_{\bf I}} = \rho_{{}_P}^{\,{}_{\bf I}} \otimes \rho_{{}_F}^{\,{}_{\bf I}}$, which is a direct product of a Gaussian state $\rho_{{}_P}^{\,{}_{\bf I}}$ of electron worldline  deviations peaked around its classical trajectory} \footnote{{The Wigner representation of our initial particle state $\rho_{{}_P}^{\,{}_{\bf I}}$ has the form $W_{{}_P}^{\,{}_{\bf I}} = {\cal N} \exp -\{G^{zz}_{ij} \tilde{z}_{}^i \tilde{z}_{}^j + G^{zp}_{ij} \tilde{z}_{}^i \tilde{p}_{}^j +G^{pp}_{ij} \tilde{p}_{}^i \tilde{p}_{ }^j\} \equiv {\cal N} \exp -{\bf Q}^\dagger {\bf G} {\bf Q} $ with momentum $\tilde{p}_{}^i$ conjugate to the deviation $\tilde{z}_{ }^i$. Here ${\cal N}$ is the normalization constant, ${\bf Q} = ({\bf z},{\bf p})$, and ${\bf G}$ is the inverse of the $6\times 6$ covariance matrix, whose elements are
initial equal-time symmetric two-point correlators $\langle\hat{z}^i, \hat{z}^j\rangle^{}_{\rm I}$, $\langle\hat{z}^i, \hat{p}^j \rangle^{}_{\rm I}$, and $\langle\hat{p}^i, \hat{p}^j \rangle^{}_{\rm I}$.}},
{and the Minkowski vacuum state $\rho_{{}_F}^{\,{}_{\bf I}} = |0^{}_M\rangle \langle 0^{}_M |$ of EM fields, which is also Gaussian.
Then from (I.2.39) or (\ref{zjExpand}), the equal-time symmetric two-point correlators of electron worldline deviation read 
\begin{equation}
\langle\hat{z}^j_{{}_{\underline{\texttt{0}}}}\big(\tau(t)\big), \hat{z}^{j'}_{{}_{ \underline{\texttt{0}}}}\big(\tau(t)\big)\rangle \equiv {\rm Tr}\, \left[  \rho_{}^{\,{}_{\bf I}} \, \{ \hat{z}^j_{{}_{\underline{\texttt{0}}}}(\tau), \hat{z}^{j'}_{{}_{\underline{\texttt{0}}}}(\tau)\} \right]
=\langle\hat{z}^j_{{}_{\underline{\texttt{0}}}}(\tau), \hat{z}^{j'}_{{}_{\underline{\texttt{0}}}}(t)\rangle^{}_P + \langle\hat{z}^j_{{}_{\underline{\texttt{0}}}}(\tau),\hat{z}^{j'}_{{}_{\underline{\texttt{0}}}}(\tau)\rangle^{}_F , \label{zizjLinear}
\end{equation}
where the $P$-part 
\begin{eqnarray}
\langle\hat{z}^j_{{}_{\underline{\texttt{0}}}}(\tau), \hat{z}^{j'}_{{}_{\underline{\texttt{0}}}}(\tau)\rangle^{}_P
&\equiv&  {\rm Tr}\,\left[ \rho_{{}_P}^{\,{}_{\bf I}} \sum_{l, l'} \left\{ \left( {\cal Z}^j_{z_{}^l}(\tau) \hat{z}_{}^l + {\cal Z}^j_{p^{}_l} (\tau) \hat{p}^{}_l \right), \left( {\cal Z}^{j'}_{z_{}^{l'}}(\tau)\hat{z}_{}^{l'} + {\cal Z}^{j'}_{p^{}_{l'}}(\tau) \hat{p}^{}_{l'} \right) \right\} \right] \nonumber\\ 
&=& \sum_{l, l'} \left[  \langle \hat{z}_{}^l, \hat{z}_{}^{l'} \rangle^{}_{\rm I} \, {\cal Z}^j_{z_{}^l}(\tau){\cal Z}^{j'}_{z_{}^{l'}}(\tau)+    \langle \hat{p}_{}^l, \hat{p}_{}^{l'} \rangle^{}_{\rm I} \, {\cal Z}^j_{p_{}^l}(\tau){\cal Z}^{j'}_{p_{}^{l'}}(\tau) \right. + \nonumber\\ && \hspace{0.85cm} \left.    \langle \hat{z}_{}^l, \hat{p}_{}^{l'} \rangle^{ }_{\rm I}\, {\cal Z}^j_{z_{}^l}(\tau){\cal Z}^{j'}_{p_{}^{l'}}(\tau)+\langle \hat{p}_{}^l, \hat{z}_{}^{l'} \rangle^{}_{ \rm I}\, {\cal Z}^j_{p_{}^l}(\tau){\cal Z}^{j'}_{z_{}^{l'}}(\tau)\right]
\label{zizjPLinear}
\end{eqnarray}
with $\langle \hat{\cal O}\rangle^{}_{\rm I} \equiv 
{\rm Tr}\, \big[\rho^{\,{}_{\bf I}}_{{}_P}  \hat{\cal O}\big]$ depends on the initial state of the particle $\rho^{\,{}_{\bf I}}_{{}_P}$ only, and the $F$-part
\begin{eqnarray}
&&\langle\hat{z}^j_{{}_{\underline{\texttt{0}}}}(\tau), \hat{z}^{j'}_{{}_{\underline{\texttt{0}}}}(\tau)\rangle^{}_F  \equiv\lim_{\tau'\to \tau} 
\sum_{\bf k} \sum_{\bf k'} \frac{1}{2} \times\nonumber\\&&
\hspace{1cm}
\left({\cal Z}^j_{(\lambda){\bf k}}(\tau) {\cal Z}^{j'*}_{(\lambda'){\bf k'}}(\tau') + {\cal Z}^{j'}_{ (\lambda){\bf k}}(\tau') {\cal Z}^{j*}_{(\lambda'){\bf k'}}(\tau) \right) \langle 0^{}_M| \hat{b}^{(\lambda)}_{\bf k} \hat{b}^{(\lambda')\dagger}_{\bf k'} |0^{}_M\rangle \label{zizjFLinear} 
\end{eqnarray}
depends on the initial state of the field $\rho_{{}_F}^{\,{}_{\bf I}}$ only [cf. (I.3.12-14)]. In (\ref{zizjFLinear}), $\sum_{\bf k} \equiv \int\frac{d^3 k}{(2\pi)^3} \sqrt{\frac{\hbar}{2\omega\varepsilon^{}_0}}$, $\langle 0^{}_M| \hat{b}^{(\lambda)}_{\bf k} |0^{}_M\rangle = 0$, and
$\langle 0^{}_M| \hat{b}^{ (\lambda)}_{\bf k} \hat{b}^{ (\lambda')\dagger}_{\bf k'} |0^{}_M\rangle = (2\pi)^3 \eta^{(\lambda)(\lambda')} \delta^3({\bf k}-{\bf k}')$. Here and below the use of Einstein notation for the polarization indices, namely, $A^{}_{(\lambda)} B_{}^{ (\lambda)} = \sum_{\lambda, \lambda'=0}^3 \eta_{}^{ (\lambda)(\lambda')} A^{}_{ (\lambda)} B^{}_{(\lambda')}$, is understood. Combining the dependence of the initial vacuum state with the fact that ${\cal Z}^j_{(\lambda){\bf k}}$ is driven by free field modes ${\cal F}^{[0](\lambda){\bf k}}_{i\mu} \big(t,\bar{\bf z}(t)\big)$ [see (I.2.52)], the above $F$-part correlators can be interpreted as generated by vacuum fluctuations of the EM fields, or in the context of open quantum systems, by the environment when electrons are considered as the system of interest in \cite{LH24}.
In (\ref{dPmqdOmegaResult}), $\langle \dot{\hat{z}}^\nu_{ {}_{\underline{\texttt{0}}}}(t),\hat{z}^j_{{}_{\underline{\texttt{0}}}}(t)\rangle$ splits in a similar way with the factorized initial state $\rho^{\,{}_{\bf I}} = \rho_{{}_P}^{\,{}_{\bf I}} \otimes \rho_{{}_F}^{\,{}_{\bf I}}$.} The partial quantum angular radiated power (\ref{dPmqdOmegaResult}) by a single electron at rest or in uniform acceleration can then be obtained simply by substituting the results for the electron correlators obtained in Paper I.

\subsection{Dipole-dipole correction}
\label{subsecRdd}

Eq. (\ref{F1mndi}) can be rewritten in the form
\begin{equation}
\hat{F}^{[1]}_{{}_{\underline{\texttt{0}}}\,\mu\nu}(x) \equiv 
   \frac{\mu^{}_0 q}{4\pi}\,\hat{\sf D}^{}_{\mu\nu j}(x)
\hat{z}^{j}_{{}_{\underline{\texttt{0}}}}\big(\tau_-(x)\big),
   \label{F1mndiD}
\end{equation}
where
\begin{equation}
  \hat{\sf D}^{}_{\mu\nu j}(x)= d^{(2)}_{\mu\nu j}(x)\partial^2_{\tau^{}_-}+
    d^{(1)}_{\mu\nu j}(x)\partial^{}_{\tau^{}_-} +d^{(0)}_{\mu\nu j}(x) \label{Djmn}
\end{equation}
with the coefficients 
\begin{eqnarray}
  d^{(2)}_{\mu \nu j}(x) &=& r_{}^{-1} \tau^{}_{-,\mu}\left[ \eta^{}_{\nu j} -
    \tau^{}_{-,j} \bar{u}^{}_\nu(\tau^{}_-) \right] -(\mu \leftrightarrow \nu),\label{d2jmn}\\
  d^{(1)}_{\mu \nu j}(x) &=& \left\{ -r_{}^{-1}\left[ \tau^{}_{-,\mu j}\bar{u}^{}_\nu(\tau^{}_-) 
    + 2\tau^{}_{-,\mu} \tau^{}_{-,j} \bar{a}^{}_\nu(\tau^{}_-) \right] \right.\nonumber\\ 
  && \left.  + r_{}^{-2}\left[ -r^{}_{,\mu}\eta^{}_{\nu j} +
    \left( r^{}_{,\mu}\tau^{}_{-,j} + \tau^{}_{-,\mu}r_{,j} \right)\bar{u}^{}_\nu(\tau^{}_-)
    \right]\right\}  -(\mu \leftrightarrow \nu),\label{d1jmn}\\ 
  d^{(0)}_{\mu\nu j}(x) &=& -\partial^{}_j\left[ 
    \partial^{}_\mu \left(\frac{\bar{u}_\nu(\tau^{}_-)}{r}\right)
    -\partial^{}_\nu \left(\frac{\bar{u}_\mu(\tau^{}_-)}{r} \right)\right]\nonumber\\
    &=& \left\{ -r^{-1}\left[ \tau^{}_{-,\mu}\tau^{}_{-,j} \dot{\bar{a}}^{}_\nu(\tau^{}_-)
    +\tau^{}_{-,\mu j} \bar{a}^{}_\nu(\tau^{}_-)\right]\right. 
    \nonumber\\&& + r^{-2}\left[\left( r_{,\mu} \tau^{}_{-,j}
    + \tau^{}_{-,\mu} r^{}_{,j}  \right)\bar{a}^{}_\nu(\tau^{}_-)
    + r^{}_{,\mu j}\bar{u}^{}_\nu(\tau^{}_-)\right]\nonumber\\&& 
    \left. -2 r^{-3} r_{,\mu} r^{}_{,j} \bar{u}^{}_\nu(\tau^{}_-)\right\} 
    -(\mu \leftrightarrow \nu). \label{d0jmn} 
\end{eqnarray}
Explicit expressions for $d^{(n)j}_{\nu\rho}\bar{n}^\rho(\tau_-)$ and $-d^{(n)j}_{\nu\rho}\bar{u}^\rho (\tau_-)/c$ in terms of $\bar{u}^\mu(\tau_-)$ and $\bar{n}^\mu(\tau_-)$ are given in (\ref{d2jnrResult})-(\ref{d0jnrResult}).
The $1/r$ terms in $d^{(n)j}_{\nu\rho}  \bar{n}^\rho$ are identical to the $1/r$ terms in $-d^{(n)j}_{\nu\rho} \bar{u}^\rho/c$ for the same $n$.

For the combined system initially in a product state $\rho^{\,{}_{\bf I}} = \rho_{{}_P}^{\,{}_{\bf I}} \otimes \rho_{{}_F}^{\,{}_{\bf I}}$ of a Gaussian wavepacket of electron worldline deviations $\rho_{{}_P}^{\,{}_{\bf I}}$ and the Minkowski vacuum $\rho^{\,{}_{\bf I}} = \rho_{{}_P}^{\,{}_{\bf I}} \otimes \rho_{{}_F}^{\,{}_{\bf I}}$ of the EM fields \cite{LH24}, the two-point functions
\begin{equation}
\langle \hat{F}^{}_{{}_{ \underline{\texttt{0}}}\,\mu\rho}(x) \hat{F}_{{}_{\underline{ \texttt{0}}}}^{{\rho}\nu}(x') \rangle = 
\langle \hat{F}^{}_{ {}_{\underline{\texttt{0}}}\,\mu\rho}(x) 
\hat{F}_{{}_{ \underline{\texttt{0}}}}^{{\rho}\nu}(x') \rangle^{}_P + 
\langle\hat{F}^{}_{{}_{\underline{\texttt{0}}}\,\mu\rho}(x)
\hat{F}_{{}_{\underline{\texttt{0}}}}^{{\rho}\nu}(x') \rangle^{}_F, \label{FFtotal}
\end{equation}
also split into two parts {[cf. eq.(\ref{zizjLinear}) and (I.3.13)]}, which are
\begin{eqnarray}
&&\langle \hat{F}^{}_{{}_{\underline{\texttt{0}}}\,\mu\rho}(x)
\hat{F}_{{}_{\underline{\texttt{0}}}}^{{\rho}\nu}(x') 
\rangle^{}_F \equiv \sum_{\bf k}\sum_{{\bf k}'} {\cal F}_{ (\lambda)\mu\rho}^{{\bf k}}(x)
{\cal F}^{{\rho}\nu *}_{(\lambda'){\bf k}'}(x') 
  \langle 0^{}_M| \hat{b}^{(\lambda)}_{\bf k}\hat{b}^{(\lambda')\dagger}_{{\bf k}'}|0^{}_M \rangle
  \label{FFF}\\
&& \langle \hat{F}^{}_{{}_{\underline{\texttt{0}}}\,\mu\rho}(x)
\hat{F}_{{}_{\underline{\texttt{0}}}}^{{\rho}\nu}(x') \rangle^{}_P \equiv 
   \sum_{i,j=1}^3 \bigg[{\cal F}_{\mu\rho}^{z_{}^i}(x)
   {\cal F}^{{\rho}\nu}_{z_{}^{j}}(x') 
    \langle \hat{z}_{}^i \hat{z}_{}^j\rangle^{}_{\rm I} +
    {\cal F}_{\mu\rho}^{z_{}^i}(x){\cal F}^{{\rho}\nu}_{p^{}_{j}}(x') 
    \langle \hat{z}_{}^i \hat{p}^{}_j\rangle^{}_{\rm I} \nonumber\\
&&  \hspace{4cm}+{\cal F}_{\mu\rho}^{p^{}_i}(x){\cal F}^{{\rho}\nu}_{z_{}^{j}}(x') 
\langle \hat{p}^{}_i \hat{z}_{}^j\rangle^{}_{\rm I} +
    {\cal F}_{\mu\rho}^{p^{}_i}(x){\cal F}^{{\rho}\nu}_{p^{}_{j}}(x') 
    \langle \hat{p}^{}_i \hat{p}^{}_j\rangle^{}_{\rm I} \bigg], \label{FFP} 
\end{eqnarray}
where ${\cal F}_{\mu\nu}^{ \Omega} \equiv \partial_\mu {\cal A}_{\nu}^{\Omega}- \partial_\nu {\cal A}_{\mu}^{\Omega}$ with the mode functions ${\cal A}_{\mu}^{\Omega} = {\cal A}_{\mu}^{[0]\Omega} + {\cal A}_{\mu}^{[1]\Omega}$ defined in (I.2.40), (I.2.45), (I.2.50) and (I.2.51) under linear approximation there, and ${\cal F}_{{\mu\nu}}^{ z_{}^i}$ and ${{\cal F}_{\mu\nu}^{p^{}_{j}}}$ from ref. \cite{LH24} are real functions. Again, the $F$-part ($P$-part) of the expectation value depends only on the initial field (particle) state. 

Similar to (\ref{F0F1}), we write 
\begin{equation}
  {\cal F}_{\mu\nu}^{\Omega}\equiv {\cal F}_{\mu\nu}^{[0]\Omega} + {\cal F}_{\mu\nu}^{[1]\Omega},
  \label{calF0F1}
\end{equation}
where
\begin{equation}
   {\cal F}^{[1]\Omega}_{\mu\nu}(x) \equiv \partial^{}_\mu {\cal A}^{[1]\Omega}_{\nu}-
   \partial^{}_\nu {\cal A}^{[1]\Omega}_{\mu}  = 
   \frac{\mu^{}_0 q}{4\pi}\,\hat{\sf D}_{\mu\nu j}(x){\cal Z}_{\Omega}^j\big(\tau_-(x)\big)
   \label{F1mnD}
\end{equation}
according to (\ref{F1mndiD}).
Then (\ref{RenAngRadPowerDD}) can be split into  
\begin{equation}
  \frac{d{P}^{}_{dd}}{d\Omega^{}_{\rm II}} = 
  \sum_{B,B'=0,1} \left[ \frac{d{P}_{ddP}^{[BB']}}{d\Omega^{}_{\rm II}}
  + \frac{d{P}_{ddF}^{[BB']}}{d\Omega^{}_{\rm II}} \right] - 
  \left[ \frac{d{P}_{ddP}^{[00]}}{d\Omega^{}_{\rm II}}  
  + \frac{d{P}_{ddF}^{[00]}}{d\Omega^{}_{\rm II}} \right], \label{dPddSplit}
\end{equation}
where $d{P}_{dd{P}}^{[BB']}/
{d\Omega^{}_{\rm II}}$ and $d{P}_{ddF}^{[BB']}/{d\Omega^{}_{ \rm II}}$ are contributed by the mode functions ${\cal F}^{[B]\Omega}_{\mu\rho}{\cal F}^{\rho\nu}_{[B']\Omega'}$ in (\ref{FFF}) and (\ref{FFP}), respectively. The `$[00]$' terms should be subtracted since they are purely contributed by the background values of the mode functions in the absence of the electrons. The corresponding subtracted expectation value of stress-energy tensor in (\ref{RenAngRadPowerDD}) or (\ref{dPdOmega}) is considered as renormalized. 

From eq.(I.2.45), ${\cal A}^{\mu}_{[0]z_{}^i}={\cal A}^{\mu}_{[0]p_{}^i} =0$, and so ${\cal F}^{\mu\rho}_{[0] z_{}^i}={\cal F}^{\mu\rho}_{ [0]p_{}^i}=0$. Then (\ref{dPddSplit}),
(\ref{RenAngRadPowerDD}), (\ref{FFP}), and (\ref{F1mnD}) yield $d{P}_{ddP}^{[01]}/ {d\Omega^{}_{\rm II}}= d{P}_{ddP}^{[10]}/{d\Omega^{}_{\rm II}}=0$, and
\begin{eqnarray}
   &&\frac{d{P}_{ddP}^{[11]}}{d\Omega^{}_{\rm II}}= {\rm Re}\lim_{r\to\infty}\frac{r^2}{\mu^{}_0} \sum_{Q=z,p} \sum_{i,i'=1}^3     \bar{u}_{}^\mu\big(\tau^{}_-(x)\big) {\cal F}_{\mu\rho }^{[1]Q_{}^i}(x) 
    {\cal F}^{\rho\nu}_{ [1]Q_{}^{i'}}(x) \bar{n}^{}_\nu\big( \tau_-(x)\big)\langle \hat{Q}_{}^i\, \hat{Q}_{}^{i'}\rangle^{}_{\rm I}   \nonumber\\
&&={\rm Re}\lim_{r\to \infty} \frac{r^2\mu^{}_0q^2}{16\pi^2} \sum_{Q=z,p}\sum_{i,i'=1}^3 \bar{u}_{}^\mu(\tau_-) \,
      \left[\hat{\sf D}^{}_{\mu\rho j}{\cal Z}_{Q_{}^i}^j\big(\tau_-\big)\right]
    \left[\hat{\sf D}^{\rho\nu}{}_{j'}{\cal Z}_{Q_{}^{i'}}^{j'}\big(\tau_-\big)\right]
    \bar{n}^{}_\nu(\tau_-) \langle \hat{Q}_{}^i\, \hat{Q}_{}^{i'}\rangle^{}_{\rm I} \nonumber\\
   &&=  \frac{3 s \bar{m}c}{8\pi} \lim_{r\to\infty} r^2 \lim_{x'\to x}
\left[{\bar{u}_{-}^\mu} \hat{\sf D}^{}_{\mu\rho j}(x)\right] 
   \left[{\bar{n}^{-}_\nu} \hat{\sf D}^{\rho\nu}{}_{j'}(x')\right] 
\langle\hat{z}_{{}_{\underline{\texttt{0}}}}^j\big( \tau^{}_-(x) \big), \hat{z}_{{}_{\underline{\texttt{0}}}}^{j'} \big(\tau^{}_-(x')\big)\rangle^{}_P,
 \label{dPdW11P}
\end{eqnarray}
with $\langle \hat{z}^{}_j (\tau^{}_-), \hat{z}^{}_{j'}(\tau'_-)\rangle^{}_F$ in (I.3.13) or (\ref{zizjPLinear}).
Similarly, we have
\begin{equation}
   \frac{d{P}_{ddF}^{[11]}}{d\Omega^{}_{\rm II}}=
   \frac{3 s \bar{m}c}{8\pi }\lim_{r\to\infty} r^2 
   \lim_{x'\to x}\left[{\bar{u}_{-}^\mu} \hat{\sf D}^{}_{\mu\rho j}(x)\right] 
   \left[{\bar{n}^{-}_\nu} \hat{\sf D}^{\rho\nu}{}_{j'}(x')\right] 
\langle\hat{z}_{{}_{\underline{\texttt{0}}}}^j\big( \tau^{}_-(x) \big), \hat{z}_{{}_{\underline{\texttt{0}}}}^{j'} \big(\tau^{}_-(x')\big)\rangle^{}_F 
 \label{dPdW11F}
\end{equation}
with $\langle \hat{z}^{}_j (\tau^{}_-), \hat{z}^{}_{j'}(\tau'_-)\rangle^{}_F$ in (I.3.14) or (\ref{zizjFLinear}).

Nevertheless, in Minkowski vacuum, ${\cal F}^{[0]\mu\nu}_{(\lambda){\bf k}}$ is nonzero and proportional to $e^{-i \omega t + i{\bf k\cdot x}}$ as in (I.3.7) and (I.4.29). So, with (\ref{FFF}) and (\ref{F1mnD}), we have nonvanishing interference terms such as
\begin{eqnarray}
&&\frac{d{P}_{ddF}^{[01]}}{d\Omega^{}_{\rm II}} =
  \lim_{r\to\infty} \lim_{x'\to x} \frac{r_{}^2}{\mu^{}_0} \sum_{\bf k}\sum_{{\bf k}'} (2\pi_{}^3)\delta^3({\bf k}-{\bf k}') 
  \bar{u}_{}^\mu\big(\tau^{}_-(x) \big) {\cal F}^{[0](\lambda){\bf k}}_{\mu\rho}(x) {\cal F}^{[1] \rho\nu *}_{ (\lambda) {\bf k}'}(x')\bar{n}^{}_\nu\big(\tau^{}_-(x')\big) \nonumber\\  
&& = \frac{\hbar q}{64\pi^4\varepsilon^{}_0}
  \lim_{r\to\infty} r^2 \lim_{x'\to x} \bar{u}^{}_\mu \big(\tau_-(x)\big) 
  \left[\bar{n}_{}^\rho\big(\tau_-(x')\big) {\hat{\sf D}^{}_{\nu\rho}{}_{}^j} 
  (x')\right]
    \big[\partial^\mu {\cal Y}^\nu_j(x,x') - \partial^\nu {\cal Y}^\mu_j(x,x') \big] , 
   \nonumber\\  
\label{dP01}
\end{eqnarray}
where
\begin{eqnarray}
  {\cal Y}^\nu_j(x,x') = \int \frac{d^3 k}{\omega}\,e^{-i\frac{\omega}{c}x_{}^0 + i {\bf k\cdot x}
   - \frac{\epsilon}{2}\omega}\,\, \epsilon^{\nu}_{(\lambda){\bf k}}\,
   {\cal Z}^{(\lambda){\bf k}*}_j\big(\tau^{}_-(x')\big) \label{Thetanj}
\end{eqnarray}
with the polarization vector $\epsilon^{\nu}_{(\lambda){\bf k}}$ \cite{LH24}.
Writing $d{P}_{ddF}^{{[01]}}/ d\Omega^{}_{\rm II} = \lim_{x'\to x} F(x,x')$, we have $d{P}_{ddF}^{{[10]}}/d\Omega^{}_{\rm II} = \lim_{x'\to x} F_{}^*(x',x)$.

\section{Quantum radiation of single electrons at rest}
\label{SecQRadRest}

The quadrupole-monopole part of the quantum radiated power (\ref{dPmqdOmegaResult}) by an electron at rest ($\bar{a}^\mu_-=0)$ vanishes because of the $(\bar{a}^-_\nu - \bar{a}^{}_{\bar{n}}\bar{n}^-_\nu)$ factor of the classical field strength [see (\ref{ClFmunuN}) and (\ref{ClUFmunu})]. The only thing left to be checked is the dipole-dipole correction to the radiated power. 

Consider an electron situated at the origin with the classical worldline $\bar{z}_{}^\mu = (ct,{\bf 0})$, and $\bar{v}^\mu(t) = c \,\delta^\mu_0$ for all $t$.
From (\ref{Rmudef}), one has 
$\tau^{}_- =t-\frac{r}{c}\equiv t_-$, $\bar{\gamma}^{}_- \equiv \bar{\gamma}(\tau^{}_-) =1$, $\bar{u}^\mu_- \equiv \bar{u}_{}^\mu(\tau^{}_-) = c \,\delta^\mu_0$, and $\bar{n}^\mu_- \equiv \bar{n}_{}^\mu(\tau^{}_-) =(0, \hat{\bf r})$, where $\hat{r}_{}^j \equiv x_{}^j/r$ with $r = |\bf x|$ is the unit three-vector pointing from the origin to the position of the observer ${\bf x}$.
In this case, the coefficients of (\ref{Djmn}) read
\begin{eqnarray}
   &&\hat{d}^{(2)}_{0ij} = \frac{1}{c r}\left( \eta^{}_{ij}-\hat{r}^{}_i \hat{r}^{}_j\right),
     \hspace{.5cm} \hat{d}^{(1)}_{0ij} = \frac{r}{c}\hat{d}^{(0)}_{0ij} = 
     \frac{1}{r^2}\left( \eta_{ij}-3\hat{r}^{}_i \hat{r}^{}_j\right),\label{d2lojRest}\\
   &&\hat{d}^{(2)}_{lij} = \frac{r}{c}\hat{d}^{(1)}_{lij} = 
     \frac{1}{c r}\left(\eta_{jl} \hat{r}^{}_i -\eta_{ji} \hat{r}^{}_l\right),\hspace{.5cm}
     \hat{d}^{(0)}_{lij} =0. \label{d2ljiRest}
\end{eqnarray}
The fact that the classical monopole fields 
$\bar{F}_{}^{0i} = \frac{E^i}{c}=\frac{q}{4\pi \varepsilon^{}_0 r^2}$
and $\bar{F}_{ji}=0$ are produced by the electron at rest indicates that they are the backreaction of the electron to the field at the classical level. Now, on top of that, the mode functions ${\cal A}^\mu_{z_{}^l}$ and ${\cal A}^\mu_{p^{}_l}$ in (\ref{Amu1Om}) with ${\cal Z}^j_{z_{}^l}$ and ${\cal Z}^j_{p^{}_l}$ in (I.3.9) yield
\begin{eqnarray}
  {\cal F}^{0i}_{z_{}^l} &=& -\frac{\mu^{}_0 qmc}{4\pi \bar{m}r^3}\left(\delta_l^i-
      3 \hat{r}_l \hat{r}^i \right)\, \theta\left( t-\frac{r}{c}-t^{}_0\right), \label{F0jzlrest}\\
  {\cal F}_{ji}^{z_{}^l} &=& 0,\\
  {\cal F}^{0i}_{p^{}_l} &=& -\frac{\mu^{}_0 qc}{4\pi \bar{m}r^3}\left(\eta^{li}-
      3 \hat{r}^l \hat{r}^i \right)(t-t^{}_0) \, \theta\left( t-\frac{r}{c}-t^{}_0\right),\\
  {\cal F}_{ji}^{p^{}_l} &=& \frac{\mu^{}_0 q}{4\pi \bar{m}r^2}\left( \hat{r}^{}_i \eta^{}_{jl}
  - \hat{r}^{}_j \eta^{}_{il} \right)\, \theta\left( t-\frac{r}{c}-t^{}_0\right).\label{Fjiplrest}
\end{eqnarray}
This is the backreaction of the electron to the fields at the quantum level. 
If {the expectation value of $\hat{z}_{}^l$ at the initial moment $t^{}_0$ with respect to the initial state,   $\langle\hat{z}_{}^l\rangle^{}_{\rm I}\equiv \langle \hat{z}_{}^l(t_0)\rangle$}, were nonzero, taking the expectation values of both sides of eq. (\ref{F1mndi}) and then introducing the expansions (I.2.39) and (I.2.40) would lead to
\begin{equation}
 \langle \hat{F}^{0i}_{{}_{\underline{\texttt{0}}}}(x) \rangle \sim
 {\cal F}^{0i}_{z_{}^l} \langle\hat{z}_{}^l\rangle^{}_{\rm I} =
 \frac{q \langle\hat{z}_{}^l\rangle^{}_{\rm I}}{4\pi \varepsilon^{}_0 r^3}
    \left( 3\hat{r}_l \hat{r}^i - \delta_l^i\right), \label{FieldConfP}
\end{equation}
which has the same form as a classical electric field produced by the static dipole moment $q\langle\hat{z}_{}^l\rangle^{}_{\rm I}$ \cite{Gr17}. Indeed, the $\hat{z}^i$ expansion of Dirac delta function $\delta^3({\bf x} - {\bf z}) = \delta^3({\bf x} - \bar{\bf z}) + \hat{z}^i \partial_i \delta^3({\bf x} - \bar{\bf z}) + \cdots$ about the classical trajectory $x=\bar{\bf z}$ is equivalent to a multipole expansion. Note that the above expectation value of the field strength diverges at $x_{}^j=\bar{z}_{}^j =0$, rather than $x^j= \bar{z}_{}^j + \langle \hat{z}^j\rangle^{}_{\rm I}$ where the peak of the wavepacket is located. This shows that only in the far zone ($r\gg \langle\hat{z}_{}^l\rangle^{}_{\rm I})$ the `field configurations' in our effective theory such as (\ref{FieldConfP}) are reliable for calculating the expectation value of the fields.

Substituting (\ref{F0jzlrest})-(\ref{Fjiplrest}) into (\ref{FFP}), we immediately find $\lim_{r\to\infty} r^2 \langle \hat{F}^{}_{{}_{ \underline{ \texttt{0}}}\, \mu\rho}\hat{F}_{ {}_{ \underline{\texttt{0}}}}^{{\rho} \nu} \rangle^{}_P=0$, while $\langle \hat{z}^{j}_{{}_{ \underline{\texttt{0}}}},\hat{z}^{j}_{ {}_{ \underline{ \texttt{0}}}}\rangle^{}_P$ in (\ref{dPdW11P}) is growing in time. So, the $P$-part of the field correlator does not contribute to any radiated power. With regard to the $F$-part, eq. (\ref{dP01}) with (\ref{Thetanj}) and (\ref{d2ljiRest}) in this case reads 
\begin{eqnarray}
\frac{d{P}_{ddF}^{[01]}}{d\Omega^{}_{\rm II}} &=&  \frac{\hbar
      q}{64\pi^4 \varepsilon^{}_0}\lim_{r\to \infty} r^2 \lim_{x'\to x} c
  \int_0^\infty \frac{\omega d\omega}{c^3} \int_0^\pi \sin\tilde{\theta} d\tilde{\theta}\int_0^{2\pi} d\tilde{\varphi}\, e^{-i\omega t + i{\bf k \cdot x}-\omega \frac{\epsilon}{2}}
  \times\nonumber\\ && 
  \left[ i\frac{\omega}{c}\epsilon^j_{(\lambda)    {\bf k}}-ik_{}^j \epsilon^0_{(\lambda){\bf k}}\right] \left[\delta_j^l-\hat{r}'{}^l\hat{r}'_j\right]
    \left[\frac{1}{cr'}\partial^2_{t'} +\frac{1}{r'{}^2} \partial^{}_{t'}\right]
    {\cal Z}^{(\lambda){\bf k}*}_l\left(t'-\frac{r}{c}\right),
    \label{dP01ddFrest1}
\end{eqnarray}
{where ${\bf k} = \frac{\omega}{c} ( \sin\tilde{\theta} \cos\tilde{\varphi}, \cos\tilde{\theta}\sin\tilde{\varphi}, \cos\tilde{\theta})$ with $\tilde{\theta}=0$ in the direction of ${\bf x}$ such that ${\bf k\cdot \bf x}=\frac{\omega}{c} r \cos\tilde{\theta}$.
Inserting the mode function ${\cal Z}^{(\lambda){\bf k}}_l$ found in (I.3.10) together with (I.3.11) to the above expression, namely,
\begin{eqnarray}
  {\cal Z}^{(\lambda){\bf k}*}_l (T) = \frac{qc}{\bar{m}} {\cal E}^{(\lambda){\bf k}*}_{l0} (1+is\omega) e^{-\omega\epsilon/2}\left[ -\frac{1}{\omega^2}\big( e^{i\omega T}-e^{i\omega t^{}_0}\big) +\frac{i}{\omega}e^{i \omega t^{}_0}(T-t^{}_0)\right]
\end{eqnarray}
with ${\cal E}^{\bf k}_{(\lambda)l0}$ defined in (I.3.8) and the renormalized mass $\bar{m}$ in (I.2.19),
and applying the identities $\epsilon^0_{ (\lambda){\bf k}}= \delta^0_\lambda$, ${\cal E}^{(0)\bf k}_{l0} = \eta_{}^{ (0)(0)}{\cal E}^{\bf k}_{ (0)l0} =-ik^{}_l$, and 
the left equation of (I.3.17) to see
\begin{eqnarray}
&&\left[ i\frac{\omega}{c} \epsilon^j_{(\lambda){\bf k}} -ik_{}^j \epsilon^0_{(\lambda){\bf k}}\right] \left[ \delta_j^l -\hat{r}'{}^l \hat{r}'_j\right]{\cal E}^{(\lambda){\bf k}*}_{l0}
\nonumber\\ &=& 
\left[ \frac{i\omega}{c}\left(-\frac{i\omega}{c} \delta^j_l \right) -ik_{}^j \left( -i k^{}_l \right) \right] \left[ \delta_j^l -\hat{r}'{}^l \hat{r}'_j\right]\nonumber\\
&=& \frac{\omega^2}{c^2} + \left( k_{}^j \hat{r}'_j \right)^2, 
\end{eqnarray}
while setting ${\bf x}'={\bf x}$ and perform the $t'$-differentiations, 
eq. (\ref{dP01ddFrest1}) then becomes 
\begin{eqnarray}
\frac{d{P}_{ddF}^{[01]}}{d\Omega^{}_{\rm II}} &=&  
\frac{\hbar q^2}{64\pi^4 \varepsilon^{}_0 c \bar{m}} \lim_{r\to \infty} r^2 \lim_{t'\to t}\int_0^\infty \omega d\omega \int_{-1}^1 d\cos\tilde{\theta} \int_0^{ 2\pi} d\tilde{\varphi} \, e^{-i\omega\left(t - \frac{r}{c}\cos\tilde{\theta}\right)- \omega\epsilon}
\times\nonumber\\ && \hspace{1cm} \frac{\omega^2}{c^2}\left(1+\cos^2\tilde{\theta}\right)(1+ is\omega)
  \left[ \frac{1}{c r}e^{i\omega(t'-\frac{r}{c})}- \frac{i}{r^2\omega}\left( 
    e^{i\omega(t'-\frac{r}{c})}-e^{i\omega t'_0}\right)\right]\nonumber\\ &=& 
\frac{3\hbar s}{16\pi^2} 
  \lim_{r\to \infty} r^2 \lim_{t'\to t}\int_0^\infty d\omega\, e^{-\omega \epsilon}\times
  \nonumber\\&& \hspace{1cm} 
  2e^{-i\omega t}\left[\left(\frac{ic}{r\omega}-\frac{ic^3}{r^3\omega^3} \right)
  \left( e^{-i\omega\frac{r}{c}}-e^{i\omega\frac{r}{c}} \right)+
  \frac{c^2}{r^2\omega^2}\left( e^{-i\omega\frac{r}{c}}-e^{i\omega\frac{r}{c}} \right) \right]\times \nonumber\\ && \hspace{1cm}  (1+is\omega) 
  \left[ \frac{\omega^3}{c r}e^{i\omega \left( t'- \frac{r}{c}\right)}-\frac{i\omega^2}{r^2}\left(
  e^{i\omega \left( t'- \frac{r}{c}\right)}-e^{i\omega t'_0}\right) \right],
\end{eqnarray} }
where we have used $\mu^{}_0 \varepsilon^{}_0=c^{-2}$. Thus, for finite $t-t'_0 >0$, 
\begin{eqnarray}
\frac{d{P}_{ddF}^{[01]}}{d\Omega^{}_{\rm II}} 
&=& \frac{3\hbar s}{8\pi^2}
  \lim_{r\to \infty} r^2 \lim_{t'\to t} \nonumber\\
  &&\left\{ \frac{1}{r^2}\left[2\left(\frac{1}{(t^{}_{+-})^3}-\frac{1}{(t^{}_{--})^3} \right) 
   +6s\left(\frac{1}{(t^{}_{+-})^4}-\frac{1}{(t^{}_{--})^4} \right)\right]\right.+\nonumber\\
  &&\frac{c}{r^3}\left[ \frac{2}{(t^{}_{+-})^2}-\frac{1}{(t^{}_{+0})^2}+
    \frac{1}{(t^{}_{-0})^2}+2s \left( \frac{2}{(t^{}_{+-})^3}-\frac{1}{(t^{}_{+0})^3}+
    \frac{1}{(t^{}_{-0})^3}\right)\right]+\nonumber\\
  &&\frac{c^2}{r^4}\left[ \frac{2}{t^{}_{+-}}-\frac{1}{t^{}_{+0}}-
    \frac{1}{t^{}_{-0}}+ s\left( \frac{2}{(t^{}_{+-})^2}-\frac{1}{(t^{}_{+0})^2}-
    \frac{1}{(t^{}_{-0})^2}\right)\right]+\nonumber\\
  && \left. \frac{c^3}{r^5}\left[ 
  \ln\frac{t^{}_{--}\,\,t^{}_{+0}}{t^{}_{+-}\,\,t^{}_{-0}} +s\left(
  \frac{1}{t^{}_{+-}}-\frac{1}{t^{}_{--}}-\frac{1}{t^{}_{+0}}+\frac{1}{t^{}_{-0}}  \right)\right]\right\}\nonumber\\
  &=& \frac{3\hbar s}{8\pi^2}
  \left( -\frac{2}{\epsilon_1^3} - s \frac{6}{\epsilon_1^4}\right), \label{dP01ddFrest}
\end{eqnarray} 
where $t^{}_{+-} \equiv t_+ - t'_- = \epsilon_1 +\frac{2r}{c}$, $t^{}_{--} \equiv t_- - t'_- = t-t'=\epsilon_1$, $t^{}_{+0} \equiv t_+ - t'_0 = t-t'_0+\frac{r}{c}$, and $t^{}_{-0} \equiv t_- - t'_0 = t-t'_0-\frac{r}{c}$ with $t^{}_\pm \equiv t \pm \frac{r}{c}$ and $t'_\pm \equiv t' \pm \frac{r}{c}$. {Here, $\epsilon^{}_1 \equiv \tau^{}_- - \tau'_- = t^{}_- -t'_-\to 0+$ could be interpreted as the uncertainty of time-tagging to the history of an electron wavepacket \cite{LH24}.}
Since one can write $d{P}_F^{{[01]}}/ d\Omega^{}_{\rm II} = \lim_{x'\to x}F(x,x')$ and $d{P}_F^{{[10]}}/ d\Omega^{}_{\rm II}= \lim_{x'\to x}F_{}^*(x',x)$ for some function $F(x,y)$, we have
\begin{equation}
\frac{d{P}_{ddF}^{{[10]}}}{d\Omega^{}_{\rm II}} = 
  \frac{3\hbar s}{8\pi^2} 
  \left( \frac{2}{\epsilon_1^3} - s \frac{6}{\epsilon_1^4}\right).
\label{dP10ddFrest}
\end{equation}
Furthermore, inserting (I.3.23) into (\ref{dPdW11F}), we find
\begin{eqnarray}
  \frac{d{P}_{ddF}^{[11]}}{d\Omega^{}_{\rm II}}
&=& \frac{3 \hbar s^2}{4\pi^2}\lim_{r\to \infty} r^2 
  \left\{\frac{6}{r^2 \epsilon_1^4}-\frac{2c}{r^3}\left[ 
    \frac{1}{(\eta^{}_- -\epsilon^{}_0)^3}+\frac{1}{(\eta^{}_- -\epsilon^{}_1)^3} \right]-
  \right. \nonumber\\ && \hspace{1cm}\frac{c^2}{r^4}\left[ \frac{1}{\epsilon_0^2} +
  \frac{2}{(\eta^{}_- -\epsilon^{}_1)^2}+\frac{2\epsilon^{}_1}{(\eta^{}_- -\epsilon^{}_1)^3}
  -\frac{1}{(\eta^{}_- -\epsilon^{}_0)^2} \right] -\nonumber\\
  && \hspace{1cm}\left. \frac{c^3}{r^5}\left[ \frac{1}{\epsilon^{}_0} - \frac{1}{\epsilon_1}+
  \frac{\eta^{}_-}{\epsilon_0^2}+\frac{1}{\eta^{}_- -\epsilon^{}_0} -
  \frac{2}{\eta^{}_- -\epsilon^{}_1} -\frac{\epsilon^{}_1}{(\eta^{}_- -\epsilon^{}_1)^2}
  \right]\right\}  \nonumber\\
  &=& \frac{9 \hbar s^2}{2\pi^2 \epsilon_1^4} 
  \label{dPdW11Frest}
\end{eqnarray}
for $\eta^{}_- \equiv \tau^{}_- - \tau^{}_0 = t-\frac{r}{c} - t_0 > 0$ 
{and  $\epsilon^{}_0 \equiv \tau'_0 - \tau^{}_0=t'_0-t^{}_0 
$ representing the uncertainty of the initial moment of an electron wavepacket (see section 3.2.2 in \cite{LH24}).} 
Adding (\ref{dP01ddFrest}), (\ref{dP10ddFrest}) and (\ref{dPdW11Frest}) together, we find that the total $F$-part of the dipole-dipole contribution
$d{P}_{ddF}^{[11]}/d\Omega^{}_{\rm II} + d{P}_{ddF}^{[01]}/ d\Omega^{}_{\rm II}+d{P}_{ddF}^{ [10]} /d\Omega^{}_{\rm II}$ is vanishing while $\langle \hat{z}^{j}_{{}_{\underline{ \texttt{0}}}},\hat{z}^{j}_{ {}_{ \underline{\texttt{0}}}} \rangle^{}_F$ in (\ref{dPdW11F}) is growing in time. Thus we conclude that the quantum radiation by an electron at rest is exactly zero, the same as its classical radiation.

As stated in ref. \cite{LH07}, $d{P}_{ddF}^{[01]}/ d\Omega^{}_{\rm II} + d{P}_{ddF}^{[10]}/ d\Omega^{}_{\rm II}$ can be interpreted as the interference of vacuum fluctuations $\sim {\cal F}^{[0]\mu\nu }_{(\lambda){\bf k}}$ and the retarded fields $\sim {\cal F}^{[1]\mu\nu}_{ (\lambda){\bf k}}$ driven by vacuum fluctuations. In the above result one can see that the radiated power $d{P}_{ddF}^{[11]}/d\Omega^{}_{\rm II} \propto s^2$ by an electron at rest is exactly canceled by those $s^2$ terms in $d{P}_{ddF}^{[01]}/ d\Omega^{}_{\rm II} + d{P}_{ddF}^{[10]}/ d\Omega^{}_{\rm II} $, which corresponds to the radiation reaction (the $O(s)$ terms) in ${\cal Z}^{[1]j}_{(\lambda){\bf k}}$. Same conclusion is reached from an analysis of quantum radiation emitted by a harmonic atom {(an Unruh-DeWitt harmonic-oscillator detector \cite{Unr76, DeW79}) at rest explicitly in ref. \cite{HH19}, 
or in \cite{LH06} in the zero-acceleration ($a\to 0$) limit.} 

Although the radiation is zero, the bound fields corresponding to the $O(r^{-3})$ terms in $\bar{u}_\mu \langle T^{\mu\nu}\rangle \bar{n}_\nu$ do depend on the presence of the electron. Considering the fact that the $\bar{\alpha}$ terms in (\ref{QTmnop}) contribute to the bound fields in the region of finite $r$, the bound fields here are gauge dependent.


\section{Quantum radiation of uniformly accelerated electrons}
\label{SecQRadUAC}

Consider a single electron accelerated in a uniform electric field $\bar{F}^{01}_{[0]} = {\cal E}/c$ along the worldline 
\begin{equation}
  \bar{z}_{}^\mu(\tau) = \left( \frac{c}{\alpha}\sinh\alpha \tau, \frac{c}{\alpha}\cosh\alpha\tau,0,0 \right)  \label{zmuUAC}
\end{equation}
with $\alpha = a/c = q{\cal E}/({\bar{m}} c)$, as described in section 4 of ref. \cite{LH24}.
Then, an observer located at $x$ in the R or F wedge (figure  1 in ref. \cite{Lin03}) will see the retarded field produced by the electron at the retarded time
\begin{equation}
  \tau^{}_-(x) = -\frac{1}{\alpha}\ln\left[ \frac{\alpha}{2c|V|}\left(X-UV+\rho^2+
    \frac{c^2}{\alpha^2}\right)\right],  \label{tauMinus}
\end{equation}
where $U=x^0-x^1$, $V=x^0+x^1$, $\rho = \sqrt{x_2^2+x_3^2}$, and
\begin{equation}
  X(x) \equiv \sqrt{\left(-UV+\rho^2+\frac{c^2}{\alpha^2} \right)^2+\frac{4c^2}{\alpha^2}UV} 
    = \frac{2cr}{\alpha} 
  \label{X2cra}
\end{equation}
with $r$ defined in (\ref{Rmudef}).
Now the four velocity of the electron is 
\begin{equation}
\bar{u}_{}^\mu(\tau)=\partial^{}_\tau\bar{z}^\mu(\tau)=
  \left( c\cosh\alpha\tau,c\sinh\alpha\tau,0,0\right), \label{UAC4vec}
\end{equation} 
accordingly we parameterize the spacelike vector $\bar{n}_{}^\mu$ in (\ref{Rmudef}) as
\begin{equation}
 \bar{n}_{}^\mu(\tau)= \left(\cos\theta\,\sinh\alpha\tau, \cos\theta\,\cosh\alpha\tau, 
\sin\theta\,\cos\varphi, \sin\theta\,\sin\varphi \right). \label{nbarUACdef}
\end{equation}
We also have the four acceleration 
\begin{equation}
  \bar{a}_{}^\mu(\tau) = \partial^{}_\tau{\bar{u}^\mu} (\tau)=
  \left( a\sinh\alpha\tau,a\cosh\alpha\tau,0,0\right), \label{UAC4acc}
\end{equation}
yielding $\bar{a}^{}_{\bar{n}}\equiv \bar{a}^{}_\mu \bar{n}_{}^\mu = a \cos\theta$.

\subsection{Quadrupole-monopole corrections}
\label{SecUACqm}

For a uniformly accelerated electron, we have $\langle \hat{z}^i_{{}_{\underline{\texttt{0}}}}(\tau^{}_-) \hat{z}^j_{{}_{ \underline{\texttt{0}}}}(\tau^{}_-)\rangle \propto \eta_{}^{ij}$ (see section 4.3 in ref. \cite{LH24}). Inserting (\ref{UAC4vec})-(\ref{UAC4acc}) into (\ref{dPmqdOmegaResult}), we get
\begin{equation}
    \frac{d{P}^{}_{qm}}{d\Omega^{}_{\rm II}} =\frac{d{P}^{}_{mq}}{d\Omega^{}_{\rm II}} = 
    \frac{d{P}^{\mathbb{L}}_{qm}}{d\Omega^{}_{\rm II}}+
    \frac{d{P}^{\mathbb{T}}_{qm}}{d\Omega^{}_{\rm II}},
\end{equation}
where the {`longitudinal' deviation (in the direction of linear acceleration, see sections 4.1.1 and 4.3.1 in \cite{LH24})} contributes
\begin{eqnarray}
  \frac{d{P}^{\mathbb{L}}_{qm}}{d\Omega^{}_{\rm II}} &=& \frac{3s\bar{m}}{8\pi}
    \sin^2\theta\left\{ -\frac{1}{2} \frac{R_-^1}{r}\cosh\alpha\tau^{}_- 
    \left[ \alpha\partial_{\tau^{}_-}^3 +3\alpha^2\cos\theta\partial_{\tau^{}_-}^2 +
    \alpha^3 (3\cos^2\theta-1)\partial^{}_{\tau^{}_-}\right] +\right.\nonumber\\
  &&\hspace{.3cm} \left. \frac{3}{2} \left[\frac{R_-^1}{r}\right]^2 
    \left[ \alpha^2 \partial_{\tau^{}_-}^2 + 4\alpha^3\cos\theta\partial_{\tau^{}_-} + 
    \alpha^4(5 \cos^2\theta-1)\right]\right\}
\langle\hat{z}^1_{{}_{\underline{\texttt{0}}}}(\tau^{}_-),
\hat{z}^1_{{}_{\underline{\texttt{0}}}}(\tau^{}_-)\rangle
\label{dPqmLdOmega}
\end{eqnarray}
with $R_-^1/r = \sinh\alpha\tau^{}_- + \cosh\alpha\tau^{}_- \cos\theta$, and the {`transverse' deviations (see sections 4.1.2 and 4.3.2 in \cite{LH24})} contribute 
\begin{eqnarray}
  &&\frac{d{P}^{\mathbb{T}}_{qm}}{d\Omega^{}_{\rm II}} = \frac{3s\bar{m}}{8\pi}
    \sin^2\theta\left\{ -\frac{1}{2} \cos\theta \left[ \alpha\partial_{\tau^{}_-}^3 
+3\alpha^2\cos\theta\partial_{\tau^{}_-}^2 +
    \alpha^3 (3\cos^2\theta-1)\partial^{}_{\tau^{}_-}\right]+\right.\nonumber\\
  &&\hspace{.3cm} \left. \frac{3}{2} \sin^2\theta 
    \left[ \alpha^2 \partial_{\tau^{}_-}^2 + 4\alpha^3\cos\theta\partial_{\tau^{}_-} + 
    \alpha^4(5 \cos^2\theta-1)\right]\right\} 
    \left[ \cos^2\varphi \langle\hat{z}^2_{{}_{\underline{\texttt{0}}}}(\tau^{}_-),
\hat{z}^2_{{}_{\underline{\texttt{0}}}}(\tau^{}_-)\rangle + \right. \nonumber\\
  &&\hspace{9cm} \left. \sin^2\varphi \langle\hat{z}^3_{{}_{\underline{\texttt{0}}}}(\tau^{}_-),
\hat{z}^3_{{}_{\underline{\texttt{0}}}}(\tau^{}_-)\rangle \right], \label{dPqmTdOmega}
\end{eqnarray}
where the particle correlators $\langle\hat{z}^j_{{}_{\underline{\texttt{0}}}}(\tau^{}_-), \hat{z}^j_{{}_{\underline{\texttt{0}}}}(\tau^{}_-)\rangle$ for a uniformly accelerated electron have been obtained in ref. \cite{LH24}. The leading-order result for $\langle\hat{z}^2_{ {}_{\underline{ \texttt{0}}}}(\tau^{}_-), \hat{z}^2_{ {}_{ \underline{\texttt{0}}}}(\tau^{}_-) \rangle^{}_F$ is explicitly given in appendix \ref{ApxzTzTF}, and $\langle\hat{z}^3_{ {}_{ \underline{\texttt{0}}}}(\tau^{}_-), \hat{z}^3_{{}_{ \underline{\texttt{0}}}}(\tau^{}_-) \rangle^{}_F = \langle\hat{z}^2_{ {}_{\underline{ \texttt{0}}}}(\tau^{}_-), \hat{z}^2_{ {}_{ \underline{\texttt{0}}}}(\tau^{}_-) \rangle^{}_F$ by symmetry of the Minkowski vacuum. Below we denote $d{P}^{\mathbb{L,T}}_{qmP}/ d\Omega^{}_{\rm II}$ and $d{P}^{\mathbb{ L,T}}_{qmF}/d\Omega^{}_{\rm II}$ as the $P$-part and $F$-part of the above expressions, which are proportional to $\langle .. \rangle^{}_P$ and $\langle .. \rangle^{}_F$, respectively.

With the parameter values in section \ref{SecNumRe}, at early times [$\eta^{}_- \sim O(\epsilon^{}_0)$], the closed-form result (I.4.48) and (I.4.51) yield
\begin{equation}
\frac{d{P}^{\mathbb{L}}_{qm}}{d\Omega^{}_{\rm II}} \approx -\frac{3s\bar{m}}{8\pi}
    \sin^2\theta \cos\theta \, \frac{\hbar s\alpha^4}{\pi \bar{m}}\left[
    \frac{1}{(e^{\alpha(\tau_- - \epsilon^{}_0)}-1)^3}+
    \frac{1}{(e^{\alpha(\tau_- - \epsilon^{}_1)}-1)^3}\right],
\end{equation}
which is $O(10^{-32})$ W and mainly contributed by $\partial_{\tau^{}_-}^3 \langle\hat{z}^1_{{}_{\underline{\texttt{0}}}}(\tau^{}_-), \hat{z}^1_{{}_{\underline{\texttt{0}}}}(\tau^{}_-)\rangle^{}_F$, 
while (I.4.61) and (\ref{zTzTFcloseform}) for transverse deviations yield
\begin{equation}
  \frac{d{P}^{\mathbb{T}}_{qm}}{d\Omega^{}_{\rm II}} \approx -\frac{9 s \alpha^2}{32\pi\bar{m}}
  \sin^2\theta  \cos 2\theta \Big(\langle p_2^2\rangle^{}_{\rm I}\cos^2\varphi + 
  \langle p_3^2\rangle^{}_{\rm I}\sin^2\varphi \Big) ,
\end{equation}
which is $O(10^{-29})$ W 
and is mainly contributed by $\alpha^2\partial_{ \tau^{}_-}^2 \langle\hat{z}^{\sf T}_{ {}_{\underline{\texttt{0}}}}(\tau^{}_-), \hat{z}^{\sf T}_{{}_{\underline{\texttt{0}}}}(\tau^{}_-)\rangle^{}_P$, ${\sf T}=2,3$ in (\ref{dPqmTdOmega}). Thus the quadrupole-monopole corrections to the radiated power is dominated by the $P$-part of $d{P}^{\mathbb{T} }_{qm}/d\Omega^{}_{\rm II}$ at early times in this parameter range. 

In the long-time regime ($\alpha\eta^{}_- \gg 1$ while $s\alpha^2\eta^{}_- \ll 1$), we have
\begin{equation}
\frac{d{P}^{\mathbb{T}}_{qm}}{d\Omega^{}_{\rm II}} 
  \approx \frac{9s}{16\pi\bar{m}}\alpha^4
\sin^4\theta \big( 5\cos^2\theta -1 \big)
\eta_- ^2 \Big(\langle p_2^2\rangle^{}_{\rm I}\cos^2\varphi + 
\langle p_3^2\rangle^{}_{\rm I}\sin^2\varphi \Big) ,
\label{dPLmqdOmLTT}
\end{equation}
which is mainly contributed by the $\alpha^4 \langle \hat{z}^{\sf T}_{{}_{\underline{\texttt{0}}}}(\tau^{}_-),\hat{z}^{\sf T}_{{}_{\underline{ \texttt{0}}}}(\tau^{ }_-)\rangle^{}_P$ in (\ref{dPqmTdOmega}).  
At the same time, $\langle \hat{z}^1_{{}_{\underline{\texttt{0}}}}(\tau^{}_-),\hat{z}^1_{{}_{\underline{\texttt{0}}}}(\tau^{}_-)\rangle$ goes to a constant of $\tau^{}_-$, denoted by $\langle \hat{z}^1_{{}_{\underline{\texttt{0}}}}\hat{z}^1_{{}_{\underline{\texttt{0}}}}\rangle$, such that 
\begin{equation}
\frac{d{P}^{\mathbb{L}}_{qm}}{d\Omega^{}_{\rm II}} 
  \approx 
  \frac{9s\bar{m}}{16\pi}\alpha^4
    \sin^2\theta \big( 5\cos^2\theta -1 \big)
    \left(\sinh\alpha\tau^{}_- + \cosh\alpha\tau^{}_- \cos\theta\right)^2
\langle\hat{z}^1_{{}_{\underline{\texttt{0}}}}\hat{z}^1_{{}_{\underline{\texttt{0}}}}\rangle ,
\label{dPLmqdOmLTL}
\end{equation}    
which grows exponentially as $\alpha\tau^{}_-$ increases and eventually dominates the quadrupole-monopole corrections. Below in the dipole-dipole correction we will also see similar secular growths, which should not be construed as a pathological behavior of our quantum theory, as we will argue in the end of section \ref{LongTimeRadofUAC}.

\subsection{Dipole-dipole correction}

The dipole-dipole correction to the angular radiated power emitted by a classical point-like UAC can also be divided into the contributions by the longitudinal deviation and by the transverse deviations.

\subsubsection{Contribution by longitudinal deviation}
\label{Z1Contribution}

From (\ref{d2jnrResult})-(\ref{d0jnrResult}), the coefficients (\ref{d2jmn})-(\ref{d0jmn}) contracted by $\bar{n}^\rho_- \equiv \bar{n}^\rho(\tau^{}_-)$ for ${\cal Z}^{\Omega}_1$ in (\ref{F1mnD}) read
\begin{eqnarray}
&& d^{(2)}_{\nu \rho 1} \bar{n}^\rho_- = 
   \frac{1}{cr}\sin\theta \cosh\alpha\tau^{}_- \bar{n}^\perp_\nu(\tau^{}_-), \label{d21nurho}\\
&& d^{(1)}_{\nu \rho 1} \bar{n}^\rho_- = 
   \frac{\alpha}{cr}\sin\theta \left(2\sinh \alpha\tau^{}_- + 
   3\cos\theta \cosh\alpha\tau^{}_- \right) \bar{n}^\perp_\nu(\tau^{}_-)  
   +O\left( r^{-2}\right), \\ 
&& d^{(0)}_{\nu \rho 1} \bar{n}^\rho_- 
   = \frac{3\alpha^2}{cr} \sin\theta \cos\theta \left(\sinh \alpha\tau^{}_- + 
   \cos\theta \cosh\alpha\tau^{}_- \right) \bar{n}_\nu^\perp(\tau^{}_-)
   +O\left(r^{-2}\right), \label{d01nurho} 
\end{eqnarray}
and $d^{{(n)}}_{\nu\rho 1}
\bar{u}^\rho_-/c = -{d^{(n)}_{\nu\rho 1}} \bar{n}^\rho_- + O(r^{-2})$.
Here, the spacelike vector $\bar{n}^\mu_\perp$ is defined as
\begin{equation}
\bar{n}^\mu_\perp(\tau)\equiv\left(\sin\theta \sinh\alpha\tau, \,\sin\theta\cosh\alpha\tau,
  \,-\cos\theta\cos\varphi,\,-\cos\theta\sin\varphi\right),
\end{equation}
which has the properties
$\bar{n}^\mu_\perp(\tau) \bar{n}_\mu(\tau) = 0$,
$\bar{n}^\mu_\perp(\tau) \bar{u}_\mu(\tau) = 0$, and
$\bar{n}^\mu_\perp(\tau) \bar{n}_\mu^\perp(\tau) = 1$
\footnote{In our calculation, we always fix $\theta'=\theta$ and $\varphi'=\varphi$, so that $\bar{n}^\mu_\perp(\tau^{}_-) \bar{n}_\mu^\perp(\tau'_-) = \sin^2\theta + \cos^2\theta \cosh \alpha(\tau^{}_- - \tau'_-) = 1 + O(\epsilon_1^2)$.
}.

Suppose that the expectation values $\langle \hat{z}^1\rangle^{}_{\rm I}$, $\langle \hat{p}^1\rangle^{}_{\rm I}$, $\langle \hat{z}^1 \hat{p}^1 \rangle^{}_{\rm I}$, $\langle \hat{p}^1 \hat{z}^1 \rangle^{}_{\rm I}$ with respect to our initial state are vanishing. With these conditions,
the $P$-part of the particle correlator $\langle\hat{z}_{{}_{ \underline{\texttt{0}}}}^1\big( \tau^{}_- \big), \hat{z}_{{}_{ \underline{\texttt{0}}}}^1\big(\tau'_-\big) \rangle^{}_P$ and the $F$-part correlator $\langle\hat{z}_{{}_{ \underline{\texttt{0}}}}^1\big( \tau^{}_-\big), \hat{z}_{{}_{ \underline{\texttt{0}}}}^1\big(\tau'_-\big)\rangle^{}_F$ in linear approximation have been given in (I.4.48) and (I.4.51), respectively. 
Inserting them with (\ref{d21nurho})-(\ref{d01nurho}) into (\ref{dPdW11P}) and (\ref{dPdW11F}), we immediately obtain their contributions to $d{P}^{[11]}_{ddP}/d\Omega^{}_{\rm II}$ and $d{P}^{[11]}_{ddF}/d\Omega^{}_{\rm II}$, 
namely,
\begin{eqnarray}
\frac{d{P}^{\mathbb{L}[11]}_{ddP,F}}{d\Omega^{}_{\rm II}} &=&
   \frac{3 s \bar{m}c}{8\pi} \lim_{r\to\infty} r^2 \lim_{x'\to x}
   \left[{\bar{u}_{-}^\mu} \hat{\sf D}^{}_{\mu\rho 1}(x)\right] 
   \left[{\bar{n}^{-}_\nu} \hat{\sf D}^{\rho\nu}{}_1(x')\right] 
    \langle\hat{z}_{{}_{\underline{\texttt{0}}}}^1\big(\tau^{}_-(x)\big),
    \hat{z}_{{}_{\underline{\texttt{0}}}}^1 \big(\tau^{}_-(x')\big)\rangle^{}_{P,\, F} .
\nonumber\\ 
\label{dP11FPL}
\end{eqnarray}

To calculate the contribution of ${\cal Z}^{(\lambda){\bf k}}_{1}$ to $d{P}^{[01]}_{ ddF}/d\Omega^{}_{\rm II}$, inserting ${\cal Z}^{(\lambda){\bf k}}_{1}$ given in (I.4.13) into (\ref{Thetanj}), after some algebra, we get 
\begin{eqnarray}
  {\cal Y}^\nu_{1} &=& \int_0^\infty \frac{\omega d\omega}{c^3}\int_0^{2\pi} d\tilde{\varphi}\int_0^{\pi} \sin\tilde{\theta} d\tilde{\theta} \, e_{}^{-i\frac{\omega}{c}x_{}^0+i{\bf k\cdot x}-\frac{\epsilon}{2}\omega} 
   \, \frac{qc}{\bar{m}} \left(-ik^{}_1 \delta_0^\nu - \frac{i\omega}{c}\delta_1^\nu \right) 
   \times\nonumber\\ && \int_{\tau'_0}^{\tau'_-} d\tilde{\tau} K^{}_\parallel(\tau'_-,\tilde{\tau})
   \bar{\gamma}(\tilde{\tau})\left(1+s\partial^{}_{\tilde{\tau}} \right)
   e^{-i k^{}_\mu \bar{z}_{}^\mu(\tilde{\tau})-\frac{\epsilon}{2}\omega} , 
  \label{Thetan1}
\end{eqnarray}
where $K^{}_\parallel(\tau'_-,\tilde{\tau})$ and $\bar{\gamma}(\tilde{\tau})$ are given in (I.4.10) and (I.4.4), respectively. 
Introduce $\varphi^{}_{\bf k}(\kappa)$ from (I.4.36), and modify eq.(91) in ref.\cite{LH06} to
\begin{equation}
  \int \frac{c^3d^3 k}{(2\pi)^3 2\omega}\varphi^*_{\bf k}(\kappa)
  e^{-i\omega t +i{\bf k}\cdot{\bf x}} = \frac{-i c^2}{(2\pi)^2 \alpha X}\left(
  \frac{e^{-i\kappa\tau^{}_-}-e^{-i\kappa \xi}e^{-i\kappa\tau^{}_+}}{1-e^{-2\pi \kappa/\alpha}} 
  \right) \label{IntroVarphi}
\end{equation}
with $\xi = 0$ and $-i\pi/\alpha$ for $x$ in the R and F wedges, respectively,
while $\tau^{}_+$ is defined as \cite{Lin03, LH06}
\begin{equation}
  \tau^{}_+(x) = \frac{1}{\alpha}\ln\left[ \frac{\alpha}{2c|U|}\left(X-UV+\rho^2+
    \frac{c^2}{\alpha^2}\right)\right] \label{tauPlus}
\end{equation}
(cf. figure 1 in ref. \cite{Lin03}),
we find that (\ref{Thetan1}) with (\ref{IntroVarphi}) becomes
\begin{eqnarray}
&&  {\cal Y}^\nu_{1} = \frac{4iq\pi}{\bar{m}\alpha^2}\int_{-\infty}^\infty d\kappa 
    (1+is\kappa)\left( \delta_0^\nu\partial_{}^1+\delta_1^\nu\partial_{}^0 \right)
    \frac{e^{-i\kappa\tau^{}_-}- e^{-i\kappa(\tau^{}_+ +\xi)}}
    {X\left(1-e^{-2\pi \kappa/\alpha}\right)} \times \nonumber\\ && \hspace{3cm}   
    \int_{\tau'_0}^{\tau'_-}d\tilde{\tau} \left( \tanh \alpha\tau'_- 
    \cosh\alpha\tilde{\tau}-\sinh\alpha\tilde{\tau}\right) e^{i\kappa\tilde{\tau}},
\nonumber\\
&&= \frac{2iq\pi}{\bar{m}\alpha^2 X}\int\frac{d\kappa(1 + is\kappa)}{1-e^{-2\pi\kappa/\alpha}}
  \left[ -\frac{\delta^\nu_0 X_{,}{}^1+\delta^\nu_1 X_{,}{}^0}{X}
  \left( e^{-i\kappa\tau^{}_-}-e^{-i\kappa(\tau^{}_++\xi)} \right)\right.\nonumber\\&&\hspace{2cm}
    -i\kappa \left( \delta^\nu_0 \tau^{}_{-,}{}^1+\delta^\nu_1 \tau^{}_{-,}{}^0 \right)
      e^{-i\kappa\tau^{}_-}
    +i\kappa \left( \delta^\nu_0 \tau^{}_{+,}{}^1+\delta^\nu_1 \tau^{}_{+,}{}^0\right)
      e^{-i\kappa(\tau^{}_++\xi)} 
    \bigg]\times \nonumber\\ 
&& \hspace{2cm} \left[ \sum_{\mathbb{B}=+,-} \left[ \tanh\alpha\tau'_--(\mathbb{B}1) \right]\,\,
  \frac{e^{(i\kappa+\mathbb{B}\alpha)\tau'_-}-e^{(i\kappa+\mathbb{B}\alpha)\tau'_0}}{i\kappa+\mathbb{B}\alpha}
  \right]. 
\label{Thetan1UACkint}
\end{eqnarray}
This gives ${\cal Y}^2_1 ={\cal Y}^3_1=0$ and 
\begin{eqnarray}
{\cal Y}^{\sf L}_1 
&=& -\frac{q\pi}{\bar{m}c\alpha} \partial_{}^{\bar{\sf L}}
    \left(\frac{\Gamma^{}_+ -\Gamma^{}_-}{r}\right)
   \label{Thetan1UAC}
\end{eqnarray}
where ${\sf L}=0,1$, $\bar{\sf L}=1-{\sf L}$, and
\begin{eqnarray}
&&  \Gamma^{}_\pm(x,x') \equiv \sum_{\mathbb{B}=+,-}
  \left[ \tanh\alpha\tau'_- - \left( \mathbb{B}1 \right)\right]\times
  \nonumber\\  && 
  \left\{e^{\mathbb{B}\alpha\tau'_-} \left[g^{}_\mathbb{B}\left( {\rm T}_{\pm -}\right)- 
      s \dot{g}^{}_\mathbb{B}\left( {\rm T}_{\pm -}\right)\right] - 
    e^{\mathbb{B}\alpha\tau'_0} \left[g^{}_\mathbb{B}\left( {\rm T}^{}_{\pm 0}\right)- 
      s \dot{g}^{}_\mathbb{B}\left( {\rm T}^{}_{\pm 0}\right)\right]\right\}
  \label{Gammapm}
\end{eqnarray}
with
${\rm T}^{}_{--}\equiv\tau^{}_- -\tau'_-$, ${\rm T}^{}_{-0}\equiv\tau^{}_- -\tau'_0$,
${\rm T}^{}_{+-}\equiv\tau^{}_+ +\xi -\tau'_-$, ${\rm T}^{}_{+0}\equiv\tau^{}_++\xi -\tau'_0$,
and
\begin{eqnarray}
  g^{}_+(T) &\equiv& \int 
  \frac{ i d\kappa}{1-e^{-2\pi\kappa/\alpha}} \,\,
  \frac{e^{-i\kappa T}}{i\kappa + \alpha} = -1-e^{\alpha T}\ln \left( 1-e^{-\alpha T}\right),
  \label{gpdef}\\
  g^{}_-(T) &\equiv& \int 
  \frac{ i d\kappa}{1-e^{-2\pi\kappa/\alpha}} \,\,
  \frac{e^{-i\kappa T}}{i\kappa - \alpha} = -1-e^{-\alpha T}\left[ 
    i\pi + \alpha T + \ln \left( 1-e^{-\alpha T}\right) \right], \label{gmdef}
\end{eqnarray}
for $T>0$. Note that $g^{}_-(T) = g^*_+(-T)$.
The definition (\ref{Rmudef}) implies that $r = \left( V e^{-\alpha\tau^{}_-}+U e^{\alpha\tau^{}_-}\right)/2$ for our uniformly accelerated electrons with the specific worldline $\bar{z}_{}^\mu(\tau^{}_-)$. Together with (\ref{X2cra}), one can see that
\begin{eqnarray}
V=e^{\alpha\tau_-}\left(r+r\cos\theta + \frac{c}{\alpha}\right),\hspace{.7cm}
  U=e^{-\alpha\tau_-}\left(r-r\cos\theta - \frac{c}{\alpha}\right), \label{VUoftauM}
\end{eqnarray}
for $x$ in the F and P wedges. 
Compare (\ref{tauPlus}) with (\ref{tauMinus}), we have
\begin{equation}
  \tau^{}_+ = - \tau^{}_- + \frac{1}{\alpha}\ln \frac{|V|}{|U|} 
 \label{tauPofrtheta}
\end{equation}
Inserting (\ref{VUoftauM}) into (\ref{tauPofrtheta}), one has
\begin{equation}
  \tau^{}_+ 
  = \tau^{}_- + \frac{1}{\alpha}\ln 
    \frac{1+\cos\theta + \frac{c}{\alpha r}}{\left| 1-\cos\theta -\frac{c}{\alpha r}\right|},
 \label{tauPofrtheta2}
\end{equation}
which implies
\begin{equation}
  e^{\alpha\tau^{}_+(x)} \to e^{\alpha\tau^{}_-(x)}\cot^2 \frac{\theta(x)}{2}
\end{equation}
as $r\to\infty$ with $\tau^{}_-$ fixed.
Almost all observation points $x$ in this far zone are in the F wedge (where $\xi=-i\pi/\alpha$) except the point exactly at $\theta=0$, which will be neglected below because its measure is zero while the value of $d{P}^{\mathbb{L}[01]}_{ddF}/d\Omega^{}_{\rm II}$ in (\ref{dP01FL}) below will be regular at $\theta=0$ after multiplied by $\sin^2\theta$. 

Eqs. (\ref{tauPofrtheta}) and (\ref{VUoftauM}) also imply $\tau^{}_{+,\mu} = - \tau^{}_{-,\mu} + O(r^{-1})$ for sufficiently large $r$ while $\theta\not= 0$ or $\pi$ [cf. eq. (\ref{Tauplusderiv})]. This gives
\begin{eqnarray}
&&\partial^{}_\mu \partial^{}_\nu \left( \frac{\Gamma^{}_+ -\Gamma^{}_-}{r}\right) = 
  \frac{1}{c^2 r} \left( \bar{n}^{}_\mu +\frac{\bar{u}^{}_\mu}{c}\right) 
  \left( \bar{n}^{}_\nu +\frac{\bar{u}^{}_\nu}{c}\right)
  \bigg[ \partial^{2}_{\tau^{}_+}\Gamma_+ -\partial^{2}_{\tau^{}_-}\Gamma_- \nonumber\\
&& \hspace{.5cm}\left. -3 \frac{\bar{a}^{}_{\bar{n}}}{c} \big( \partial^{}_{\tau^{}_+}\Gamma_+ + 
  \partial^{}_{\tau^{}_-}\Gamma_- \big) +
  \left( \frac{\dot{\bar{a}}^\rho}{c} \bar{n}^{}_\rho + 3\frac{\bar{a}_{\bar{n}}^2}{c^2} 
  - \frac{\bar{a}_\mu\bar{a}^\mu}{c^2}\right) \big(\Gamma_+ - \Gamma_- \big)\right]+
  O \left( \frac{1}{r^2} \right) \nonumber\\
\label{dmudnuGpGm}
\end{eqnarray}
according to (\ref{DtauDmu})--(\ref{DDrDmuDnu}).
For our uniformly accelerated electrons, $\dot{\bar{a}}^\rho \bar{n}^{}_\rho =0$, $\bar{a}^{}_{\bar{n}}(\tau) \equiv \bar{a}_\mu(\tau) \bar{n}^\mu(\tau) = a \cos\theta$, and $\bar{a}_\mu(\tau)\bar{a}^\mu(\tau) = a^2$. Thus, from (\ref{Thetan1UAC}) and (\ref{dmudnuGpGm}), we have
\begin{eqnarray}
&&\bar{u}^{}_\mu(\tau_-)\left[ \partial_{}^\mu{\cal Y}^\nu_1(x,x')-
  \partial_{}^\nu{\cal Y}^\mu_1(x,x')\right]
  =
\frac{q\pi}{\bar{m}c^2\alpha r}
\bar{n}^\nu_\perp(\tau^{}_-) \sin\theta 
\bigg[\partial^{2 }_{\tau^{}_+}\Gamma^{}_+ -\partial^{2}_{\tau^{}_-}\Gamma^{}_- \nonumber\\
&& \hspace{2cm} -3\alpha\cos\theta \big( \partial^{}_{\tau^{}_+}\Gamma^{}_+ + \partial^{}_{\tau^{}_- }\Gamma^{}_- \big) + \alpha^2 (3\cos^2\theta -1) \big( \Gamma^{}_+ -\Gamma^{}_- \big)\bigg]. \label{udT1dT1}  
\end{eqnarray}
Inserting (\ref{udT1dT1}) 
and (\ref{d21nurho})-(\ref{d01nurho}) into  (\ref{dP01}), we find that the contribution of ${\cal Z}^{(\lambda){\bf k}}_{1}$ to $d{P}^{[01]}_{ddF}/d\Omega^{}_{\rm II}$ is
\begin{eqnarray}
\frac{d{P}^{\mathbb{L}[01]}_{ddF}}{d\Omega^{}_{\rm II}} &\equiv& 
  \frac{\hbar q}{64\pi^4\varepsilon^{}_0}
  \lim_{r\to\infty} r^2 \lim_{x'\to x} 
  \left[\bar{n}_{}^\rho\big(\tau'_-\big) {\hat{\sf D}^{}_{\nu\rho}{}_{}^1}
  (x')\right]
    \bar{u}^{-}_\mu \big[ \partial^\mu{\cal Y}^\nu_1(x,x') -\partial^\nu {\cal Y}^\mu_1(x,x')\big]
    \nonumber\\
&=& \frac{3\hbar s}{32\pi^2\alpha} \sin^2\theta 
    \lim_{x'\to x} \bar{n}^\nu_\perp(\tau'_-)\bar{n}_\nu^\perp(\tau^{}_-)\times
    \nonumber\\&& \hspace{.5cm} \Big[ \cosh\alpha\tau'_-\partial^2_{\tau'_-} + 
    \alpha\left( 2\sinh\alpha\tau'_-+3\cos\theta\cosh\alpha\tau'_-\right)\partial^{}_{\tau'_-}
    \nonumber\\ &&\hspace{.6cm}
    +3\alpha^2\cos\theta\left(\sinh\alpha\tau'_-+\cos\theta\cosh\alpha\tau'_-\right)\Big]
    \Big[ \partial^{2}_{\tau^{}_+}\Gamma_+ -\partial^{2}_{\tau^{}_-}\Gamma_- 
    \nonumber\\&& \hspace{.6cm} 
    -3\alpha\cos\theta\big(\partial^{}_{\tau^{}_+}\Gamma_+ +\partial^{}_{\tau^{}_-}\Gamma_-\big)
    + \alpha^2 \left( 3\cos^2\theta-1\right)\big(\Gamma_+ - \Gamma_- \big)\Big],    
     \label{dP01FL}
\end{eqnarray}
up to $O(\alpha\epsilon^{}_0, \alpha \epsilon^{}_1)$.

\subsubsection{Early-time behavior}

Suppose the electron-field coupling is switched on at $\tau^{}_0=0$. 
At $\tau'_- \to \tau'_0 = \epsilon^{}_0$, we have
\begin{equation}
  \frac{d{P}^{\mathbb{L}}_{dd}}{d\Omega^{}_{\rm II}} \approx 
  \frac{9\hbar s\alpha^3}{8\pi^2 (1-e^{\alpha(\tau^{}_--\epsilon_0)})^2}\sin^2\theta\cos\theta,
\label{dPddLdOmegaEarlyApprox}
\end{equation}
which is mainly contributed by the interference terms $d{P}^{\mathbb{L}[01]}_{ddF}/ d\Omega^{}_{\rm II} + d{P}^{\mathbb{L}[10]}_{ddF}/ d\Omega^{}_{\rm II}$. 
At the initial moment $\tau^{}_- \to \epsilon^{}_0 + \epsilon^{}_1(0)$ with $\epsilon^{}_1(0)=
\epsilon'_1/\gamma(0) = \epsilon'_1$, it is approximately 
\begin{equation}
  \frac{d{P}^{\mathbb{L}}_{dd}}{d\Omega^{}_{\rm II}} \to
  \frac{9\hbar s\alpha}{8\pi^2 \epsilon'_1{}^2}\sin^2\theta\cos\theta   \label{dPLddInitApprox}
\end{equation}
while we always take $\epsilon'_1 = \epsilon^{}_0$. In the TEM in ref. \cite{TE89}, $\epsilon^{}_0 \approx 1.4 \times 10^{-14}$ s and $\alpha \approx 2.9 \times 10^8$ ${\rm s}^{-1}$
\cite{LH24}, and so this transient radiated power contributed by $d{P}^{\mathbb{L}}_{ddF}/d\Omega^{}_{\rm II}$ ($\sim 10^{-23}$ W) would be much greater than $d{P}^{\mathbb{L}}_{ddP}/d\Omega^{}_{\rm II}$ ($\sim 10^{-32}$ W) in the beginning, though it is still much smaller than the classical Larmor radiation ($\sim 10^{-20}$ W) without considering the classical transient radiation as the electron started to accelerate after birth. 

\subsubsection{Long-time behavior}
\label{LongTimeRadofUAC}

The contribution of ${\cal Z}^1_{z_{}^j}$ and ${\cal Z}^1_{p^{}_j}$ to $d{P}/d\Omega^{}_{\rm II}$ when
$\alpha\tau^{}_- \gg 1$ and $\epsilon^{}_1(\tau^{}_-)=\epsilon'_1/\bar{\gamma}(\tau^{}_-) \to 0$ is approximately
\begin{eqnarray}
&&\frac{d{P}^{\mathbb{L}}_{ddP}}{d\Omega^{}_{\rm II}} \approx
\frac{27 s \alpha^2}{32\pi \bar{m}}
   \sin^2\theta \cos^2\theta (1+\cos\theta)\Big\{ 4 
   \langle p_1^2 \rangle^{}_{\rm I}\big[ 2-\tanh \alpha\tau^{}_0 - 
\tanh\alpha(\tau^{}_0+\epsilon^{}_0) \big]+\nonumber\\
&&\hspace{.5cm}e^{2\alpha\tau^{}_-} (1+\cos\theta)\Big[ 
\alpha^2\bar{m}^2\langle\overline{\hat{z}_1^2}\rangle^{}_{\rm I}+\langle p_1^2 \rangle^{}_{\rm I}\big(1-\tanh\alpha\tau^{}_0\big) 
  \big(1-\tanh\alpha(\tau^{} +\epsilon^{}_0)\big)\Big] \Big\} \label{dPj1Presult}  
\end{eqnarray}
{from (I.4.48).}
Here the information on the initial conditions encoded in $\epsilon^{}_0$, $\langle p_1^2 \rangle^{}_{\rm I}$ , and {$\langle \overline{ \hat{z}_1^2} \rangle^{}_{\rm I}\equiv \frac{m^2}{\bar{m}^2}\langle z_1^2\rangle^{}_{\rm I}$} never vanishes.
For the $F$-part, from (\ref{dP11FPL}) and (\ref{dP01FL}), we find
\begin{eqnarray}
&&\frac{d{P}^{\mathbb{L}}_{ddF}}{d\Omega^{}_{\rm II}} \approx
  \frac{9 \hbar s^2 \alpha^4}{64\pi^2}\sin^2\theta\bigg\{ 
   1 - 64 \cos\theta - 149 \cos^2\theta - 48 \cos^3\theta + 44 \cos^4\theta +
   \nonumber\\ &&  
  \left[ \frac{12}{\alpha^2 \epsilon_0^2}(1+e^{-2\alpha\tau^{}_0}) - 
   24\big[\ln (\alpha \epsilon^{}_0)-\alpha \eta^{}_-\big] - 5 e^{-2\alpha \tau^{}_0} + 
 48 \ln \left(\cos\frac{\theta}{2}\right)\right] \cos^2\theta \sin^2\theta + \nonumber\\
 && e^{2\alpha\eta^{}_-}\left(-\frac{6}{\alpha^2 \epsilon_0^2} +\frac{5}{2} \right)\cos^2\theta(1+\cos\theta)^2  \bigg\} + 
   O\big( \alpha\epsilon^{}_0, \alpha\epsilon^{}_1, s^3\alpha^3\big) \label{dPj1Fresult}
\end{eqnarray}
($\eta^{}_- \equiv \tau^{}_- - \tau^{}_0$) for $e^{2\alpha\eta^{}_-}\gg 1$ and $\alpha\epsilon^{}_0 \ll 1$. With the parameter values in section \ref{SecNumRe}, in particular $\tau^{}_0=0$, we have
\begin{eqnarray}
&&\frac{d{P}^{\mathbb{L}}_{ddP}}{d\Omega^{}_{\rm II}} \approx
\frac{27 s \alpha^2}{32\pi \bar{m}}\Big[ \alpha^2\bar{m}^2\langle\overline{\hat{z}_1^2}\rangle^{}_{\rm I}+ \langle p_1^2 \rangle^{}_{\rm I} \Big] e^{2\alpha\tau^{}_-} 
   \sin^2\theta \cos^2\theta (1+\cos\theta)^2
\end{eqnarray}
dominates over
\begin{equation}
\frac{d{P}^{\mathbb{L}}_{ddF}}{d\Omega^{}_{\rm II}} \approx 
  -\frac{27\hbar s^2 \alpha^4}{32\pi^2\alpha^2 \epsilon_0^2}e^{2\alpha\tau^{}_-}
  \sin^2\theta \cos^2\theta (1+\cos\theta)^2  
\end{equation}
in the long-time regime.

Similar to the quadrupole-monopole correction (\ref{dPLmqdOmLTL}), in the last lines of (\ref{dPj1Presult}) and (\ref{dPj1Fresult}), one can see that the values of the results for $\alpha\tau_-\gg 1$ grow exponentially like $\pm e^{2\alpha\tau^{}_-}$ for $\alpha\tau^{}_- \gg 1$. 
The secular exponential growths here are produced by the zeroth derivatives in $d{P}^{\mathbb{L}[11]}_{ddP,F}/d\Omega^{}_{\rm II}$ in (\ref{dP11FPL}), namely,
\begin{equation}
  \frac{\mu^{}_0 q^2}{16\pi^2} \,\lim_{r\to \infty} r^2 \lim_{x'\to x} \eta^{\mu\nu} 
  \left[ \bar{u}^\sigma(\tau^{}_-)d^{(0)1}_{\mu\sigma}(x)\right] 
\left[\bar{n}^\rho(\tau'_-)d^{(0)1}_{\nu\rho}(x') \right] 
  \langle \hat{z}^1_{{}_{\underline{\texttt{0}}}}(\tau^{}_-),
\hat{z}^1_{{}_{\underline{\texttt{0}}}}(\tau'_-)\rangle , 
\label{dPL11dOIIzeroD}
\end{equation}
with $\langle \hat{z}^1_{{}_{ \underline{\texttt{0}}}}(\tau^{}_-), \hat{z}^1_{{}_{ \underline{\texttt{0}}}}(\tau'_-)\rangle = \langle \hat{z}^1_{{}_{\underline{\texttt{0}}}}(\tau^{}_-), \hat{z}^1_{{}_{\underline{\texttt{0}}}}(\tau'_-)\rangle^{}_P + \langle \hat{z}^1_{{}_{ \underline{\texttt{0}}}}(\tau^{}_-), \hat{z}^1_{{}_{ \underline{\texttt{0}}}}(\tau'_-)\rangle^{}_F$ going to the constant $\langle \hat{z}^1_{{}_{\underline{ \texttt{0}}}}\hat{z}^1_{{}_{ \underline{\texttt{0}}}}\rangle$ introduced in (\ref{dPLmqdOmLTL}). 
As for the remaining terms in $d{P}^{\mathbb{L}[11]}_{ddP,F}/d\Omega^{}_{\rm II}$, those in the form of $\eta^{\mu\nu}\left[ \bar{u}^\sigma(\tau^{}_-)d^{{(n)}1}_{\mu\sigma}(x)\partial^{{n}}_{\tau^{}_-} \right] \left[\bar{n}^\rho(\tau'_-)d^{{(n')}1}_{\nu\rho}(x')\partial^{{n'}}_{\tau'_-}\right]$
$ \langle \hat{z}^1_{{}_{\underline{\texttt{0}}}}(\tau^{}_-)\, \hat{z}^1_{{}_{\underline{\texttt{0}}}}(\tau'_-)\rangle$ with both ${n, n'} \ge 1$ decay out at late times, while those with ${n}=0$ or ${n'}=0$ but not ${n=n'=0}$ go to constants of time, and these constants may depend on the initial conditions.

The secular exponential growths similar to (\ref{dPLmqdOmLTL}) and (\ref{dPL11dOIIzeroD}) are not rare in nonequilibrium quantum field theory \cite{BdV03,GMPW23}. 
One may wonder if they are due to quantization of a field theory. The answer is no. As shown in appendix \ref{SecClRadUAC}, for 
{an ensemble} of classical point-charges shifted from $\bar{z}_{}^\mu(\tau)$ in (\ref{zmuUAC}) in the $x^1$ direction, the sum of their classical radiated power weighted by a normalized Gaussian function of the longitudinal deviation, centered at (\ref{zmuUAC}) with the variance equal to $\langle \hat{z}^1_{{}_{\underline{\texttt{0}}}}(\tau^{}_-), \hat{z}^1_{{}_{\underline{\texttt{0}}}}(\tau'_-)\rangle$, has almost the same growing behavior {in (\ref{dPdOmclDelta1Series}) and} in the time interval shown in figure \ref{FigQRadbyz1z1P} (left). Thus, such a behavior is not of quantum origin. It simply signals the failure of the series approximation in terms of multi-point correlators. 

{Moreover, if we go to higher orders, the series result in terms of the two-point correlators of the longitudinal deviations would be asymptotic rather than perturbative, as usually happening in quantum field theory. Nevertheless, 
our classical result (\ref{dPWdOmGauss}), (\ref{dPWdOmGaussSeries}) and (\ref{ExpectdPW}) suggests that these 
ill behaviors could be suppressed at late times after resummation \cite{Ra82}. Indeed, (\ref{dPdOmclDelta1Series}) is the Borel-Le Roy series \cite{Ha49} of (\ref{dPWdOmGaussSeries}), and (\ref{dPdOmclDelta1Exact}) is nothing but the Borel-Le Roy function, whose expectation value  (\ref{dPWdOmGauss}), the Borel-Le Roy sum, is finite at most of $\theta$, as in (\ref{ExpectdPW}). 
From the discussion below (\ref{ExpandEVinMW}) for the classical case, we expect that our quantum result would also have remaining divergences at $\theta=0$ and $\pi$, which could be regularized by considering a finite duration of acceleration instead of eternal uniform acceleration. Then the regularized resummed result would be finite at all $\theta$ and could approximate the late-time behavior of the quantum radiation emitted by an accelerated electron, as the classical one in figure \ref{FigQRadbyz1z1P} (right).} 

There are some $e^{-2\alpha \tau^{}_0}$ factors in the $F$-part contribution (\ref{dPj1Fresult}), which will become very large if $\tau^{}_0$ is negatively large. {Fortunately, the magnitude of (\ref{dPj1Fresult}) would not overtake the $P$-part contribution (\ref{dPj1Presult}), because similar behaviors in  (\ref{dPj1Presult}), though implicitly, can also be seen as follows.} Suppose an electron gun is uniformly accelerated and an electron is emitted at some moment $\tau^{}_0=\tau(t^{}_0) < 0$. Then we expect $\langle (\hat{z}_1)^2 \rangle^{}_{\rm I} = (c\epsilon_0)^2 =[ c\epsilon'_0/\gamma(\tau^{}_0)]^2 < (c\epsilon'_0)^2$ with a constant time scale $\epsilon'_0$ for an electron gun at rest, reflecting length contraction of the wavepacket in the longitudinal direction. This implies $\langle (\hat{p}_1)^2\rangle^{}_{\rm I} \sim \hbar^2/(4\langle (\hat{z}_1)^2 \rangle^{}_{\rm I})\propto \bar{\gamma}^2 (\tau^{}_0) = \cosh^2\alpha \tau^{}_0$ for a wavepacket of minimal uncertainty, $\langle \overline{\hat{z}_1^2} \rangle^{}_{\rm I}\langle \overline{\hat{p}_1^2} \rangle^{}_{\rm I} = \big( -i\langle [ \hat{z}^1, \hat{p}_1 ]\rangle^{}_{\rm I}\, \big)^2/4 =\hbar^2/4$ [see (I.3.50-52)]. 
From (I,4.49) the $\langle \hat{p}_1^2 \rangle^{}_{\rm I}$ terms in (\ref{dPj1Presult}), and so the value of the whole (\ref{dPj1Presult}), would be about $4 e^{2\alpha |\tau^{}_0|}$ times greater than the ones with $\tau^{}_0=0$. This helps to keep the dominance of $d{P}^{\mathbb{L}}_{ddP}/ d\Omega^{}_{\rm II}$ in (\ref{dPj1Presult}) over $d{P}^{\mathbb{L}}_{ddF}/d\Omega^{}_{\rm II}$ in (\ref{dPj1Fresult}) for large negative $\tau^{}_0$.

\subsubsection{Contribution by transverse deviations} 
\label{ZTContribution}

Since $\bar{u}^{\sf T}_- =0$, ${\sf T}=2,3$, the coefficients (\ref{d2jnrResult})-(\ref{d0jnrResult}) for ${\cal Z}^{\Omega}_{\sf T}$ in (\ref{F1mnD}) reduce to
\begin{eqnarray}
&& d^{(2){\sf T}}_{\nu \rho} \bar{n}^\rho_- = 
   \frac{1}{cr} \left( \delta^{\sf T}_\nu - \bar{n}_\nu^- \bar{n}^{\sf T}_- \right), 
   \label{d2Tnurho}\\
&& d^{(1){\sf T}}_{\nu \rho} \bar{n}^\rho_- = \frac{1}{cr} \left[  
   \frac{\bar{a}^-_{\bar{n}}}{c} \delta_\nu^{\sf T}+
   \left( 2 \frac{\bar{a}_\nu^-}{c} -3\frac{\bar{a}^-_{\bar{n}}}{c} \bar{n}_\nu^-\right)
   \bar{n}_-^{\sf T}\right] +O\left( r^{-2}\right), \\
&& d^{(0){\sf T}}_{\nu \rho} \bar{n}^\rho_- 
   = \frac{3}{cr} \frac{\bar{a}^-_{\bar{n}}}{c} \left( \frac{\bar{a}^-_\nu}{c}-
   \frac{\bar{a}^-_{\bar{n}}}{c}\bar{n}^-_\nu \right)\bar{n}^{\sf T}_-
   +O\left(r^{-2}\right), \label{d0Tnurho}
\end{eqnarray}
and $d^{(n){\sf T}}_{\nu\rho}\bar{u}^\rho_-/c =-d^{(n){\sf T}}_{\nu\rho}\bar{n}^\rho_- +O(r^{-2})$.
Then ${\cal Z}^{\sf T}_\Omega$ contributes
\begin{equation}
\frac{d{P}^{\mathbb{T}[11]}_{ddP,F}}{d\Omega^{}_{\rm II}} \equiv 
  \frac{3 s \bar{m}c}{8\pi} \lim_{r\to\infty} r^2 \sum_{{\sf T}=2,3}\lim_{x'\to x}
  \left[{\bar{u}_{-}^\mu} \hat{\sf D}^{}_{\mu\rho{\sf T}}(x)\right] 
  \left[{\bar{n}^{-}_\nu} \hat{\sf D}_{}^{\rho\nu{\sf T}}(x')\right] 
  \langle\hat{z}^{}_{\sf T}\big(\tau^{}_-\big),
  \hat{z}_{}^{\sf T} \big(\tau'_-\big)\rangle^{}_{P,\, F}
\label{dPTTFPL}
\end{equation}
since $\langle\hat{z}_{}^i (\tau),\hat{z}_{}^j(\tau)\rangle^{}_{P, F} \propto \eta_{}^{ij}$ here. Inserting (I.4.61) into the above expression, the result for the $P$-part is relatively simple:
\begin{eqnarray}
\frac{d{P}^{\mathbb{T}}_{ddP}}{d\Omega^{}_{\rm II}} = 
  \frac{d{P}^{\mathbb{T}[11]}_{ddP}}{d\Omega^{}_{\rm II}}  &=& 
\frac{3 s \alpha^2}{8\pi\bar{m}} \Big\{ 9\alpha^2 m^2 \left( \langle \overline{\hat{z}_2^2} \rangle^{}_{\rm I} \cos^2 \varphi + \langle \overline{\hat{z}_3^2} \rangle^{}_{\rm I} \sin^2 \varphi \right) \cos^2\theta \sin^4 \theta  \nonumber\\ &&  
+\left(\langle p_2^2 \rangle^{}_{\rm I} \cos^2 \varphi+ \langle p_3^2 \rangle^{}_{\rm I} \sin^2 \varphi \right)\Big[ 4-11 \cos^2\theta+9\cos^4\theta \nonumber\\ &&+
6 \alpha\eta^{}_- \cos\theta(2-5\cos^2\theta+3\cos^4\theta)
+ 9 \alpha^2\eta_-^2 \cos^2 \theta \sin^4\theta \Big] \nonumber\\
&& +O(s\alpha, \alpha \epsilon^{}_0, \alpha \epsilon^{}_1)\Big\}.
\label{dPddPTdOmega}
\end{eqnarray}
There are power-law growing terms $(\alpha\eta^{}_-)^n$ up to $n=6$ in the higher orders of $s\alpha$, which have been neglected above since they are not significant until $\alpha\eta_- \sim 1/(s \alpha)$, or $\eta^{}_- \equiv \tau^{}_- - \tau^{}_0 \sim 10^6$ s with the parameter values in section \ref{SecNumRe}, {which is way longer than the time scales of the experiments in TEM.}
Regarding the $F$-part, our result for $d{P}^{\mathbb{T} [11]}_{ddF}/d\Omega^{}_{\rm II}$ with (\ref{zTzTFcloseform}) may be too complicated to offer a clear picture. Thus we only present their approximated forms as well as the numerical results in the early-time and long-time regimes below.

For $d{P}^{[01]}_{ddF}/ d\Omega^{}_{\rm II}$ and $d{P}^{[10]}_{ddF}/d\Omega^{}_{\rm II}$, inserting ${\cal Z}^{(\lambda){\bf k}}_{\sf T}$ given in (I.4.26), and applying the identities (I.3.17) and (I.4.57) as well as 
\begin{eqnarray}
  &&\epsilon^0_{(\lambda){\bf k}} {\cal E}^{(\lambda){\bf k}*}_{j0} = -i k^{}_j = 
      {\cal E}^{(0){\bf k}*}_{j0} = -{\cal E}^{{\bf k}*}_{(0)\,j0}, \hspace{1cm}
  \epsilon^j_{(\lambda){\bf k}} {\cal E}^{(\lambda){\bf k}*}_{j'0} = 
      -i \frac{\omega}{c}\delta^j_{j'} \label{eEj0}\\
  &&\epsilon^0_{(\lambda){\bf k}} {\cal E}^{(\lambda){\bf k}*}_{{\sf T} 1} = 0,\hspace{1cm}
  \epsilon^1_{(\lambda){\bf k}} {\cal E}^{(\lambda){\bf k}*}_{{\sf T} 1} = -i k^{}_{\sf T},
  \hspace{1cm}
  \epsilon^{{\sf T}'}_{(\lambda){\bf k}} {\cal E}^{(\lambda){\bf k}*}_{{\sf T}1} = 
      i k^{}_1 \delta^{{\sf T}'}_{\sf T}, \label{eET1}
\end{eqnarray}
which can be derived from (I.3.8) and (I.4.28), we have
\begin{eqnarray}
&&{\cal Y}^\nu_{\sf T} = 
    \int_0^\infty \frac{\omega d\omega}{c^3}\int_0^{2\pi} d\tilde{\varphi} \int_0^{\pi} \sin \tilde{\theta} d \tilde{\theta} \, e_{}^{-i\frac{\omega}{c}x_{}^0+i {\bf k \cdot x}-\frac{\epsilon}{2}\omega} \, \frac{qc}{\bar{m}} \int_{\tau'_0}^{\tau'_-} d\tilde{\tau}K^{}_\perp(\tau'_-,\tilde{\tau})
(1+s\partial^{}_{\tilde{\tau}})\nonumber\\ &&\hspace{.7cm}\left\{ \left[ 
\left(-ik^{}_{\sf T}\delta_0^\nu-\frac{i\omega}{c}\delta_{\sf T}^\nu\right)\cosh\alpha\tilde{\tau}+ \left(ik^{}_1\delta_{\sf T}^\nu-ik^{}_{\sf T}\delta_1^\nu\right)\sinh\alpha\tilde{\tau}\right]
e^{i k^{}_\mu \bar{z}^\mu(\tilde{\tau})-\frac{\epsilon}{2}\omega}\right\}
\end{eqnarray}
with ${\sf T}=2,3$ and $K^{}_\perp$ given in (I.4.23). 
Following the same pathway as in section \ref{Z1Contribution}, we obtain
\begin{eqnarray}
&&{\cal Y}^\nu_{\sf T} =-\frac{q\pi}{s\bar{m} \alpha^2  c} 
   \sum_{\mathbb{B}=+,-} 
   \Big[ \left(\delta_0^\nu+\mathbb{B}\delta_1^\nu\right) \partial^{}_{\sf T}
   -\delta_{\sf T}^\nu \left( \partial^{}_{0}+\mathbb{B} \partial^{}_{1}\right) \Big]
   \left( \frac{\tilde{\Gamma}^{}_{+\mathbb{B}} - \tilde{\Gamma}^{}_{-\mathbb{B}}}{r}\right), 
   \label{ThetanTUAC}   
\end{eqnarray}
where 
\begin{eqnarray}
&&\tilde{\Gamma}^{}_{\pm\mathbb{B}} \equiv 
   e^{\mathbb{B}\alpha\tau'_-}\Big\{ \left(1+\mathbb{B} s\alpha\right)
    \left[ g^{}_{\mathbb{B}}\left({\rm T}_{\pm -}\right)-
           g^{}_{\mathbb{B}+}\left({\rm T}_{\pm -}\right)\right] -
     s\left[ \dot{g}^{}_{\mathbb{B}}\left({\rm T}_{\pm -}\right)-
           \dot{g}^{}_{\mathbb{B}+}\left({\rm T}_{\pm -}\right) \right] \Big\}
           -\nonumber\\ 
&&  e^{\mathbb{B}\alpha\tau'_0}\left\{ \left(1+\mathbb{B}s\alpha\right) 
    \left[ g^{}_{\mathbb{B}}\left({\rm T}_{\pm 0}\right)-
 e^{-\frac{s\alpha}{\varsigma}\eta'_-}g^{}_{\mathbb{B}+}\left({\rm T}_{\pm 0}\right)\right]
    -s\left[ \dot{g}^{}_{\mathbb{B}}\left({\rm T}_{\pm 0}\right)-
    e^{-\frac{s\alpha}{\varsigma}\eta'_-} \dot{g}^{}_{\mathbb{B}+}\left({\rm T}_{\pm 0}
    \right)\right] \right\} \nonumber\\
\end{eqnarray}
with $\varsigma \equiv 1+s^2\alpha^2$, ${\rm T}^{}_{\pm -}$ and ${\rm T}^{}_{\pm 0}$ given below (\ref{Gammapm}), $g^{}_\pm$ given in (\ref{gpdef}) and (\ref{gmdef}), and 
\begin{eqnarray}
  g^{}_{++}(T) &\equiv& \int_{-\infty}^{\infty} \frac{ i d\kappa}{1-e^{-2\pi\kappa/\alpha}} \,\,
  \frac{e^{-i\kappa T}}{i\kappa + \alpha +\frac{s\alpha^2}{\varsigma}} =
  \frac{1}{1+\frac{s \alpha}{\varsigma}}\left[ F_{1|+}(T)-1\right], \\
  g^{}_{-+}(T) &\equiv& \int_{-\infty}^{\infty} \frac{ i d\kappa}{1-e^{-2\pi\kappa/\alpha}} \,\,
  \frac{e^{-i\kappa T}}{i\kappa - \alpha +\frac{s\alpha^2}{\varsigma}}\nonumber\\
  &=& -\frac{1}{1-\frac{s \alpha}{\varsigma}} F_{-1|+}(T) -
  \frac{2i\pi e^{-\alpha\left(1-\frac{s\alpha}{\varsigma}\right)T}}
       {1-e^{2i\pi\left(1-\frac{s\alpha}{\varsigma}\right)}}
\end{eqnarray}
with $F_{n|\pm}(T) \equiv {}_2F_1 \left( 1, n\pm \frac{s\alpha}{\varsigma}, 
n+1\pm\frac{s\alpha}{\varsigma}, e^{-\alpha T} \right)$ for $T>0$.

Inserting (\ref{ThetanTUAC}) and (\ref{d2Tnurho})-(\ref{d0Tnurho}) into (\ref{dP01}), we obtain
\begin{eqnarray}
&&\frac{d{P}^{\mathbb{T}[01]}_{ddF}}{d\Omega^{}_{\rm II}} = 
  \frac{\hbar q}{64\pi^4\varepsilon^{}_0}
  \lim_{r\to\infty} r^2 \lim_{x'\to x} \sum^{}_{{\sf T}=2,3} \left[
  \bar{n}_{}^\rho\big(\tau'_-\big) {\hat{\sf D}^{}_{\nu\rho}{}_{}^{\sf T}} 
  (x')\right]\bar{u}^{-}_\mu 
  \big[ \partial^\mu {\cal Y}^\nu_{\sf T} - \partial^\nu {\cal Y}^\mu_{\sf T}\big]
   \nonumber\\
&=& \frac{3\hbar}{32\pi^2 \alpha^2} 
  \lim_{x'\to x}
  \sum_{\mathbb{B}=+,-} e^{-\mathbb{B}\alpha\tau^{}_-}(\mathbb{B}1-\cos\theta) \bigg[ 
  \left(\mathbb{B}1-\cos\theta -2\cos\theta\sin^2\theta
  \sinh^2\frac{\alpha\epsilon^{}_1}{2} \right)\partial_{\tau'_-}^2   \nonumber\\ 
&& \hspace{.5cm} +\alpha \left( (\mathbb{B}1-\cos\theta)(2+3\cos\theta) + 2(3 \cos^2\theta -2)
    \sin^2\theta\sinh^2\frac{\alpha\epsilon^{}_1}{2}\right) \partial^{}_{\tau'_-}\nonumber\\
&&\hspace{.5cm} +3\alpha^2 \cos\theta \sin^2\theta \left( 1+ 2
    \sin^2\theta\sinh^2\frac{\alpha\epsilon^{}_1}{2} \right)\bigg] \nonumber\\ 
&& \hspace{.5cm} 
  \bigg[ \ddot{\tilde{\Gamma}}^{}_{+\mathbb{B}}-\ddot{\tilde{\Gamma}}^{}_{-\mathbb{B}}
  -3\alpha\cos\theta \left( \dot{\tilde{\Gamma}}^{}_{+\mathbb{B}}+
  \dot{\tilde{\Gamma}}^{}_{-\mathbb{B}}\right)+\alpha^2(3\cos^2\theta -1)
  \left(\tilde{\Gamma}^{}_{+\mathbb{B}}-\tilde{\Gamma}^{}_{-\mathbb{B}}\right) \bigg]
\label{dPT01FdOmegaFinal}
\end{eqnarray}
with $\epsilon^{}_1 = \tau^{}_- - \tau'_-$, after some algebra.

\subsubsection{Early-time behavior}

With the parameter values in section \ref{SecNumRe}, when $\tau'_-=\tau^{}_--\epsilon^{}_1 \sim O(\epsilon^{}_0)$ while $\tau'_- >\epsilon_0$, 
we have
\begin{eqnarray}
  \frac{d{P}^{\mathbb{T}}_{ddP}}{d\Omega^{}_{\rm II}} &\approx&
  \frac{3 s \alpha^2}{8\pi\bar{m}}\Big[ 9\alpha^2 m^2 
  \left( \langle \overline{\hat{z}_2^2} \rangle^{}_{\rm I}\cos^2\varphi + \langle \overline{\hat{z}_3^2} \rangle^{}_{\rm I}\sin^2\varphi \right)
    \cos^2\theta \sin^4 \theta \nonumber\\ &&+
  \left(\langle p_2^2\rangle^{}_{\rm I}\cos^2\varphi +
  \langle p_3^2\rangle^{}_{\rm I}\sin^2\varphi \right) 
  \left( 4-11\cos^2\theta+9\cos^4\theta \right) \Big]
\label{dPddPTdOmegaEarlyApprox}
\end{eqnarray}
from (\ref{dPddPTdOmega}), and
\begin{eqnarray}
  \frac{d{P}^{\mathbb{T}}_{ddF}}{d\Omega^{}_{\rm II}} &\approx& 
  \frac{d{P}^{\mathbb{T}[01]}_{ddF}}{d\Omega^{}_{\rm II}}+
  \frac{d{P}^{\mathbb{T}[10]}_{ddF}}{d\Omega^{}_{\rm II}}
  \approx \frac{3\hbar s\alpha^3 }{8\pi^2 (e^{\alpha\tau'_-}-1)^2} \cos\theta (3\cos^2\theta-1),
\label{dPddFTdOmegaEarlyApprox}
\end{eqnarray}
which is dominated by the $O(s \alpha^3)$ 
part of the $\partial^{}_{\tau'_-}$ terms in (\ref{dPT01FdOmegaFinal}). 
At the initial moment $\tau'_-\to\tau'_0=\epsilon_0$, 
\begin{equation}
\frac{d{P}^{\mathbb{T}}_{ddF}}{d\Omega^{}_{\rm II}} 
  \to 
\frac{3\hbar s\alpha}{8\pi^2 \epsilon_0^2}\cos\theta (3\cos^2\theta-1),
\label{dPddTdOmegaEarlyInit}
\end{equation}
for $\alpha \epsilon^{}_0\ll 1$. In the TEM in ref. \cite{TE89}, the early-time transient radiated power contributed by $d{P}^{\mathbb{T}}_{ddP}/d\Omega^{}_{\rm II}$ ($\sim 10^{-29}$ W) is dominated by the $\langle p_{\sf T}^2\rangle^{}_{\rm I}$ terms in (\ref{dPddPTdOmegaEarlyApprox}) and much smaller than $d{P}^{\mathbb{T}}_{ddF}/d\Omega^{}_{\rm II}$ [ $\sim 10^{-23}$ W, which is roughly one third of the magnitude of $d{P}^{\mathbb{L} }_{ddF}/d\Omega^{}_{\rm II}$ dominating (\ref{dPLddInitApprox})], while the latter is still much less than the classical Larmor radiation ($\sim 10^{-20}$ W). 
The transient radiation (\ref{dPddTdOmegaEarlyInit}) [$\propto \cos\theta (3\cos^2\theta-1)$] has the maximum absolute values at $\theta=0$ and $\pi$, where the blind spots of the classical Larmor radiation are located.

\subsubsection{Long-time behavior}

In the long-time regime but not at very late times ($e^{-\alpha\eta^{}_-} \ll 1$ while $e^{-s \alpha^2\eta^{}_-}\approx 1$), the expression (\ref{dPddPTdOmega}) for $d{P}^{\mathbb{T}}_{ddP}/d\Omega^{}_{\rm II}$ still works with the parameter values in section \ref{SecNumRe} where $s\alpha \sim 10^{-15} \ll 1$. As for the $F$-part, the combined angular radiated power contributed by ${\cal Z}^2_{(\lambda){\bf k}}$ and ${\cal Z}^3_{(\lambda){\bf k}}$ {is approximately} 
\begin{eqnarray}
 &&\frac{d{P}^{\mathbb{T}}_{ddF}}{d\Omega_{\rm II}}=  
  \frac{d{P}^{\mathbb{T}[11]}_{ddF}}{d\Omega_{\rm II}} + \frac{d{P}^{\mathbb{T}[01]}_{ddF}}
  {d\Omega_{\rm II}}+\frac{d{P}^{\mathbb{T}[10]}_{ddF}}{d\Omega_{\rm II}} 
  \nonumber\\ && {\approx} 
  -\frac{3 \hbar s \alpha^3}{32\pi^2} \bigg\{ 16 - 15 \cos \theta + 6 (-8 + \pi^2)\cos^2\theta 
  + 52 \cos^3\theta + 12 (3 - \pi^2) \cos^4\theta \nonumber\\ && \hspace{.5cm}
  - 37 \cos^5\theta + 6 \pi^2 \cos^6\theta  
  - 24 (2 \cos\theta - 5 \cos^3\theta + 3 \cos^5\theta )
    \log\left(\cos^2 \frac{\theta}{2}\right) \nonumber\\ && \hspace{.5cm}
   - 36 (\cos^2\theta - 2\cos^4\theta + \cos^6\theta) {\rm Li}^{}_2\left(
   \frac{\cos\theta-1}{\cos\theta+1}\right)\bigg\} + 
   O(s^2\alpha^4, \alpha\epsilon^{}_0, \alpha\epsilon^{}_1),
\label{dPddFTdOmegaLT}
\end{eqnarray} 
with the polylogarithm function ${\rm Li}^{}_n(z)$.
The above result is independent of time [see the behavior of the orange curves in the lowest two plots of the middle column in figure \ref{dPFPdthetaTauAT}]. The neglected $s^2\alpha^4$-terms in (\ref{dPddFTdOmegaLT}) behave like $(\alpha\eta^{}_-)^3$, 
which will overtake the above quasi-stationary value at later time: For the parameter values in section \ref{SecNumRe}, this occurs at about O($100\tau^{}_a$), when even higher-order corrections may become more important and the above expression of $O(s)$ will not be a good approximation.

\subsection{Numerical results}
\label{SecNumRe}

Below we look at the cases with cylindrical symmetry about the $x_{}^1$-direction. We consider a Gaussian wavepacket of electron created and started to be accelerated at the moment $t^{}_0 =\tau^{}_0=0$ with $\langle \overline{\hat{z}_2^2}\rangle^{}_{\rm I} = \langle \overline{\hat{z}_3^2}\rangle^{}_{\rm I} $ and $\langle p_2^2\rangle^{}_{\rm I} = \langle p_3^2\rangle^{}_{\rm I} $. We will show a few examples of the distribution of radiated power in polar angle $\theta$, 
\begin{equation}
  \frac{d{P}}{d\theta} = \int_0^{2\pi} d\varphi \frac{d{P}}{d\Omega^{}_{\rm II}}.
\end{equation}
with the parameter values of the TEM in \cite{TE89}.  Other parameter values are the same as those in sections 3.2.2 and 4.4 of ref. \cite{LH24}.

\subsubsection{Cancellation of divergences}

Before performing numerical calculations, one needs to analytically identify the potentially divergent terms in each component in the limits of $s$, $\alpha$, or $\epsilon^{}_1\to 0$ to check if all these divergences cancel and our finite results are reliable.
This is because in $d{P}^{\mathbb{L},\mathbb{T}}_{qmP,F} / d\Omega^{}_{\rm II}$ and $d{P}^{\mathbb{L},\mathbb{T}[BB']}_{ddP,F}/ d\Omega^{}_{\rm II}$, there are some large terms of the order of $(\alpha s)^{-n}$, $n>0$ ($\alpha s \sim 10^{-15}$ to $10^{-16}$ in TEM). When they are summed together, the result can be much smaller than each of these large terms in absolute value, and so one must be careful in numerical calculations to reduce the effect of round-off errors. Moreover, as we have argued in ref. \cite{LH24}, the regulator $\epsilon^{}_1$ should be time dependent in the lab frame. Specifically, $\epsilon^{}_1(t) = \epsilon'_1/\bar{\gamma}(t)$ (with constant $\epsilon'_1$) goes to zero as the electron speed approaches the speed of light and $\bar{\gamma}$ goes to infinity, which can make some terms in the F-parts of (\ref{dPqmLdOmega}), (\ref{dPqmTdOmega}), and (\ref{dP11FPL}), as well as (\ref{dP01FL}) divergent at late times. 

We have examined that {all the $s^{-n}$, $\epsilon_1^{-n}$ ($n=1,2,3,4$), and $\ln \alpha\epsilon^{}_1$ divergences occurring in different terms of $d{P}^{\mathbb{L}[BB']}_{ddP,F}/d\Omega^{}_{\rm II}$ (contributed by the longitudinal deviation) cancel when added together, similar to the case of electrons at rest in section \ref{SecQRadRest}. So do those divergences in $d{P}^{\mathbb{T}[BB']}_{ddP,F}/d\Omega^{}_{\rm II}$ (contributed by the transverse deviations) to the leading order when $\varsigma = 1 + s^2\alpha^2$ is replaced by $1$ (see the discussion below eq. (I.4.33) \cite{LH24}). The remaining $\epsilon_1^{-n}$ `divergences' in $d{P}^{\mathbb{T}[BB']}_{ddP,F}/d\Omega^{}_{\rm II}$ of $O(s^2\times s^2 \alpha^2)$ are actually negligibly small with the parameter values in this section. A more complete check for cancellation needs the $O(s^2)$ terms of $\langle \hat{z}_{}^{\sf T}, \hat{z}_{}^{\sf T}\rangle^{}_F$ neglected in eqs. (I.4.60) and (\ref{zTzTFcloseform}), which is beyond the approximation in this paper.}
The $\epsilon_1^{-n}$ divergences also cancel in $d{P}^{}_{qm}/d\Omega^{}_{\rm II}$ and $d{P}^{}_{mq}/d\Omega^{}_{\rm II}$, while there are $\ln \alpha \epsilon^{}_1(\tau^{}_-)$ terms that survive in $d{P}^{}_{qm}/d\Omega^{}_{\rm II}$ and $d{P}^{}_{mq}/d\Omega^{}_{\rm II}$. Fortunately, those terms never dominate because each $\ln \alpha \epsilon^{}_1(\tau^{}_-)$ is associated with a factor $(\sinh\alpha\tau^{}_- + \cosh\alpha\tau^{}_- \cos\theta)^2/\cosh^2 \alpha \tau^{}_-$ and so the combination behaves like $\ln \gamma(\tau^{}_-) \sim \tau^{}_-$ in the long-time regime, which is not significant compared with the exponentially growing terms in that regime. 

\subsubsection{Early-time behaviors}

\begin{figure}[ht]
\includegraphics[width=4.9cm]{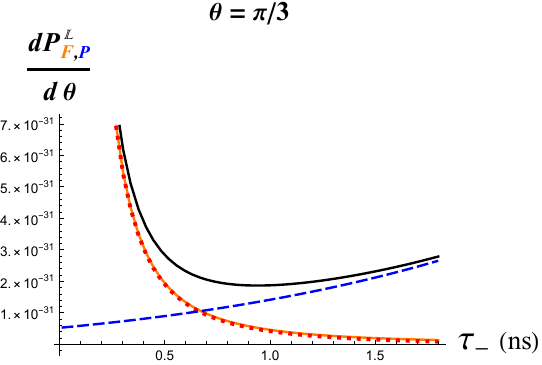}
\includegraphics[width=4.9cm]{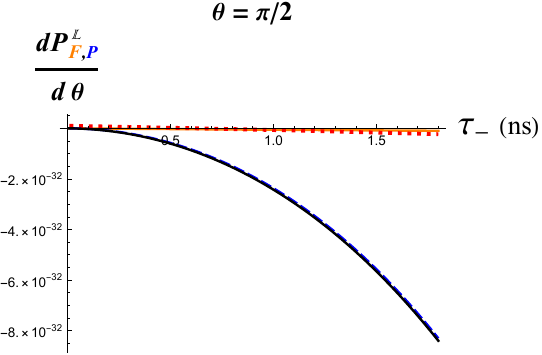}
\includegraphics[width=4.9cm]{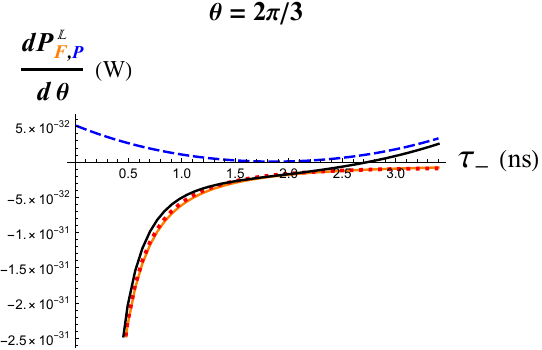} \\
\includegraphics[width=4.9cm]{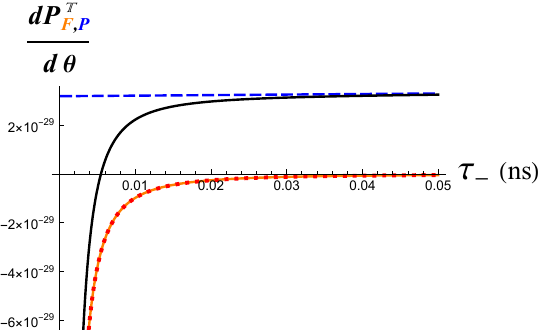}
\includegraphics[width=4.9cm]{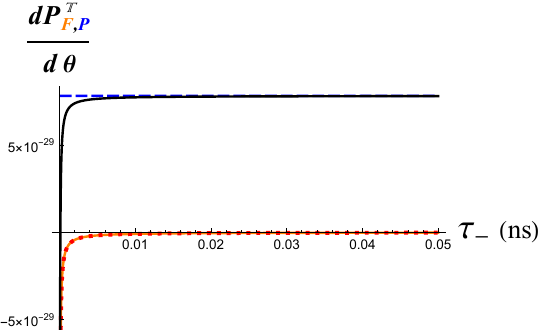}
\includegraphics[width=4.9cm]{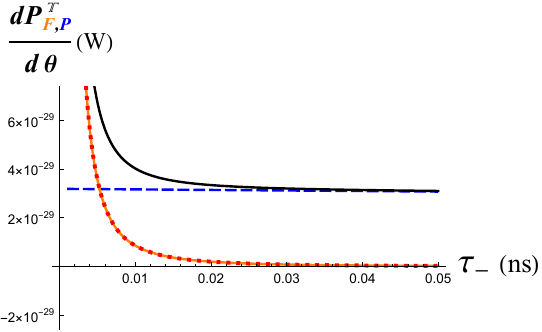}
\caption{Early-time radiation contributed by the longitudinal (upper row) and transverse (lower) deviations at $\theta= \pi/3$ (left), $\pi/2$ (middle), and $2\pi/3$ (right). The blue dashed and orange curves represent the $P$-part and $F$-part of the angular radiated power, respectively, and the black curves represent the sums of the $P$- and $F$-parts.
Early-time transient behaviors are dominated by $d{P}^{{\mathbb L},{\mathbb T}[01]}_{ddF}/d\theta+ d{P}^{{\mathbb L},{\mathbb T}[10]}_{ddF}/d\theta$ (red dotted) except those around the angles where $d{P}^{\mathbb L}_F/d\theta =0$ or $d{P}^{\mathbb T}_F/d\theta =0$ at the initial moment.}
\label{RadTauThetaEarly}
\end{figure}

At early times, the quantum radiation produced by the longitudinal deviation is dominated by the transient in the $F$-part, i.e. $d{P}^{\mathbb{L}}_F/{d\theta}$, 
except for the angles around $\theta=\pi/2$ where $d{P}^{\mathbb{L}}_F/{d\theta} = 0$ [figure \ref{RadTauThetaEarly} (upper row)]. At angles where $d{P}^{\mathbb{L}}_F/ {d\theta}$ is not too small, $d{P}^{\mathbb{L}}_F/{d\theta}$ is dominated by the interference terms $d{P}^{\mathbb{L}[01]}_{ddF}/{d\theta} + d{P}^{\mathbb{L}[10]}_{ddF}/ {d\theta}$. 
It decays as $e^{-2\alpha \tau^{}_-}$ and loses dominance after $\tau^{}_- \sim 10^{-10}$ s when $d{P}^{\mathbb{L}}_P/{d\theta}$ takes over [e.g., after $\tau^{}_- \approx 0.7$ ns $\approx 50000 \epsilon^{}_0 \approx \tau^{}_a/2$ for $\theta=\pi/3$ in figure \ref{RadTauThetaEarly} (upper left), where $\tau_a = 1.50 
\times 10^{-9}$ s in the electron's proper time ($t_a=1.54 
\times 10^{-9}$ s in the laboratory time) is the estimated time \cite{LH24} that the electron exits the acceleration tube in ref. \cite{TE89}].

For the quantum radiation contributed by the transverse deviations, $d{P}^{\mathbb{T}}_F /{d\theta}$ also dominates over $d{P}^{\mathbb{T}}_P /{d\theta}$ initially [figure \ref{RadTauThetaEarly} (lower row)]. Then, as the proper time $\tau^{}_-$ increases, $d{P}^{\mathbb{T}}_F/{d\theta}$ decays as $e^{-2\alpha \tau^{}_-}$ and its absolute value becomes less than $d{P}^{ \mathbb{T}}_P/{d\theta}$ after $\tau^{}_- \sim 10^{-11}$ s [e.g., after $\tau^{}_- \approx 0.007$ ns $\approx 500\epsilon_0 \approx \tau^{}_a/210$ for $\theta=\pi/3$, see figure \ref{RadTauThetaEarly} (lower left)]. This time scale is shorter than its longitudinal counterpart.

Such a large transient radiated power can be seen in similar systems. For example, there is a contribution proportional to $\frac{a e^{-\gamma\eta^{}_-}}{(e^{a\eta_-}-1)}\cos\Omega\eta^{}_-$ by $\Theta_{--}$ given in eq.(A3) of ref. \cite{LH06} to the early-time radiation of an Unruh-DeWitt harmonic-oscillator detector. 
This may be considered as an impact of suddenly switching on the coupling between single electrons and infinitely many degrees of freedom of the EM fields, while an electron at rest does not have this transient radiation -- similar transient behaviors occur only for the bound field [e.g., the $1/[r^3(t^{}_{-0})^{2}]$ and $1/[r^4 t^{}_{-0}]$ terms in (\ref{dP01ddFrest})].

\subsubsection{Middle-time behaviors}

\begin{figure}[ht]
\includegraphics[width=4.9cm]{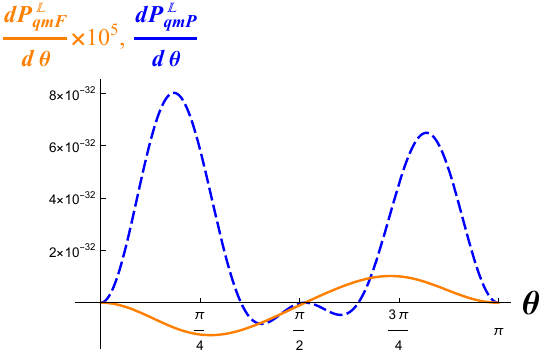}
\includegraphics[width=4.9cm]{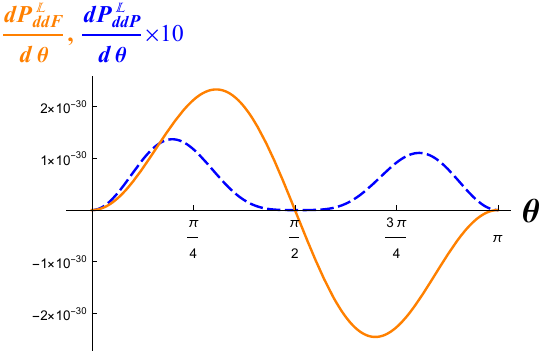}
\includegraphics[width=4.9cm]{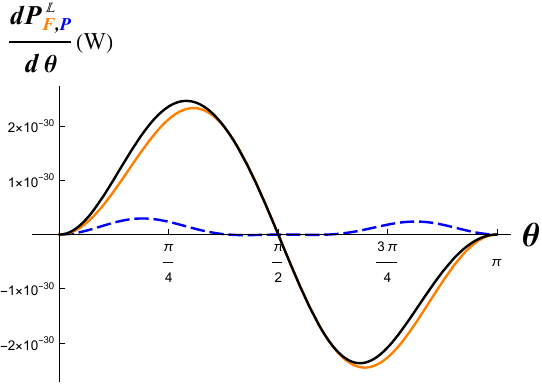}\\
\includegraphics[width=4.9cm]{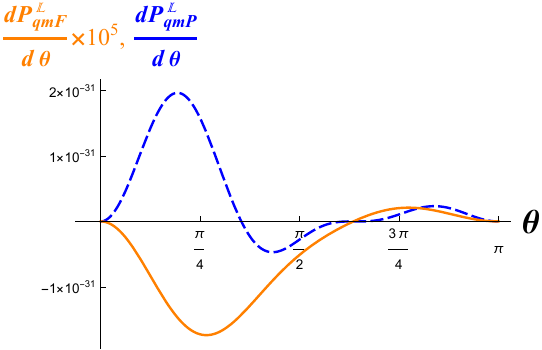}
\includegraphics[width=4.9cm]{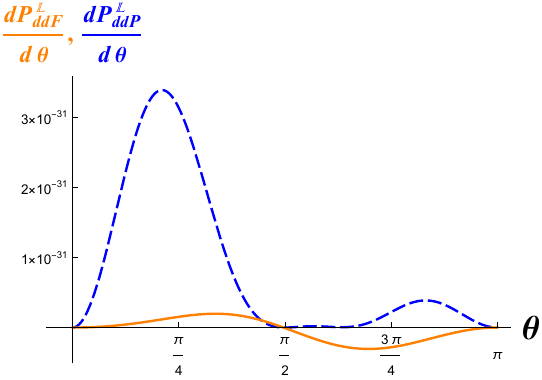}
\includegraphics[width=4.9cm]{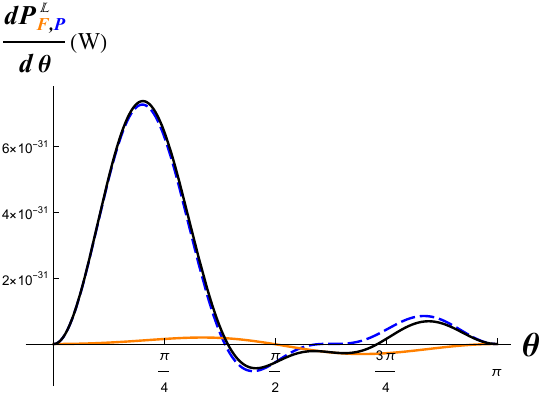}\\
\includegraphics[width=4.9cm]{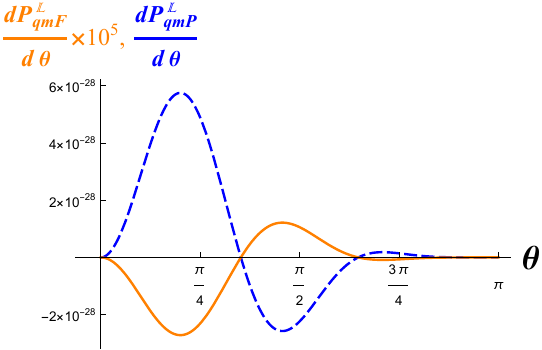}
\includegraphics[width=4.9cm]{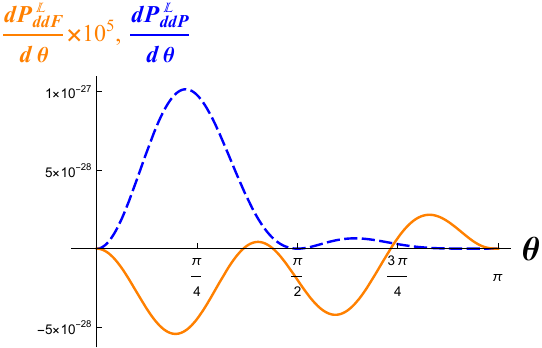}
\includegraphics[width=4.9cm]{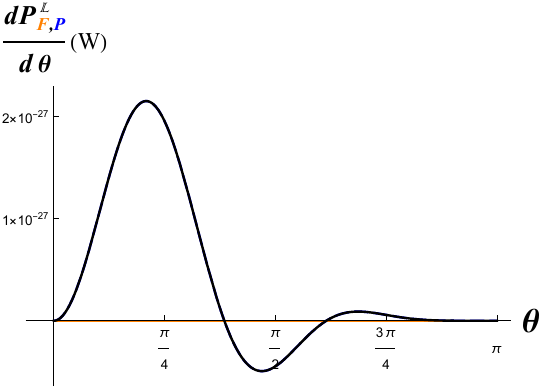}\\
\includegraphics[width=4.9cm]{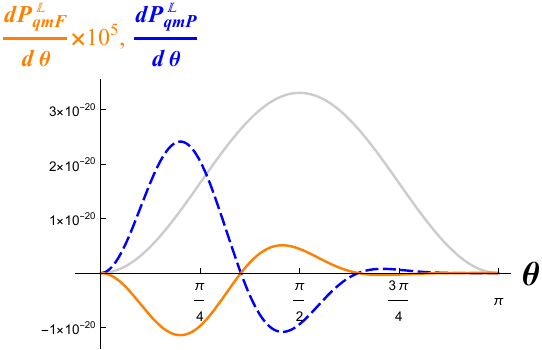}
\includegraphics[width=4.9cm]{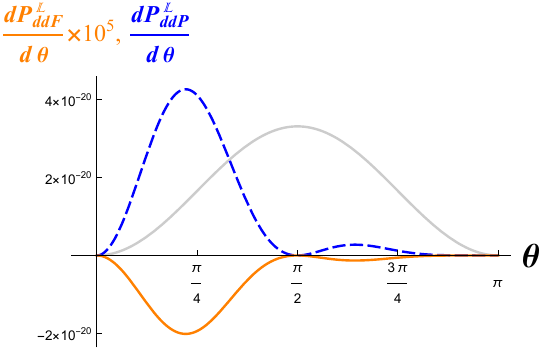}
\includegraphics[width=4.9cm]{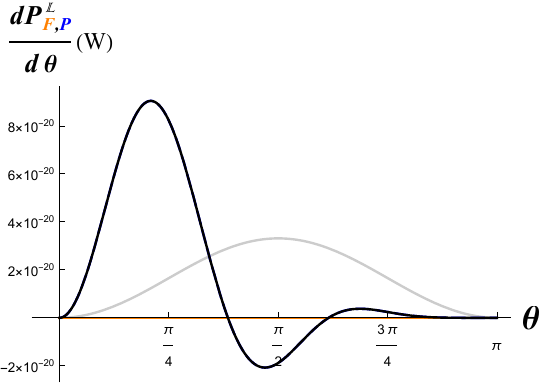}
\caption{From top row to bottom we show the angular distributions of quantum radiated power contributed by the longitudinal deviation at $\tau^{}_- = 0.1 \tau^{}_a$, $\tau^{}_a$, $10\tau^{}_a$, and $30\tau^{}_a$ with $\tau^{}_a = 1.50 \times 10^{-9}$ s. The blue dashed and orange curves represent the $P$-part and $F$-part of the angular radiated power, respectively, and the black curves in the right plots represent the sums of the $P$- and $F$-parts. The gray curves in the bottom row represent the classical radiated power ${d\bar{P}}/d\theta$ in (\ref{ClassicaldRdOmega}) emitted by a point-like UAC.}
\label{dPFPdthetaTauAL}
\end{figure}

\begin{figure}[ht]
\includegraphics[width=4.9cm]{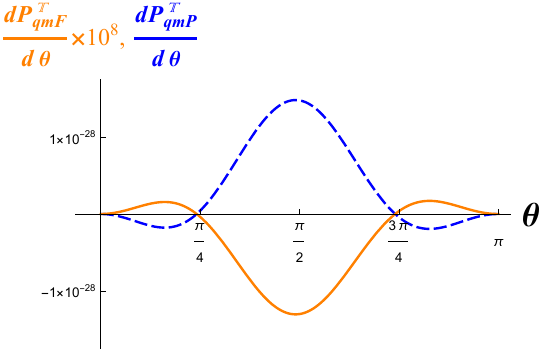}
\includegraphics[width=4.9cm]{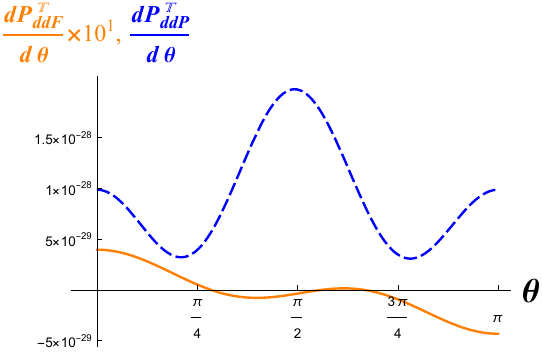}
\includegraphics[width=4.9cm]{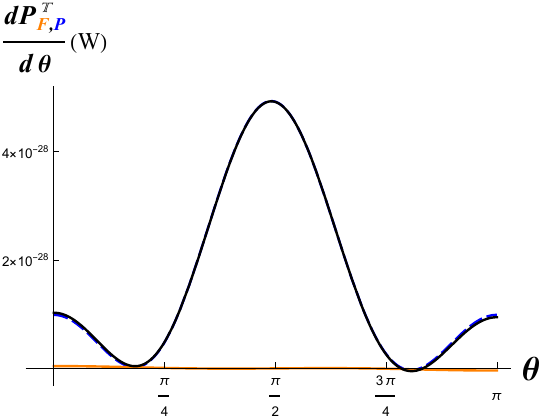}\\
\includegraphics[width=4.9cm]{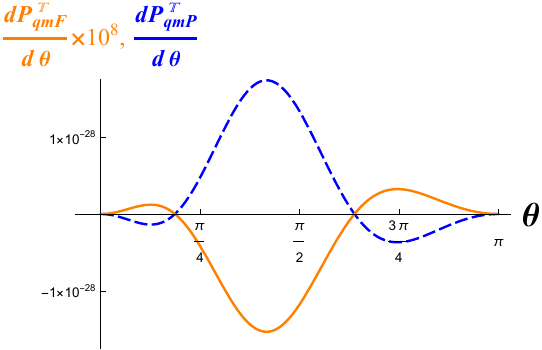}
\includegraphics[width=4.9cm]{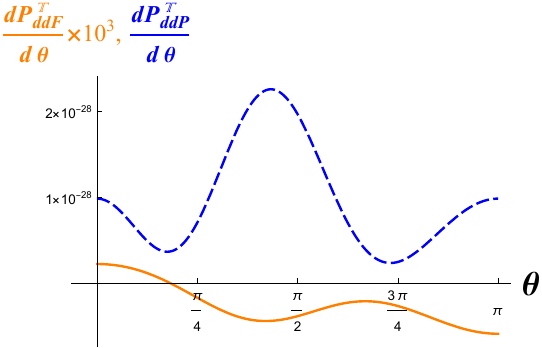}
\includegraphics[width=4.9cm]{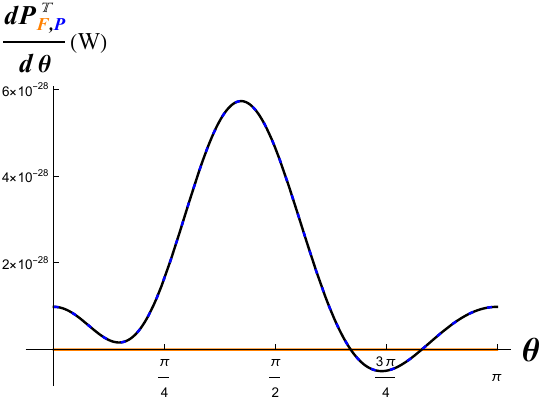}\\
\includegraphics[width=4.9cm]{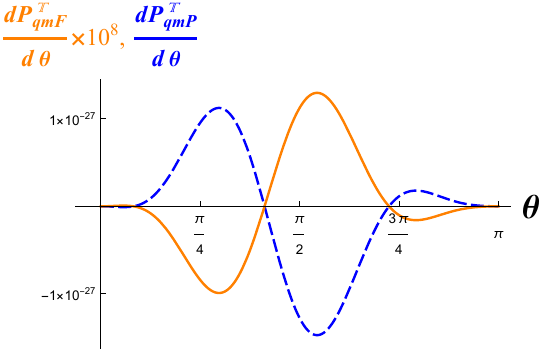}
\includegraphics[width=4.9cm]{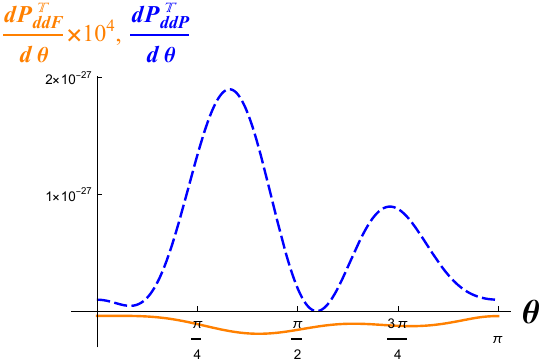}
\includegraphics[width=4.9cm]{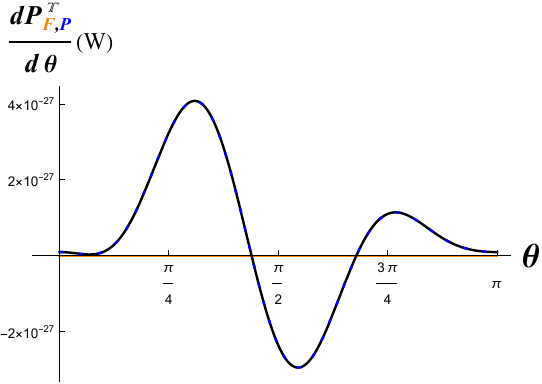}\\
\includegraphics[width=4.9cm]{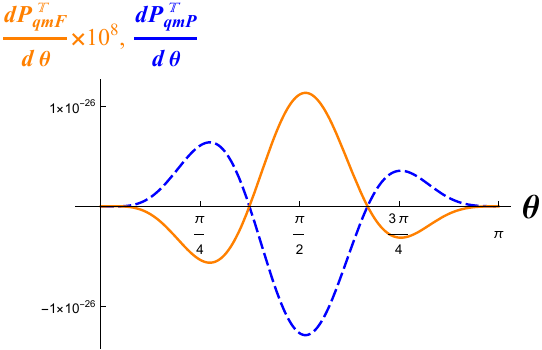}
\includegraphics[width=4.9cm]{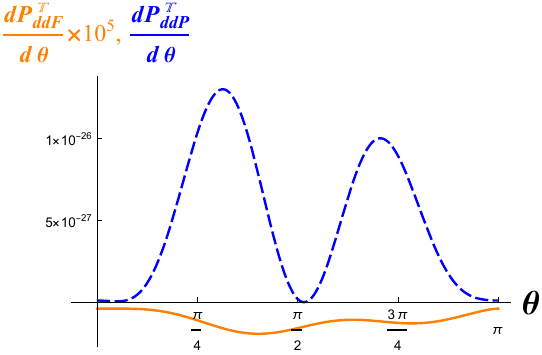}
\includegraphics[width=4.9cm]{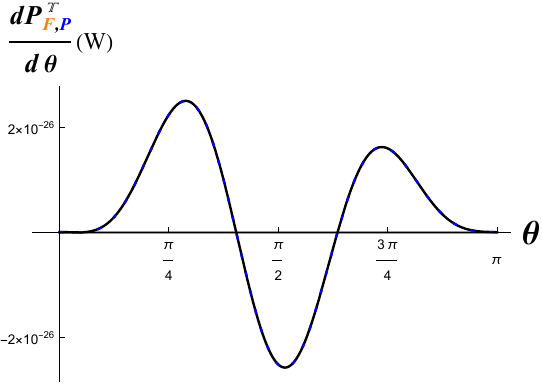}
\caption{The angular distributions of quantum radiated power contributed by the transverse deviations at $\tau^{}_- = 0.1 \tau^{}_a$, $\tau^{}_a$, $10\tau^{}_a$, and $30\tau^{}_a$ with $\tau^{}_a = 1.50 \times 10^{-9}$ s. The blue dashed and orange curves represent the $P$-part and $F$-part of the angular radiated power, respectively, and the black curves in the right plots represent the sums of the $P$- and $F$-parts.}
\label{dPFPdthetaTauAT}
\end{figure}

\begin{figure}[ht]
\includegraphics[width=6cm]{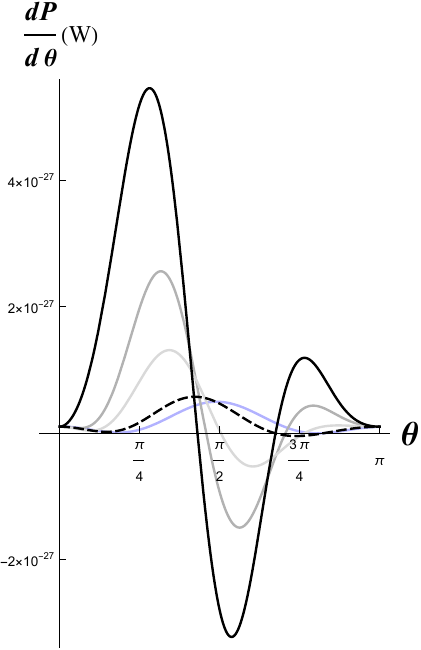}
\caption{Angular distributions of total quantum radiated power at $\tau^{}_- = 0.1 \tau^{}_a$ (light blue), $\tau^{}_a$ (black dashed), $4\tau^{}_a$ (light gray), $7\tau^{}_a$ (dark gray), and $10\tau^{}_a$ (black), which are dominated by $d{P}^{\mathbb{T} }_P /d\theta$ (right column in figure \ref{dPFPdthetaTauAT}). The magnitudes of $dP/d\theta$ around $\theta=0$ and $\pi$ are almost constant through this period.}
\label{dPdthetaTauA}
\end{figure}

In figures \ref{dPFPdthetaTauAL} and \ref{dPFPdthetaTauAT}, we present the contributions to the angular radiated power by the longitudinal and transverse deviations, respectively, at $\tau_- = 0.1 \tau^{}_a$, $\tau^{}_a$, $10 \tau^{}_a$, and $30 \tau^{}_a$ with $\tau_a = 1.50 
\times 10^{-9}$ s.
Both the contributions from the $F$- and $P$-parts of the correlators of electron worldline deviations can be negative in some angular intervals between $\theta=0$ and $\pi$. 

Comparing figure \ref{dPFPdthetaTauAL} with figure \ref{dPFPdthetaTauAT}, one can see that $d{P}^{\mathbb{T}}/d 
{\theta}$ dominates over $d{P}^{\mathbb{L}}/ d 
{\theta}$ from $\tau^{}_- \sim 0.1\tau^{}_a$ to $\tau^{}_- \sim 10\tau^{}_a$, while both contributions are much smaller than the classical radiated power. One can also see that $d{P}^{\mathbb{T}}/d 
{\theta}$, is dominated by its $P$-part throughout this period, so the total quantum radiated power $d{P}/d\theta = d{P}^{\mathbb{L}}/d\theta + d{P}^{\mathbb{T}}/d\theta$ is dominated by $d{P}^{\mathbb{T} }_P /d\theta$ until $\tau^{}_- \sim 10 \tau^{}_a$ (figure \ref{dPdthetaTauA}), after which it enters the long-time regime.  {Note that the magnitudes of the correction $dP/d\theta$ around $\theta=0$ and $\pi$ are almost constant through this middle-time regime while the peak magnitudes at other angles increase considerably, as shown in figure \ref{dPdthetaTauA}.}

\subsubsection{Long-time behaviors}
\label{SecLongitudLongt}

\begin{figure}[ht]
\includegraphics[width=4.9cm]{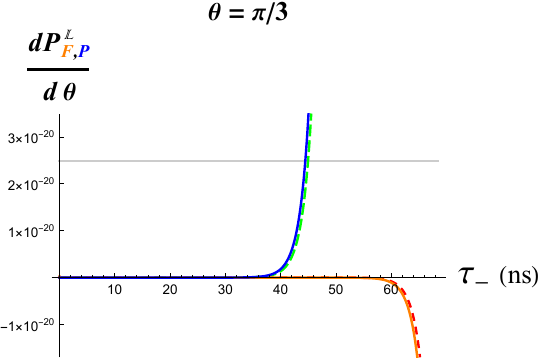}
\includegraphics[width=4.9cm]{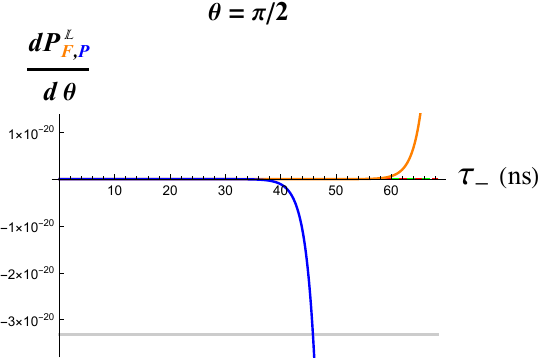}
\includegraphics[width=4.9cm]{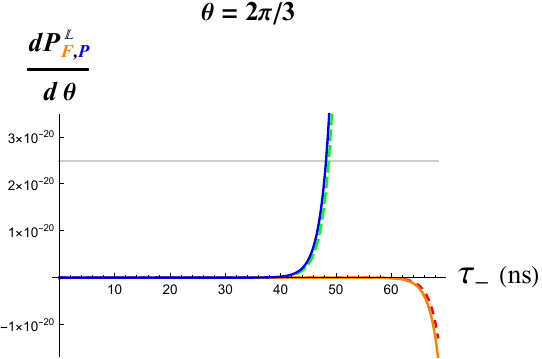}\\
\includegraphics[width=4.9cm]{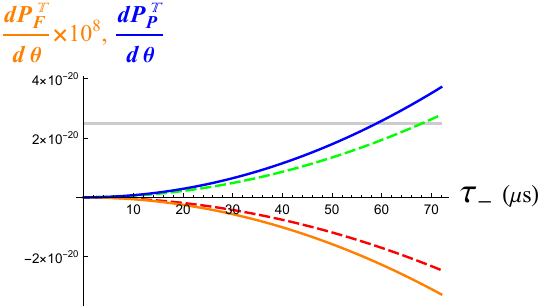}
\includegraphics[width=4.9cm]{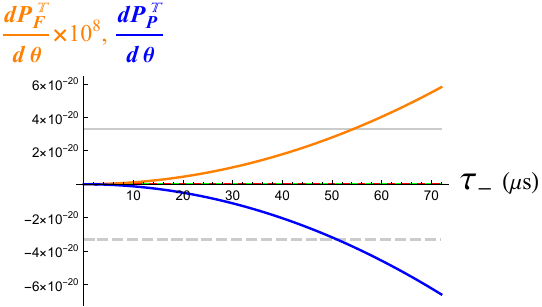}
\includegraphics[width=4.9cm]{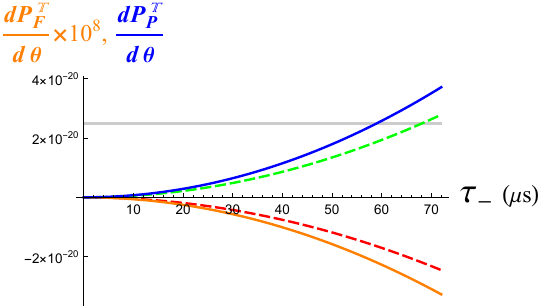}
\caption{Long-time behavior of the angular radiated power contributed by the longitudinal (upper row, in ns) and transverse deviations (lower row, in $\mu{\rm s}$). The red- and green-dashed curves represent the dipole-dipole contributions $d{P}^{\mathbb{L, T}}_{ddF}/d\theta$ and $d{P}^{\mathbb{L, T}}_{ddP}/d\theta$, and the gray lines represent the classical angular radiated power $d\bar{P}/d\theta$ emitted by a point-like UAC.}
\label{RadTauThetaLong}
\end{figure}

The quadratic growth in $d{P}^{\mathbb T}_{P}/d\theta$ contributed by the transverse deviations becomes significant as early as $\tau^{}_- \sim O(10^{-9})$ s $\sim \tau^{}_a$, while $d{P}^{\mathbb T}_F/d\theta$ temporarily becomes a stationary function of $\theta$ in (\ref{dPddFTdOmegaLT}) in the beginning of the long-time regime [figure \ref{dPFPdthetaTauAT} (third and fourth row)]. Then $d{P}^{\mathbb T}_P/d\theta$ overtakes the classical radiated power $d\bar{P}/d\theta$ emitted by a point-like UAC at about $O(10^{-5})$ s [e.g., $\tau^{}_- \approx 60$ $\mu{\rm s} \approx 40000 \tau^{}_a$ for $\theta = \pi/3$ in figure \ref{RadTauThetaLong} (lower left)].

For the contribution by the longitudinal deviation, the secular exponential growth in (\ref{dPj1Presult}) and (\ref{dPj1Fresult}) for  $d{P}^{\mathbb L}_P/ d\Omega_{\rm II}$ and $d{P}^{\mathbb L}_F/ d\Omega_{\rm II}$, respectively, becomes significant in the long-time regime. In figure \ref{RadTauThetaLong} (upper row), one can see that this occurs around $\tau^{}_- \sim O(10)$ ns $\sim 10 \tau^{}_a$, which is much earlier than the similar time scale for the contribution by the transverse deviations. As shown in figure \ref{RadTauThetaLong} (upper left), quantum radiated power exceeds the classical one after about 45 ns at $\theta=\pi/3$, and the same time scale at other $\theta$. This marks the end-point of validity of our leading-order approximation in this paper. Fortunately, our estimates around $\tau^{}_a$ in ref. \cite{LH24} and in this paper are still safe. 

To estimate the behavior of angular radiated power after the time scale of O(10) ns,  we could replace the classical correlators $\langle \Delta^2\rangle$ in (\ref{dPdOmClIntegral}) ($W = \sqrt{2\langle \Delta^2\rangle}$ )  and $\langle\Delta'^2\rangle$ in (\ref{dPdOmClIntegral2}) ($W' = \sqrt{2\langle \Delta'^2\rangle}$ ) by the late-time values of $\langle (\hat{z}_{}^i)^2\rangle$, since the interference terms $dP^{[01]}_{ddF}/d\Omega^{}_{\rm II}$ and $dP^{[10]}_{ddF}/ d\Omega^{}_{\rm II}$ are not important from the middle-time regime on.

\subsubsection{Identifying the Unruh effect in quantum radiation?}

The Unruh effect states that a uniformly accelerated particle or atom in the Minkowski vacuum of a quantum field will experience quantum fluctuations of the field with a thermal spectrum. 
In ref. \cite{CT99}, Chen and Tajima proposed to detect the Unruh effect from quantum radiation in the blind spots of classical radiation around the polar angles $\theta=0$ and $\pi$.
In the cases considered in figures \ref{dPFPdthetaTauAL}-\ref{dPdthetaTauA}, one can see that  {while quantum corrections to the radiation in these two blind spots are almost constant through the middle-time period (figure \ref{dPdthetaTauA}), they} are fully contributed by the transverse deviations (figure \ref{dPFPdthetaTauAT}), whose dominant part $d{P}^{\mathbb{T} }_P/d\theta$ is contributed by the $P$-part of the transverse-deviation correlators which is unfortunately irrelevant to the Unruh effect. Although $d{P}^\mathbb{T}_F/ d\theta$ corresponding to the initial state of the fields is also nonvanishing in these blind spots, its magnitude is much smaller, and the Unruh effect is not significant in the result consisting of the transverse-deviation correlators with the $P$-part and $F$-part added together, as shown in figure  3 of ref. \cite{LH24}. Thus, even in the blind spots of the classical Larmor radiation, the background can be much greater than the signal of the Unruh effect.

Of course, the above setting with electrons accelerated by a uniform electric field is different from the one suggested in ref. \cite{CT99}, where electrons would be driven by an intense laser field. To address the latter, a closer look at the quantum radiation emitted by single-electron wavepackets in oscillatory motion is needed \cite{DLHM13, Lin17}

\section{Summary}
\label{SecSumOut}

In this work we have calculated the angular radiated power emitted by a single-electron wavepacket at rest or uniformly, linearly  accelerated in the Minkowski vacuum of EM fields, and compared our results with those in classical electrodynamics, as described in appendix \ref{SecClRadUAC}. We found that the nonlinear cubic terms in the action are necessary to produce a correct classical correspondence, as detailed in sections \ref{SecS3S2S1} and \ref{subsecRmq}.
With the parameter values of single-electron experiments in TEM [accelerating electric field $\approx$ 0.5 MV/m, {initial widths of the electron wavepacket $\big( \langle \overline{\hat{z}_1^2} \rangle^{}_{\rm I}\big)^{1/2} \approx 1.7\,\mu{\rm m}$ and $\big( \langle \overline{ \hat{z}_2^2} \rangle^{}_{\rm I}\big)^{1/2} = \big(\langle \overline{\hat{z}_3^2} \rangle^{}_{\rm I}\big)^{1/2} \approx 5$ nm in the longitudinal and transverse directions, respectively}], we found that the $F$-part contribution relevant to the initial vacuum state of the EM fields only dominates at early-times [within 1 ns], as shown in figures \ref{RadTauThetaEarly} to \ref{dPdthetaTauA}.

In section \ref{SecQRadRest} we showed that the leading-order quantum radiation of a single-electron wavepacket at rest is exactly zero, 
although the wavepacket is spreading in time.  
For a single-electron wavepacket in uniform acceleration, its quantum correction to the classical angular radiated power emitted by a point-like UAC is small within O(10) ns in a linear accelerator with electric field of the same order as in TEM. The secular exponential growth of the leading-order correction, however, will overtake the classical radiated power of point-like UAC at some later time, as shown in figure \ref{RadTauThetaLong}.

We found that such a secular growth is not of quantum origin.
As demonstrated in appendix \ref{SecClRadUAC}, the leading-order correction of the radiation emitted by a Gaussian-distributed ensemble of longitudinally-deviated classical UACs to the one emitted by a point-like UAC exhibits a similar secular exponential growth. In particular, the classical leading-order correction to the angular radiated power is 
{very close to} 
our result with the quadrudpole-monopole and dipole-dipole contributions in the long-time regime added together. This gives a classical interpretation for the secular growth of our result.

Our calculation in classical electrodynamics suggests that the secular growths in our leading-order corrections (\ref{dPL11dOIIzeroD}) and (\ref{dPLmqdOmLTL}) also signal the failure of polynomial approximation at the level of action $S$. When the leading-order correction is comparable with the classical radiated power by a point-like UAC, introducing higher-order corrections would make the approximation even worse. Fortunately, since the interference terms $dP^{[01]}_{ddF}/ d\Omega^{}_{\rm II}$ and $dP^{[10]}_{ddF}/ d\Omega^{}_{\rm II}$ in (\ref{dPddSplit}) and (\ref{dP01}) are negligible after the early stage, we may insert the two-point correlators of electron from our effective quantum theory into the {variances} 
$W^2/2$ and $W'^2/2$ of non-perturbative result (\ref{dPdOmClIntegral}) and (\ref{dPdOmClIntegral2}) in classical electrodynamics as an estimate. Then, at late times, the estimated non-perturbative result of angular radiated power would reduce to zero in most of $\theta$ rather than diverge. That is to say, the correction by a Gaussian wavepacket could at most cancel the classical radiation emitted by the point-charge along $\bar{z}^\mu$ {in most of $\theta$}. 

From our calculation and results, some additional noteworthy points can be made:

First, we found that (\ref{GaussInt}) in classical electrodynamics is an asymptotic expansion, similar to the cases in quantum field theory \cite{Ra82}. 

Second, our calculations of emitted radiation pertain to the far zone, where radiation is well defined, as opposed to the near zone. Like the multipole fields in classical electrodynamics \cite{Gr17, Ja99, Fr05,Sc98}, the configurations of the mode functions of the fields in our effective theory are not reliable for calculating the expectation value of the fields in the near zone.

Third, the intensity of the part of quantum radiation relevant to the Unruh effect (the $F$-part) is much smaller than the background, at least in the experiments of the scale of TEM. We thus concur with the suggestions of Sch\"utzhold, Schaller and Habs \cite{SSH06, SSH08} that a better place to look for traces of Unruh effect is in the correlations between the radiated photons instead of the intensity.

Finally, in relation to experimental detection of acceleration quantum radiation, we are looking further into the parameter regime(s) where detection of the Unruh effect from emitted quantum radiation could be significant. We will also continue our studies of electrons in circular motion \cite{BL83, AS07, Sc54, Lin03c} and oscillatory motion \cite{DLHM13, Lin17}. 
To get more theoretical insight, we are comparing our calculations with those using the worldline influence functional method \cite{JH05}.

\begin{acknowledgments}
SYL is supported by the National Science and Technology Council of Taiwan under grant No. NSTC 114-2112-M-018-001. SYL thanks the hospitality of Kazuhiro Yamamoto during his visit at the Department of Physics and the Quantum and Spacetime Research Institute, Kyushu University, Japan, where a good part of this work was conducted. SYL also thanks Kazuhiro Yamamoto, Jorma Louko and Chao-Hsuan Tsai for helpful discussions. BLH appreciates the warm hospitality of Chong-Sun Chu and Kin-Wang Ng during his visits to the National Tsing Hua University and the Institute of Physics,  Academia Sinica, Taiwan, R.O.C.
The authors have benefited from the activities of COST Action CA23115: Relativistic Quantum Information, funded  by COST (European Cooperation in Science and Technology).
\end{acknowledgments}

\appendix

\section{Classical radiation by uniformly accelerated charges}
\label{SecClRadUAC}

From (I.2.11) and (I.2.13), one can see that the inhomogeneous solution of the classical EM vector field observed at $x$ can be expressed as
\begin{equation}
  \bar{A}^\mu_{[1]}(x) = \frac{\mu^{}_0 q c}{4\pi} \frac{\bar{v}_{}^\mu(t_-(x))}{R(x)}
  = \frac{\mu^{}_0 q}{4\pi} \frac{\bar{u}_{}^\mu(\tau^{}_-(x))}{r(x)}
\end{equation} 
because of (\ref{Rxcrgamma}) and $\bar{u}_{}^\mu(\tau) = \partial^{}_\tau \bar{z}^\mu(\tau) = 
\bar{\gamma}(\tau(t)) \bar{v}_{}^\mu(\tau(t))$. So the classical inhomogeneous EM field-strength reads
\begin{equation}
  \bar{F}^{[1]}_{\mu\nu}(x) = 
  \partial^{}_\mu \bar{A}_\nu^{[1]}- \partial_\nu \bar{A}_\mu^{[1]} = 
  \frac{\mu^{}_0 q}{4\pi} \left[ 
    \frac{\dot{\bar{u}}^{}_\nu(\tau^{}_-)\tau^{}_{-, \mu}}{r} - 
    \frac{\bar{u}^{}_\nu(\tau^{}_-) r^{}_{,\mu}}{r^2}  
    - (\mu \leftrightarrow \nu) \right]. \label{ClFmunu}
\end{equation}
Inserting (\ref{DtauDmu}) and (\ref{rmuResult}) into the above expression, it is straightforward to see that
\begin{eqnarray}
\bar{F}^{\rho\nu}_{[1]}\bar{n}^{-}_\nu &=& \frac{\mu^{}_0 q}{4\pi}\left( 
    \frac{\bar{a}_{-}^\rho-\bar{a}^{-}_{\bar{n}} \bar{n}_{-}^\rho}{cr} + 
    \frac{\bar{u}_{-}^\rho}{r^2} \right), \label{ClFmunuN} \\
\frac{\bar{u}_{-}^\mu}{c} \bar{F}_{\mu\rho}^{[1]} &=& \frac{\mu^{}_0 q}{4\pi}\left( 
    \frac{\bar{a}^{-}_\rho-\bar{a}^{-}_{\bar{n}} \bar{n}^{-}_\rho}{cr} - 
    \frac{c \bar{n}^{-}_\rho}{r^2} \right), \label{ClUFmunu}
\end{eqnarray}
where $f^{}_- \equiv f(\tau^{}_-)$, $\bar{a}_{}^\mu = \partial^{}_\tau \bar{u}_{}^\mu$, and $\bar{a}^{}_{\bar{n}} \equiv \bar{a}^{}_\mu \bar{n}_{}^\mu$.
Suppose $\bar{F}^{\rho\nu}_{ [0]}\bar{n}_\nu$ goes to zero faster than $1/r$ as $r\to\infty$. Then one has
\begin{equation}
  \lim_{r\to\infty} r^2\frac{\bar{u}_{-}^\mu}{\mu^{}_0}\bar{F}_{\mu\rho}^{}\bar{F}^{\rho\nu}_{}\bar{n}^{-}_\nu =
 \lim_{r\to\infty}r^2 \frac{\bar{u}_{-}^\mu}{\mu^{}_0}\bar{F}_{\mu\rho}^{[1]}\bar{F}^{\rho\nu}_{[1]}\bar{n}^{-}_\nu
  = \frac{\mu_0 q^2}{(4\pi)^2 c} 
  \Big[ \bar{a}^-_\rho \bar{a}_-^\rho -
  \big( \bar{a}^{}_{\bar{n}} \big)^2 \Big] 
  \label{CluFFnGerenal}
\end{equation}
in the radiation zone $r\to\infty$.
A uniformly accelerated charge (UAC) has $\bar{a}_{-}^\rho \bar{a}^{-}_\rho - (\bar{a}^{-}_{\bar{n}})^2 = a^2 \sin^2\theta$ from (\ref{nbarUACdef}) and (\ref{UAC4acc}), yielding the angular radiated power (\ref{ClassicaldRdOmega}) which is independent of $\tau^{}_-$.

\begin{figure}[ht]
\includegraphics[width=9cm]{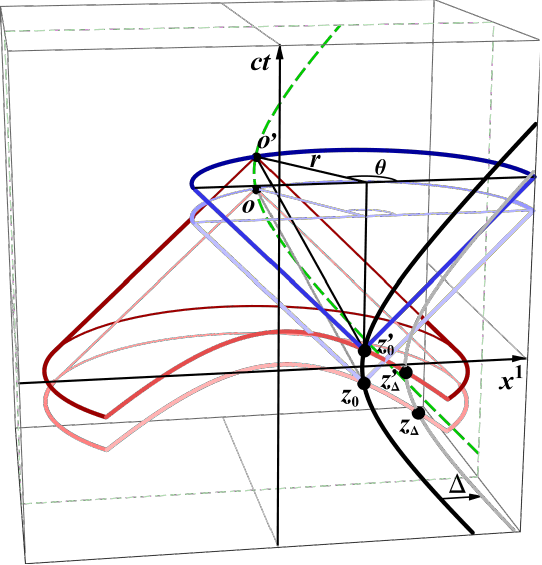} 
\caption{{A point-charge is uniformly accelerated along (\ref{zmuDelta}) with displacement $\Delta$ in the $x^1$-direction (gray worldline) from the reference worldline $\bar{z}^\mu(\tau)$ in (\ref{zmuUAC}) (black worldline) on the $x_{}^0 x_{}^1$-plane ($x_{}^0=ct$). A retardedly comoving observer (green dashed worldline) detects radiation over segment $oo'$ in the same proper time interval as $\tau^{}_- \in (-\epsilon, \epsilon)$ of segment $z^{}_0 z'_0$ [events $o'$ and $o$ are on the future lightcones of $z'_0$ (blue) and $z^{}_0$ (light blue), respectively].
The red and light-red hyperbolas are the intersections of the $x_{}^0x_{}^1$-plane and the past lightcones of events $o'$ (red) and $o$ (light red), which intersect the reference worldline (black) at $z'_0$ and $z^{}_0$, and the worldline of the shifted UAC (gray) at $z'_\Delta$ and $z^{}_\Delta$, respectively. The observer will thus see the radiated energy emitted by the shifted UAC along (\ref{zmuDelta}) during segment $z^{}_\Delta z'_\Delta$ of the gray worldline.}}
\label{FigLCplot}
\end{figure}

\subsection{Longitudinal deviation} 
\label{SecClRadWPL}

\begin{figure}[ht]
\includegraphics[width=7cm]{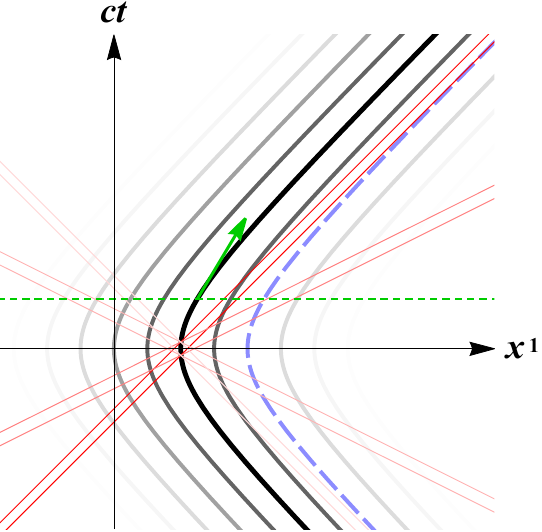}
\hspace{1cm}
\includegraphics[width=7cm]{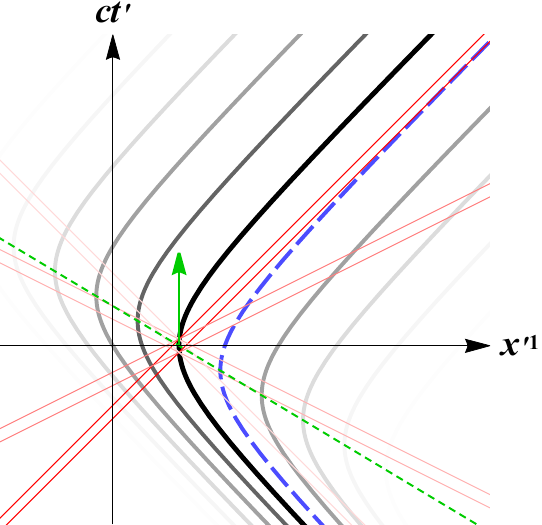}  
\caption{{(Left) Consider an ensemble of UAC worldlines (\ref{zmuDelta}) with various values of displacement $\Delta$ in the $x_{}^1$-direction in the left plot (gray and blue-dashed worldlines) from the reference worldline (black). The dark- to light-red line pairs represent the observation `windows' for the retardedly comoving observers at $\theta=0$, $\pi/3$, $2\pi/3$ and $\pi$, respectively, in the far zone. In particular, the almost straight-line pair of $\theta=2\pi/3$ (with a slope of $-1/2$) is part of the red hyperbola pair on the $x_{}^0x_{}^1$-plane in figure \ref{FigLCplot} in the vicinity of the reference worldline (intersecting $z^{}_0$ and $z'_0$) in the limit of $r\to\infty$. The green vector represents the four-velocity $\bar{u}^\mu (\tau)$ of the reference worldline at some $\tau= \tau^{}_* >0$, 
and the green-dashed line represents the contour of $\tau=\tau^{}_*$ of the worldline ensemble 
$\bar{z}_{\Delta}^\mu (\tau)$. (Right) The gray worldlines, green vector, and green-dashed line here are those in the left plot boosted from $\tau^{}_*$ back to $0$ with the reference worldline invariant in shape [see descriptions above (\ref{zmuDeltataustar})]. The blue-dashed worldline in each plot will hit the observer at $\theta=0$ at the null infinity.}}
\label{FigWLSet}
\end{figure}

{To mimic a single Gaussian wave packet of uniformly accelerated electron spreading in the direction of acceleration,} consider an ensemble of classical electron worldlines 
\begin{equation}
  \bar{z}^\mu_{\Delta_{}} (\tau) = \bar{z}^{\mu}(\tau) + (0,\Delta_{},0,0) \label{zmuDelta}
\end{equation}
of UACs, each with a different spatial shift $\Delta_{}$ in the $x_{}^1$-direction from the reference worldline $\bar{z}^{\mu}(\tau)$ in (\ref{zmuUAC}) [black ($\bar{z}_{}^\mu$) and gray ($\bar{z}^\mu_\Delta$) curves in figures \ref{FigLCplot} and \ref{FigWLSet}]. A retardedly comoving observer along the worldline $x^\mu(\tau^{}_-) = (r + \frac{c}{\alpha}\sinh\alpha \tau^{}_-, r\cos\theta + \frac{c}{\alpha}{\cosh \alpha\tau^{}_-} , r \sin\theta\cos\varphi,r \sin\theta\sin\varphi)$ 
[green curve in figure \ref{FigLCplot}] in the far zone ($r\to \infty$) detects radiation over the time interval $\tau^{}_- \in (-\epsilon, \epsilon)$ [segment $o'o$ in figure \ref{FigLCplot}]. The observer will see the collective radiated energy by the UACs in this ensemble, which is the sum of the radiated energy emitted from each worldline $\bar{z}_{\Delta}^\mu(\tau)$ during the interval $z^{}_{\Delta}$ to $z'_\Delta$ in figure \ref{FigLCplot}, along the observer's past lightcones (red cones). 

The past lightcone of the observing event $o'$ intersects the $x_{}^1 x_{}^0$-plane at
\begin{equation}
  \left[ x_{}^0 - \left(\frac{c}{\alpha}\sinh\alpha\epsilon + r\right)\right]^2 -
  \left[ x_{}^1 - \left(\frac{c}{\alpha}\cosh\alpha\epsilon + r\cos\theta\right)\right]^2
  = r^2\sin^2\theta
\end{equation}
(the red hyperbola in figure \ref{FigLCplot}). The intersection of the past lightcone of event $o$ is similar to the above one with $\epsilon$ replaced by $-\epsilon$ (the light-red hyperpola). In this `window' of detection the observer sees the spacetime between these two past lightcones. {Around $\bar{z}_{}^\mu(\pm \epsilon)$, for $r\to \infty$ and $\alpha \epsilon \ll 1$, the two hyperbolas 
can be} approximated by the straight lines
\begin{equation}
 x^0 - \frac{c}{\alpha}\sinh(\pm\alpha\epsilon) =  \cos\theta \left( x^1 - \frac{c}{\alpha}\cosh\alpha \epsilon \right) \label{tzPaLCInfiR}
\end{equation}
{in} figure \ref{FigWLSet}, where the slopes of both straight lines on the $x_{}^1 x_{}^0$-plane are $\cos\theta$.

The radiated power emitted by the same ensemble of UACs at $\tau_- = \tau^{}_* \not=0$ {(cf. figure \ref{FIGsetting})} can be obtained by performing a Lorentz boost from $\tau_- =\tau^{}_*$ back to $\tau'_- = 0$ in the new coordinates where the radiated energy in the interval $\tau'_- \in [ -\epsilon, \epsilon]$ is collected by the retardedly comoving observer. More precisely, we perform a Lorentz transform $x'^0 = \bar{\gamma}(\tau^{}_*)\big[ x_{}^0 - \frac{\bar{v}(\tau^{}_*)}{c} x_{}^1\big]$, $x'^1= \bar{\gamma}(\tau^{}_*)\big[ x_{}^1 - \frac{\bar{v}(\tau^{}_*)}{c} x_{}^0\big]$, $x'^2=x_{}^2$, {and $x'^3=x_{}^3$ 
with} $\bar{\gamma}(\tau^{}_*) = \cosh \alpha\tau^{}_*$ and $\bar{v}(\tau^{}_*)/c = \tanh\alpha\tau^{}_*$ from (I.4.4) and (I.4.5). In the new coordinates, the UAC worldlines are expressed as
\begin{equation}
  \bar{z}'{}^\mu_{\Delta_{}}(\tau') = \left( 
  \frac{c}{\alpha} \sinh\alpha \tau' - \Delta_{}\,\sinh \alpha\tau^{}_* , 
  \frac{c}{\alpha} \cosh\alpha \tau' + \Delta_{}\,\cosh \alpha\tau^{}_* , 0,0 \right)
\label{zmuDeltataustar}
\end{equation}
with $\tau' = \tau - \tau^{}_*$ [figure \ref{FigWLSet} (right)] while the observer is {going along $( r' +\frac{c}{\alpha} \sinh\alpha \tau', r'\cos\theta' + \frac{c}{\alpha} \cosh\alpha \tau', r'\sin\theta' \cos\varphi, r'\sin\theta'\sin\varphi)$ with $r'\equiv r(\cosh\alpha\tau^{}_* - \cos\theta\sinh\alpha\tau^{}_*)$ and $\sin\theta' \equiv \frac{r}{r'} \sin\theta$. 
For $\tau' = \tau'_- \approx 0$, the observer is almost} 
at rest when detecting radiation over a period like $oo'$ in figure \ref{FigLCplot}. Note that the shape of the reference worldline $\bar{z}_{}^\mu$ and the observer's $x_{}^\mu$ are invariant under these transforms.

{Rewriting $\theta'$ as $\theta$ (and $r'$ as $r$) and inserting} the above $\bar{z}'{}^0_{\Delta_{}}(\tau')$ and $\bar{z}'{}^1_{\Delta_{}}(\tau')$ into $x_{}^0$ and $x_{}^1$ of (\ref{tzPaLCInfiR}), the solutions of $\tau'$ are
\begin{equation}
  \tau^{}_{\Delta_{}}(\pm\epsilon,\theta) = \frac{1}{\alpha} \ln \frac{ b^{}_{\pm\epsilon} +
  \sqrt{ b^2_{\pm\epsilon}+\sin^2\theta}}{1-\cos\theta}\label{TauDeltapmEps}
\end{equation}
with
\begin{equation}
  b^{}_{\epsilon} \equiv 
    \sinh \alpha\epsilon -\cos\theta\,\cosh\alpha\epsilon +\frac{\alpha}{c}\Delta_{}
    \left( \sinh\alpha\tau^{}_* + \cos\theta \cosh\alpha\tau^{}_* \right),
\end{equation}
which reduces to
\begin{equation}
  \tau^{}_{\Delta_{}}(\pm\epsilon,0) = -\frac{1}{\alpha}\ln \left[  e^{\mp\alpha\epsilon}
   - \frac{\alpha}{c}\Delta_{}\, e^{\alpha\tau^{}_*} \right], \hspace{.3cm}
  \tau^{}_{\Delta_{}}(\pm\epsilon,\pi) = \frac{1}{\alpha}\ln \left[ e^{\pm\alpha\epsilon}
   - \frac{\alpha}{c}\Delta_{}\, e^{-\alpha\tau^{}_*} \right], 
\end{equation}
for $\theta = 0$ and $\pi$, respectively. 

In classical electrodynamics, 
the angular radiated power with respect to the proper time of the electron along $\bar{z}'{}^\mu_\Delta(\tau')$ in the observing window between the two lines in (\ref{tzPaLCInfiR}) is [e.g., (14.38) in ref. \cite{Ja99}]
\begin{eqnarray}
\frac{d\bar{P}^{}_{\Delta_{}}}{d\Omega^{}_{\rm II}} &\equiv& 
   \lim_{\epsilon\to 0} \frac{\mu^{}_0 q^2 a^2}{(4\pi)^2 c} \frac{1}{2\epsilon} 
   \int_{\tau^{}_{\Delta_{}}(-\epsilon,\theta)}^{\tau^{}_{\Delta_{}}(\epsilon,\theta)} 
   \bar{\gamma}(\tau')d\tau'\frac{\left|{\bf n}\times \left[\left({\bf n}-
     \frac{\bf v}{c}\right)\times \frac{1}{c}\frac{d {\bf v}}{dt}\right] \right|^2}
   {\left(1 - \frac{\bf v}{c} \cdot {\bf n}\right)^5} 
   \nonumber\\ &=& \lim_{\epsilon\to 0} \frac{\mu^{}_0 q^2 a^2}{(4\pi)^2 c} 
   \int_{\tau^{}_{\Delta_{}}(-\epsilon,\theta)}^{\tau^{}_{\Delta_{}}(\epsilon,\theta)} 
   \frac{d\tau'}{2\epsilon}\frac{\sin^2\theta}{(\cosh\alpha\tau' -\sinh\alpha\tau'\cos\theta)^{5}} 
   \label{dPdOmClDeltaInt0}\\  &=& \lim_{\epsilon\to 0} \frac{\mu^{}_0 q^2 a^2}{(4\pi)^2 c}
   \left[\frac{\delta\tau^{}_{\Delta_{}}}{2\epsilon}\right]
   \frac{\sin^2\theta}{(\cosh\alpha T^{}_{\Delta_{}} - \sinh\alpha T^{}_{\Delta_{}} \cos\theta)^{5}}
\label{dPdOmClDeltaInt}
\end{eqnarray}
with ${\bf v} = (at/\bar{\gamma}(t), 0,0) = (c\tanh\alpha\tau',0,0)$ from (I.4.5), ${\bf n}=(\cos\theta, \sin\theta\cos\varphi, \sin\theta\sin\varphi)$ from (\ref{nbarUACdef}), $T^{}_{\Delta_{}} \equiv [\tau^{}_{\Delta_{}}(+\epsilon,\theta)+\tau^{}_{\Delta_{}}(-\epsilon,\theta)]/2$, and $\delta \tau^{}_{\Delta_{}} \equiv \tau^{}_{\Delta_{}}(+\epsilon,\theta)-\tau^{}_{\Delta_{}}(-\epsilon,\theta)$.
Inserting (\ref{TauDeltapmEps}) to the above expression, we find
\begin{eqnarray}
  && \frac{d\bar{P}^{}_{\Delta_{}}}{d\Omega^{}_{\rm II}}=
  \frac{\mu^{}_0 q^2 a^2}{(4\pi)^2 c}\frac{\sin^2\theta}
   {\left[1 + \big({\cal M(\tau^{}_*,\theta)}\Delta_{} \big)^2 - 2 {\cal M}(\tau^{}_*,\theta)\Delta_{} \cos\theta \right]_{}^3} 
  \label{dPdOmclDelta1Exact}
  \\ && = 
  \frac{\mu^{}_0 q^2 a^2}{(4\pi)^2 c} \sin^2\theta \Big[1 + 6 \cos\theta\,({\cal M}\Delta_{}) 
  + 3(8\cos^2 \theta -1) ({\cal M}\Delta_{})^2 + \nonumber\\
  &&\hspace{-.5cm}
  8\cos\theta(10\cos^2\theta -3)({\cal M}\Delta_{})^3+
  6(40\cos^4\theta-20\cos^2\theta+1)({\cal M}\Delta_{})^4+\cdots 
    \Big], \label{dPdOmclDelta1Series}
\end{eqnarray} 
where the power series representation converges for ${\cal M}\Delta < 1$, and 
\begin{equation}
  {\cal M}(\tau^{}_*,\theta) \equiv\frac{\alpha}{c}\left(\sinh\alpha\tau^{}_* + \cosh\alpha\tau^{}_* \cos\theta\right)
\end{equation}
can be negative or positive depending on $\theta$, and its value for most of $\theta$ grows exponentially at late times as $\tau^{}_*$ increases.

Suppose that the $x^1$-shifted UAC worldlines $\bar{z}_{\Delta_{} }^\mu(\tau)$ in the ensemble are weighted in a normalized Gaussian spatial distribution $e^{-\Delta_{ }^2/W^2}/(\sqrt{\pi} W)$ centered at $\bar{z}^\mu(\tau)$ with variance $\langle \Delta_{}^2 \rangle = W^2/2$, and detected by similar observers at polar angles $\theta$ with respect to the reference UAC $\bar{z}^\mu(\tau)$ during $\tau^{}_- \in [-\epsilon, \epsilon]$ with $\epsilon\to 0+$. Here the tail of the Gaussian distribution goes beyond the event horizon of the worldline $\bar{z}^\mu(\tau)$, and thus Rindler coordinates may not help.  
Then the classical angular radiated power can be obtained by calculating the integral 
\begin{equation}
   \frac{d\bar{P}^{}_W}{d\Omega^{}_{\rm II}} = 
   \left< \frac{d\bar{P}^{}_{\Delta}}{d\Omega^{}_{\rm II}} \right> \equiv 
   \int_{-\infty}^\infty d\Delta \frac{e^{-(\Delta/W)^2}}{\sqrt{\pi}W} 
   \frac{d\bar{P}^{}_{\Delta}}{d\Omega^{}_{\rm II}}. \label{dPdOmClIntegral}
\end{equation}
Since $e^{-(\Delta_{}/W)^2}$ is an even function of $\Delta$, we would have
\begin{eqnarray}
   \frac{d\bar{P}^{}_W}{d\Omega^{}_{\rm II}} &=& \frac{\mu^{}_0 q^2 a^2}{(4\pi)^2 c}\sin^2\theta 
\left<\left[1+\left({\cal M}\Delta_{}\right)^2-2{\cal M}\Delta_{}\cos\theta\right]_{}^{-3}\right>  \label{dPWdOmGauss}\\ &=& \frac{3s\bar{m}}{8\pi} c^2 \alpha^2 \sin^2\theta \Big[1 + 3(8\cos^2 \theta -1) 
   {\cal M}^2 \langle \Delta_{}^2 \rangle + \nonumber\\  && \hspace{1cm}
    6(40\cos^4\theta-20\cos^2\theta+1){\cal M}^4 \langle\Delta_{}^4\rangle + \cdots \Big],
\label{dPWdOmGaussSeries}
\end{eqnarray}
if, at least, ${\cal M}^{2n} \langle \Delta^{2n}\rangle \ll 1$ for large $n$. Nevertheless, 
\begin{equation}
  \langle\Delta_{}^{2n}\rangle = \int_{-\infty}^\infty \frac{d\Delta_{} 
  }{\sqrt{\pi}W} e^{-(\Delta_{}/W)^2}\Delta_{}^{2n} = (2n-1)!! \left[\frac{W^2}{2}\right]^n =  
  (2n-1)!!\langle\Delta_{}^{2}\rangle^n , \label{GaussInt}
\end{equation} 
and one has $(2n-1)!! \approx \sqrt{2}\left(2n/e\right)^{n}$ for large $n$, so in (\ref{dPWdOmGaussSeries}) the $n$-th order term ${\cal M}^{2n}\langle \Delta^{2n}\rangle \approx \sqrt{2} \left( 2n {\cal M}^2 \langle \Delta^2 \rangle/e \right)^n$  diverges as $n\to\infty$ no matter how small ${\cal M}^2 \langle \Delta^2 \rangle$ is. Thus, (\ref{dPWdOmGaussSeries}) is an asymptotic expansion of (\ref{dPWdOmGauss}) in powers of $\langle \Delta^2 \rangle$, rather than a perturbative expansion.

\begin{figure}[ht]
\includegraphics[width=8cm]{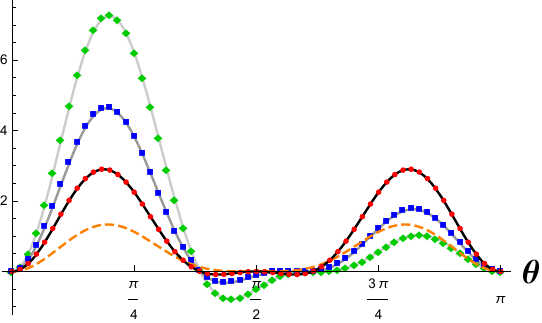}
\includegraphics[width=7cm]{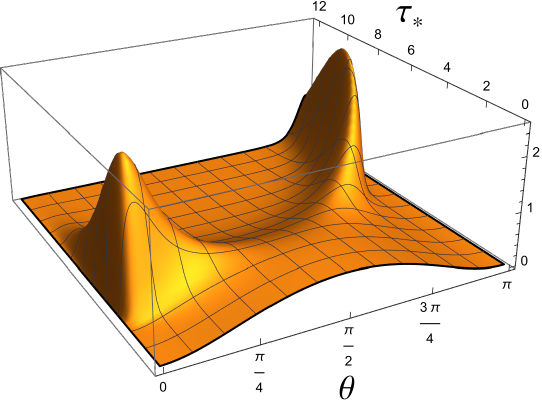}
\caption{(Left) Scaled corrections of an ensemble of UAC of $\alpha=c=1$ and $W=0.03$ in (\ref{dPdOmClIntegral}) to the angular radiated power emitted by a point charge along (\ref{zmuUAC}).
The red dots show the numerical result of the scaled correction at $\tau^{}_*=0$,
while the orange dashed curve represents the dipole-dipole correction (\ref{dPL11dOIIzeroD}) in the long-time regime
with the constant  $\langle\hat{z}^1\hat{z}^1 \rangle$ set to $\langle \Delta^2 \rangle$. 
The black curve represents the total quantum correction with the quadrupole-monopole corrections in (\ref{dPmqdOmegaResult}) added to the dipole-dipole correction. 
The blue squares and dark-gray curve represent the {classical and quantum results of $\tau^{}_* = 0.2$, respectively,} and the green diamonds and light-gray curve is for $\tau^{}_* = 0.4$.
(Right) Time evolution of scaled angular radiated power emitted by the same ensemble of UAC and modified from (\ref{dPdOmClIntegral}) with finite period of acceleration from $\tau^{}_I=-2$ to $\tau^{}_F=2$.} 
\label{FigQRadbyz1z1P}
\end{figure}

{In figure  \ref{FigQRadbyz1z1P} we present our numerical result of angular radiated power emitted by an ensemble of Gaussian distributed classical UAC of $\alpha=c=1$ and $W=0.03$. The result is independent of azimuthal angle $\varphi$ by symmetry.
The left plot shows the scaled corrections to the radiated power emitted by a single point-like UAC along $\bar{z}^\mu(\tau)$, i.e.,
$\frac{(4\pi)^2 c}{\mu^{}_0 q^2 a^2\langle\Delta^2\rangle} \int \frac{d\varphi}{2\pi} \Big( \frac{d\bar{P}^{}_W}{d\Omega^{}_{\rm II}}- \frac{d\bar{P}}{d\Omega^{}_{\rm II}} \Big)$ with 
(\ref{dPdOmClIntegral}),
(\ref{dPdOmClDeltaInt0}) and 
(\ref{ClassicaldRdOmega}).
In the regime of ${\cal M}\Delta \sim O(10^{-2}) \ll 1$ with the data points at $\theta=0$ and $\pi$ set to zero by hand,
the numerical results of $\tau^{}_*=0$ (red dots), $0.2$ (blue square dots) and $0.4$ (green triangle dots) 
fit quite well to the black, dark gray, and light gray curves, respectively, representing the $\langle \Delta^2\rangle$ term in  (\ref{dPWdOmGaussSeries}) [$\propto \sin^2\theta (8\cos^2\theta-1) {\cal M}^2 ( \tau^{}_*,\theta)$], which is identical to the sum of 
the quadrupole-monopole contributions in (\ref{dPLmqdOmLTL}) ($dP^{\mathbb{L}}_{mq}/d\Omega^{}_{\rm II} + dP^{\mathbb{L}}_{qm}/ d\Omega^{}_{\rm II}= 2 \,dP^{\mathbb{L}}_{qm}/ d\Omega^{}_{\rm II}$) and the dipole-dipole contribution in (\ref{dPL11dOIIzeroD}) with (\ref{d01nurho}) in the long-time regime for $\langle\Delta_{}^2 \rangle = \langle\hat{z}^1_{{}_{\underline{\texttt{0}}}}\hat{z}^1_{{}_{\underline{\texttt{0}}}}\rangle$. Based on this observation, the apparent exponential growth proportional to ${\cal M}^2(\tau^{}_-,\theta)$ 
in our quantum corrections can have a classical interpretation. Moreover,
the magnitudes of (\ref{dPLmqdOmLTL}) and (\ref{dPL11dOIIzeroD}) are comparable [see the black and orange-dashed curves in figure \ref{FigQRadbyz1z1P} (left)], thus the quadrupole-monopole contributions in section \ref{SecUACqm} and the associated consideration of the cubic terms (\ref{S3def}) in sections \ref{SecS3S2S1} and \ref{subsecRmq} are necessary to the leading order for the correct classical correspondence of quantum radiation emitted by single-electron wavepackets.}


The expectation value in (\ref{dPWdOmGauss}) with respect to the Gaussian distribution actually has a closed-form expression,  
\begin{eqnarray}
&& \left<\left[1+\left({\cal M}\Delta_{}\right)^2-2{\cal M}\Delta_{}\cos\theta\right]_{}^{-3}\right> \nonumber\\
&=& \frac{-1}{4\Sigma}\partial^{}_\Sigma \, \frac{-1}{2\Sigma}\partial^{}_\Sigma \int_{-\infty}^{\infty}  \frac{d\Delta}{\sqrt{\pi}W} \left. \left[\frac{e^{-(\Delta/W)^2}}{\Sigma^2 + ( {\cal M}\Delta - \cos\theta)^2 }\right] \right|^{}_{\Sigma = \sin\theta} \nonumber\\
&=& \frac{1}{4\Sigma}\partial^{}_\Sigma \, \frac{1}{2\Sigma}\partial^{}_\Sigma 
 \left. {\rm Re} \left\{\frac{\sqrt{\pi}}{{\cal M}W \Sigma} \left[ e^{-\frac{(\cos\theta + i\Sigma)^2}{({\cal M}W)^2}} + 
 \frac{2i}{\sqrt{\pi}} {\sf F}\left(\frac{\cos\theta+i\Sigma}{{\cal M}W} \right) \right] \right\} \right|^{}_{\Sigma=\sin \theta} \nonumber\\
&=&\frac{1}{16\sqrt{\pi}({\cal M}W \sin\theta)^5} {\rm Re} \,\bigg\{ 4\sqrt{\pi}{\cal M}W\sin\theta 
\big(3{\cal M}^2 W^2 -2\sin^2\theta \big)+ 2\pi\Big[ (1-e^{2i\theta})^2 +\nonumber\\ &&
2i{\cal M}^2 W^2 \sin\theta \big(3 e^{i\theta}+\sin\theta\big)+3{\cal M}^4 W^4 \Big]\left[ e^{-\frac{e^{2i\theta}}{({\cal M}W)^2}} + 
 \frac{2i}{\sqrt{\pi}} {\sf F}\left(\frac{e^{i\theta}}{{\cal M}W} \right) \right]\bigg\}
 \label{ExpectdPW}
\end{eqnarray}
with the Dawson integral ${\sf F}(z)$. For constant $W$, we have $|{\cal M}W|\to \infty$ for  $0\le \theta <\pi$ as $\alpha \tau^{}_*\to \infty$, and 
\begin{equation}
\left<\left[1+\left({\cal M}\Delta_{}\right)^2-2{\cal M}\Delta_{}\cos\theta\right]_{}^{-3}\right> \to
\frac{3\sqrt{\pi}}{8\sin^5\theta {\cal M}W} + 
O\Big( ({\cal M}W)^{-3} \Big)
\label{ExpandEVinMW}
\end{equation}
from (\ref{ExpectdPW})\footnote{For finite $\alpha\tau^{}_*$, in addition to $\theta=0$ and $\pi$, the expansion in (\ref{ExpandEVinMW}) does not work for $\theta = \theta^{}_*$ with $\theta^{}_*$ defined in (\ref{thetastar}), either, since ${\cal M}(\tau^{}_*, \theta^{}_*) = 0$.}. This indicates that 
$d\bar{P}^{}_{W} /d\Omega^{ }_{\rm II}$ in (\ref{dPWdOmGauss})   
goes to zero like $e^{-\alpha\tau^{}_*}$ for $\theta$ not too close to $0$ or $\pi$ at late times, while every non-zeroth order of its series expansion (\ref{dPWdOmGaussSeries}) in $\langle\Delta_{}^2\rangle$ diverges as $|{\cal M}|\to\infty$.

(\ref{dPdOmclDelta1Exact}) diverges at $\theta=0$ for $\Delta= \Delta^{}_*\equiv (c/\alpha)e^{-\alpha\tau^{}_*}$  because the worldline $z^\mu_{\Delta^{}_*}(\tau)$ will hit the observer at the future null infinity at $\theta=0$ (blue-dashed worldlines in figure \ref{FigWLSet}), meaning that an infinitely long segment of the worldline will be in the window of detection for the observer, i.e., between the two parallel red lines in 45 degrees in figure \ref{FigWLSet}.
A similar divergence occurs at $\theta=\pi$ for $\Delta^{}_*= (c/\alpha)e^{\alpha\tau^{}_*}$.
As we integrate an ensemble of UAC worldlines with all values of $\Delta$, the expectation value (\ref{ExpectdPW}) always suffers from this kind of divergence at $\theta=0$ and $\pi$ for all $\tau^{}_*$. 

Like other divergences produced by a UAC \cite{Bo80}, such divergences can be regularized by assuming a finite-time constant linear acceleration instead of an eternal uniform acceleration \footnote{{Here we neglect the transient radiation reaction $|\Delta \bar{v}^1| \sim O\big[s q {\cal E}/(\bar{m} \cosh^2 \alpha \tau^{}_{I,F})\big]$ during the sudden transitions from $0$ to $a$ around $\tau^{}_I$ and from $a$ to $0$ around $\tau^{}_0$ 
\cite{Sc98}.}}.
For example, suppose that the background uniform electric field is switched on only from $t(\tau^{}_I)$ to $t(\tau^{}_F)$ with $\tau^{}_I$ and $\tau^{}_F$ in the proper time of the reference point-charge or any one of the $x^1$-shifted point-charge in the ensemble 
[cf. figure \ref{FigWLSet} (left)]. When a (shifted) point-charge is not in the time interval $\tau^{}_I < \tau < \tau^{}_F$, its motion is inertial and its radiated power is zero. 
Thus, the limits of $\Delta$-integration in (\ref{dPdOmClIntegral}) should be modified to $\int_{\Delta^{ }_I}^{\Delta^{}_F} d\Delta \ldots$ for $0 \le \theta < \theta^{}_*$, $\int_{-\infty}^{\infty} d\Delta \ldots$ for $\theta =\theta^{}_*$, and $\int_{\Delta^{}_F}^{\Delta^{ }_I} d\Delta \cdots$ for $\theta^{}_* <\theta \le\pi$,
with
\begin{equation}
  \Delta^{}_{I,F}(\tau^{}_*, \theta)
  \equiv \frac{1}{{\cal M}(\tau^{}_*, \theta)}\Big[ 
  \sinh\alpha\tau^{}_{I,F} +\cos\theta \big(1-\cosh\alpha\tau^{}_{I,F}\big) \Big]
\end{equation}
such that $T^{}_{\Delta^{}_I} = \tau^{}_I$ and $T^{}_{\Delta^{}_F} = \tau^{}_F$ in (\ref{dPdOmClDeltaInt}), and
\begin{equation}
    \theta^{}_* \equiv \cos^{-1}(-\tanh \alpha\tau^{}_* )
\label{thetastar}
\end{equation}
such that ${\cal M}(\tau^{}_*, \theta^{}_*)=0$. 
This regularized integral will give a finite value around $\theta=0$ and $\pi$ [e.g., figure \ref{FigQRadbyz1z1P} (right)]. Then the numerical nonperturbative result from this regularized integral can be very close to the `perturbative' result up to the leading-order correction truncated from (\ref{dPWdOmGaussSeries}) for $\tau^{}_I \ll \tau^{}_* \ll \tau^{}_F$ and ${\cal M}(\tau^{}_*,\theta)\Delta \ll 1$. 

In figure \ref{FigQRadbyz1z1P}  (right), we show the evolution of $d\bar{P}^{}_W/ d\theta$ emitted by a similar ensemble of accelerated charges to the left plot but here with a finite period of acceleration from $\tau^{}_I=-2$ to $\tau^{}_F=2$. The value of $d\bar{P}^{}_W/ d\theta$ given in (\ref{dPWdOmGauss}) has been scaled to $\ln \big[ 1 + \frac{(4\pi)^2 c}{\mu^{}_0 q^2\alpha^2} \frac{d\bar{P}^{}_W}{d\theta}  \big]$ in this plot to reduce the contrast.
For $|{\cal M}\langle \Delta^2 \rangle| \ll 1$ ($\tau^{}_* \ll 3$), the correction is small compared with the classical point-charge radiation (cf. gray curve in the bottom-right plot of figure \ref{dPFPdthetaTauAL}), {and can be described well by the polynomial approximation (\ref{dPWdOmGaussSeries}). After $\tau^{}_* \approx 3$, the polynomial approximation fails,} and the correction near $\theta=0$ becomes significant and quickly reaches about 10 times of the peak value of the point-charge radiation at about $\tau^{}_*\approx 4$. Then the peak around $\theta=0$ drops and the peak near $\theta=\pi$ begins to rise. The latter reaches its maximum value, also about 10 times of the classical point-charge result, around $\tau^{}_* \approx 8$, then drops and fades away. This numerical result is regular everywhere for all time. The larger $|\tau^{}_F|$ and $|\tau^{}_I|$, the peaks around $\theta\approx 0$ then $\theta\approx \pi$ are higher and lasting longer, while the general behavior are similar to figure \ref{FigQRadbyz1z1P} (right).

The classical results (\ref{dPdOmclDelta1Exact}) and regularized (\ref{dPWdOmGauss}) suggest that the full nonperturbative quantum correction 
by a single-electron wavepacket concentrated around a classical trajectory $\bar{z}_{}^\mu(\tau)$
would be always finite and would fade away at late times.

\subsection{Transverse deviations}

One may consider a similar setting with classical UACs of the worldlines
\begin{equation}
  \bar{z}^\mu_{\Delta'} (\tau) = \bar{z}^{\mu}(\tau) + 
  (0,0,\Delta'\cos\varphi', \Delta'\sin\varphi'), \label{zmuDeltaT}
\end{equation}
which are shifted in the transverse directions about the reference UAC $\bar{z}^{\mu}(\tau)$ in (\ref{zmuUAC}). The past lightcone of the observing event $o'$ and $o$ in figure  \ref{FigLCplot} intersect $\bar{z}^\mu_{\Delta'} (\tau)$ at the solutions $\tau = \tau^{}_{\Delta'}(\pm\epsilon)$ to the equation 
\begin{eqnarray}
&&\left[\frac{c}{\alpha}\sinh \alpha\tau -
    \left(\frac{c}{\alpha}\sinh(\pm\alpha\epsilon) + r \right) \right]^2 -
  \left[\frac{c}{\alpha}\cosh \alpha\tau -
    \left(\frac{c}{\alpha}\cosh(\pm\alpha\epsilon) + r \right) \right]^2 = \nonumber\\
&&\left( \Delta'\cos\varphi' - r \sin\theta\cos\varphi\right)^2 + 
  \left( \Delta'\sin\varphi' - r \sin\theta\sin\varphi\right)^2 .
\end{eqnarray}
For $r\to \infty$, the above equation can be approximated by
\begin{eqnarray}
\sinh\alpha\tau^{}_{\Delta'}(\pm\epsilon)-\cos\theta \cosh\alpha\tau^{}_{\Delta'}(\pm\epsilon) &=& 
\sinh(\pm\alpha\epsilon) -
\cos\theta\cosh\alpha\epsilon + \nonumber\\  && 
\frac{\alpha}{c} \Delta' \sin \theta\cos(\varphi'-\varphi),
\end{eqnarray}
which gives
\begin{equation}
  \tau_{\Delta'}(\pm\epsilon) = \frac{1}{\alpha}\ln 
  \frac{-b'_{\pm\epsilon} + \sqrt{ ( b'_{\pm\epsilon})^2 +\sin^2\theta}}{1+\cos\theta}
\end{equation}
with 
\begin{equation}
  b'_{\epsilon} \equiv \sinh\alpha\epsilon -\cos\theta\cosh\alpha\epsilon
    + \frac{\alpha}{c}\Delta' \sin\theta\cos(\varphi'-\varphi).
\end{equation} 
Inserting $\tau_{\Delta'}(\pm\epsilon)$ to (\ref{dPdOmClDeltaInt}), we obtain 
\begin{eqnarray}
  && \frac{d\bar{P}^{}_{\Delta'}}{d\Omega^{}_{\rm II}} =
  \frac{\mu^{}_0 q^2 a^2}{(4\pi)^2 c}\frac{\sin^2\theta}
   {\left[1 + \left({\cal M}'\Delta' \right)^2 - 2 {\cal M}'\Delta' \cos\theta \right]_{}^3}
\label{dPdOmclDelta2Exact}
\end{eqnarray} 
with 
\begin{equation}
   {\cal M}' \equiv \frac{\alpha}{c}\sin\theta \cos(\varphi'-\varphi). \label{defMprime}
\end{equation}
Below we set $\varphi'=\varphi$ without loss of generality. The expression (\ref{dPdOmclDelta2Exact}) has the same form as (\ref{dPdOmclDelta1Exact}), while ${\cal M}'$ here is a constant of time, and (\ref{dPdOmclDelta2Exact}) is regular for all $\theta \in [0,\pi]$.

Suppose that this ensemble of transverse-shifted worldlines is weighted by a normalized Gaussian spatial distribution of width $W'$, such that the angular radiated power emitted by this ensemble of UACs is given as  
\begin{equation}
   \frac{d\bar{P}^{}_{W'}}{d\Omega^{}_{\rm II}} = 
   \left< \frac{d\bar{P}^{}_{\Delta'}}{d\Omega^{}_{\rm II}} \right> \equiv 
  \int_{-\infty}^\infty d\Delta'  \frac{e^{-(\Delta'/W')^2}}{\sqrt{\pi}W'} 
   \frac{d\bar{P}^{}_{\Delta'}}{d\Omega^{}_{\rm II}}. \label{dPdOmClIntegral2}
\end{equation}
The result would be the same as (\ref{dPWdOmGauss}) except that all ${\cal M}\Delta$ there are replaced by ${\cal M}'\Delta'$. 
The counterpart of (\ref{dPWdOmGaussSeries}) here is again an asymptotic expansion. 
Also, the closed form result has the same form as (\ref{ExpectdPW}), while $d\bar{P}^{}_{W'}/d\theta = \int_0^{2\pi}d\varphi (d\bar{P}^{}_{W'}/d\Omega^{}_{\rm II})$ is regular over $\theta \in [0, \pi]$ here.

\begin{figure}[ht]
\includegraphics[width=8cm]{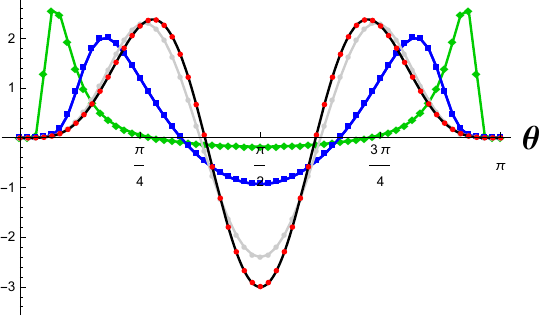}
\includegraphics[width=7cm]{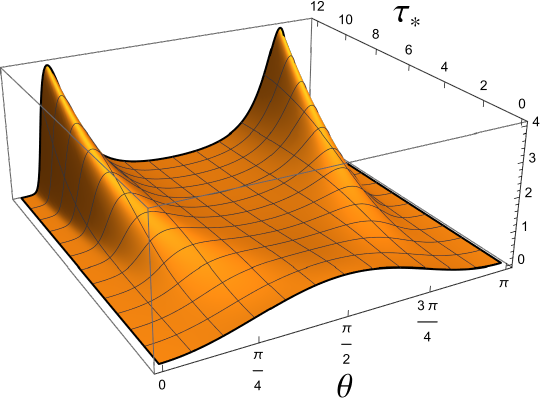}
\caption{(Left) Scaled corrections of an ensemble of UAC of $\alpha=c=1$ and $W'=0.03$ (red), $0.3$ (gray), $1$ (blue), and $3$ (green) in the counterpart of (\ref{dPWdOmGauss}) to the angular radiated power emitted by a point charge along (\ref{zmuUAC}).
The black curve represents the 
scaled sum of $dP^{\mathbb{T}}_{ qmP}/d\theta$, $dP^{\mathbb{T}}_{mpP}/d \theta$ in  (\ref{dPLmqdOmLTT}) and $dP^{\mathbb{T}}_{ddP}/ d\Omega^{}_{\rm II}$ in (\ref{dPddPTdOmega}) in the long-time regime.
(Right) Time evolution of scaled angular radiated power emitted by the same ensemble of UAC of $W'= 2\sqrt{0.015^2+ 0.08^2 \tau_*^2}$ obtained from the counterpart of  (\ref{dPWdOmGauss}).} 
\label{FigQRadbyzT}
\end{figure}

The coefficient of $\langle \Delta^2 \rangle$ in the counterpart of (\ref{dPWdOmGaussSeries})
is $\frac{3s\bar{m}}{8\pi} c^2\alpha^2 \sin^2\theta \times$ 
$3(8\cos^2\theta-1) {\cal M}'^2 = 
\frac{9s\bar{m}}{8\pi} \alpha^4  \sin^4\theta (8\cos^2\theta-1)$.
This is exactly the coefficient of the sum of  
$dP^{\mathbb{T}}_{qmP}/d\Omega^{}_{\rm II}$, $dP^{\mathbb{T}}_{mpP}/d \Omega^{}_{\rm II}$ in  (\ref{dPLmqdOmLTT}) and $dP^{\mathbb{T}}_{ddP}/ d\Omega^{}_{\rm II}$ in (\ref{dPddPTdOmega}) in the long-time regime with $\langle \Delta^2\rangle$ identified as $\eta_-^2 \big( \langle p_2^2\rangle^{}_{\rm I} \cos^2\varphi + \langle p_3^2\rangle^{}_{\rm I} \sin^2\varphi \big)/\bar{m}^2$, {which is time dependent} [see (I.4.61)]. Indeed, in  figure \ref{FigQRadbyzT} (left), when $W'$ is sufficiently small, the numerical result of the scaled correction $\frac{8\pi c}{\mu^{}_0 q^2 a^2 W'^2} \Big( \frac{d\bar{P}^{}_{W'}}{d\theta} - \frac{d\bar{P}}{d\theta}\Big)$ (red dotted) fits very well to the scaled coefficient of $\langle \Delta^2 \rangle$ (black).
In the same plot, one can see that when ${\cal M}' W'$ increases to $O(1)$, nonperturbative effect become significant: The peaks are pushed toward $\theta=0$ and $\pi$, and the peak values change.

To illustrate, we insert $W' =2 \sqrt{0.015^2 + 0.08^2 \tau_*^2}$ into (\ref{dPdOmClIntegral2}) [cf. (I.4.61)] and numerically calculate time evolution of scaled $d\bar{P}^{}_{W'}/d\theta$ emitted by the ensemble of the Gaussian distributed $x_{}^{\sf T}$-shifted UACs using the counterpart of (\ref{dPWdOmGauss}). 
The result of the scaled angular radiated power $\frac{8\pi c}{\mu^{}_0 q^2 a^2 W'^2} \frac{d\bar{P}^{}_{W'}}{ d\theta}$ is shown in figure \ref{FigQRadbyzT} (right),
where one can see that for $\tau^{}_* \ll 1$ and ${\cal M}'W' \ll 1$, the radiated power $d\bar{P}^{}_{W'}/d\theta$ emitted by the ensemble of UACs is very close to $d\bar{P}/d\theta$ emitted by the point-like reference UAC, with the scaled difference shown in the left plot. As ${\cal M}'W'$ grows to $O(1)$, the correction becomes of the same order of $d\bar{P}/d\theta$, and then keep increasing as ${\cal M}'W'$ grows further. Two peaks symmetric about $\theta=\pi/2$ emerge and moving towards $\theta=0$ and $\pi$, while the peak values roughly increases like $\tau_*^3 \sim W'^3$ ($\alpha=c=1$). Between the two peaks, the angular radiated power can be approximated by (\ref{ExpandEVinMW}) with ${\cal M}W$ replaced by ${\cal M}'W'$, which goes to zero for $\theta$ not too close to $0$ or $\pi$ as $|{\cal M}'W'| \to\infty$ at late times.

\section{Some useful formulas}
\label{ApxCoeff}

From eqs. (4.94) and (4.95) in ref. \cite{Ro65}, one has 
\begin{equation}
  \tau^{}_{-,\mu} = -\frac{1}{c}\left[ \bar{n}^{}_\mu(\tau^{}_-)+
    \frac{\bar{u}^{}_\mu(\tau^{}_-)}{c} \right]. \label{DtauDmu}
\end{equation}    
Calculating $\bar{n}^{-}_\mu R^\mu{}_{,\nu}$ with (\ref{Rmudef}) inserted, and then applying (\ref{DtauDmu}) and $\bar{n}^{-}_\mu \dot{\bar{n}}^\mu_- =
\partial^{}_{\tau^{}_-} 
\big[\bar{n}^{-}_\mu \bar{n}^\mu_-\big]/2 = 0$, one gets
\begin{equation}
  r^{}_{,\mu} = \bar{n}^{}_\mu(\tau^{}_-) +\frac{\bar{a}_{\bar{n}}(\tau^{}_-)}{c^2} R^{}_\mu 
  \label{rmuResult}
\end{equation}
with $\bar{a}_{\bar{n}} \equiv \bar{a}^{}_\nu \bar{n}_{}^\nu$ [(4.96) in ref. \cite{Ro65}].
Note that $R^{}_\mu \propto r$ from (\ref{Rmudef}).

Further calculation gives 
\begin{eqnarray}
  \tau^{}_{-,\mu\nu} &=& -\frac{1}{cr}\left( \eta^{}_{\mu\nu} -\bar{n}^{-}_\mu\bar{n}^{-}_\nu +
    \frac{\bar{u}^{-}_\mu\bar{u}^{-}_\nu}{c^2}\right)+\frac{\bar{a}^{-}_{\bar{n}}}{c^3}
    \frac{R^{}_\mu R^{}_\nu}{r^2}, \label{DDtauDmuDnu}\\
  r^{}_{,\mu\nu} &=& \left(\frac{1}{r}+\frac{\bar{a}^{-}_{\bar{n}}}{c^2}\right)
    \left(\eta_{\mu\nu}-\bar{n}^-_{\mu}\bar{n}^-_\nu+\frac{\bar{u}^-_\mu\bar{u}^-_\nu}{c^2}\right)
    + \frac{\bar{a}^-_\mu R_\nu + \bar{a}^-_\nu R_\mu}{c^2 r} \nonumber\\ &&
    - \left[ \frac{\bar{a}^-_{\bar{n}}}{c^2}+ \frac{r}{c^3}\bar{n}^-_\rho \dot{\bar{a}}_-^\rho
    + \frac{r}{c^4}\left( \bar{a}^{-}_{\bar{n}}\bar{a}^{-}_{\bar{n}}-\bar{a}^-_\rho \bar{a}_-^\rho
    \right)\right]\frac{R_\mu R_\nu}{r^2}. \label{DDrDmuDnu}
\end{eqnarray}
It is clear that $\tau^{}_{-,\mu\nu} = \tau^{}_{-,\nu\mu}$ and $r^{}_{,\mu\nu} = r^{}_{,\nu\mu}$.
Moreover, one has
\begin{eqnarray}
  \tau^{}_{-,\mu\nu\rho}(x) &=& -\left( \frac{\dot{\bar{a}}^\sigma_- \bar{n}^-_\sigma}{c^4} + 
    3 \frac{(\bar{a}_{\bar{n}}^-)^2}{c^5} -\frac{\bar{a}_\sigma^- \bar{a}^\sigma_-}{c^5} \right)
    \frac{R^{}_\mu R^{}_\nu R^{}_\rho}{r^3} + O \left( r^{-1}\right), \label{DDDtauDmuDnuDrho}\\
  r^{}_{,\mu\nu\rho}(x) &=& 
    \left\{ \frac{\ddot{\bar{a}}_-^\sigma \bar{n}^-_\sigma}{c^4}
    + \frac{\dot{\bar{a}}_-^\sigma}{c^5}\left[ \bar{a}^-_{\bar{n}} \left( 4\bar{n}^-_\sigma
     +2\frac{\bar{u}^-_\sigma}{c}  \right)-3\bar{a}^-_\sigma \right]- \right. \nonumber\\
  && \hspace{3cm}\left.\frac{2}{c^6} \bar{a}^-_\sigma \bar{a}_-^\sigma \bar{a}^-_{\bar{n}} + 
     3\frac{(\bar{a}^-_{\bar{n}})^3}{c^6} \right\}\frac{R^{}_\mu R^{}_\nu R^{}_\rho}{r^2}+ O(r^0).
    \label{DDDrDmuDnuDrho}
\end{eqnarray}
So we have 
\begin{eqnarray}
&&\hspace{-.5cm}d^{(2)j}_{\nu\rho}\bar{n}^\rho_- 
  = \frac{1}{cr}\left( \delta_\nu^j - \bar{n}_\nu^- \bar{n}^j_- + 
      \frac{\bar{u}_\nu^- \bar{u}^j_-}{c^2} \right),\label{d2jnrResult}\\
&&\hspace{-.5cm}d^{(1)j}_{\nu\rho}\bar{n}^\rho_- = \frac{1}{c r}
    \left[ 2\frac{\bar{a}^{-}_\nu}{c}  \frac{{R}_{}^j}{r}
     + \frac{\bar{a}^-_{\bar{n}}}{c} \left(\delta_\nu^j -3\bar{n}^{-}_\nu\bar{n}_{-}^j+ 
      \frac{\bar{u}^{-}_\nu \bar{u}_{-}^j}{c^2} -2\bar{n}^{-}_\nu\frac{\bar{u}_{-}^j}{c}\right)
\right]+O\big(r_{}^{-2}\big),\\ 
&&\hspace{-.5cm} d^{(0)j}_{\nu\rho}\bar{n}^\rho_- = 
   \left[ \frac{\dot{\bar{a}}^-_\nu}{c^2}
   -\frac{\dot{\bar{a}}^-_\rho}{c^2}\bar{n}_-^\rho \bar{n}^-_\nu-
   \frac{\bar{a}^-_\rho\bar{a}_-^\rho \bar{u}^-_\nu}{c^4} -
   3\frac{\left(\bar{a}^-_{\bar{n}}\right)^2}{c^3}\bar{n}^{-}_\nu +
   3\frac{\bar{a}^-_{\bar{n}} \bar{a}^-_\nu}{c^3}\right] 
    \frac{R_{}^j}{r^2} 
    + O\big( r_{}^{-2} \big) \label{d0jnrResult}
\end{eqnarray}  
and $d^{(n)j}_{\nu\rho}\frac{\bar{u}^\rho_-}{c} = -d^{(n)j}_{\nu\rho}\bar{n}^\rho_- + O\big( r_{}^{-2}\big)$ for $n=0,1,2$.
Here we have used 
\begin{eqnarray}
  0 &=& \partial^{}_\tau \big[ \bar{a}_{}^\alpha(\tau) \bar{u}^{}_\alpha(\tau) \big]
   = \dot{\bar{a}}_{}^\alpha \bar{u}^{}_\alpha + \bar{a}_{}^\alpha \bar{a}^{}_\alpha,\\
  0 &=& \partial^{2}_\tau \big[ \bar{a}_{}^\alpha(\tau) \bar{u}^{}_\alpha(\tau) \big]
   = \ddot{\bar{a}}_{}^\alpha \bar{u}^{}_\alpha + 3\dot{\bar{a}}_{}^\alpha \bar{a}^{}_\alpha .
\end{eqnarray}
since $\bar{a}_{}^\alpha \bar{u}^{}_\alpha =0$.

Also, for a uniformly accelerated charge along the worldline (\ref{zmuUAC}), from (\ref{tauPofrtheta}), one has
\begin{equation}
  \tau^{}_{+,0} = -\tau^{}_{-,0}+\frac{1}{\alpha}\left( \frac{1}{V}+\frac{1}{U}\right),
  \hspace{.7cm}
  \tau^{}_{+,1} = -\tau^{}_{-,1}+\frac{1}{\alpha}\left( \frac{1}{V}-\frac{1}{U}\right),
  \label{Tauplusderiv}
\end{equation}
{$\tau^{}_{+,{\sf T}}= -\tau^{}_{-,{\sf T}}$}, and 
\begin{eqnarray}
&&\tau^{}_{+,00} = -\tau^{}_{-,00}-\frac{1}{\alpha}\left( \frac{1}{V^2}+\frac{1}{U^2}\right),
  \hspace{.7cm}
  \tau^{}_{+,11} = -\tau^{}_{-,11}-\frac{1}{\alpha}\left( \frac{1}{V^2}+\frac{1}{U^2}\right),
  \nonumber\\
&&\tau^{}_{+,01} = -\tau^{}_{-,01}-\frac{1}{\alpha}\left( \frac{1}{V^2}-\frac{1}{U^2}\right),
  \hspace{.4cm} \tau^{}_{+,{\sf TT}'} = -\tau^{}_{-,{\sf TT}'},
  \hspace{.4cm} \tau^{}_{+,{\sf TL}}=-\tau^{}_{-,{\sf TL}}
\end{eqnarray}
for ${\sf T}, {\sf T}' = 2,3$ and ${\sf L}=0,1$. 
Inserting (\ref{VUoftauM}) into the above equations,
the derivatives of $\tau^{}_+$ can be expressed as functions of $\tau_-$, $r$, $\theta$ and $\varphi$. Then one can see that those $ V^{-n}\pm U^{-n}$ ($n=1,2$) terms are roughly proportional to $r^{-n}$ for sufficiently large $r$ and $\theta$ not exactly $0$ or $\pi$. 

\section{Two-point correlators of transverse deviations of a uniformly accelerated electron}
\label{ApxzTzTF}

From (I.4.60), to the leading order of $s$, the two-point correlators of the transverse deviations of our uniformly accelerated electron in linear approximation read
\begin{eqnarray}
&&\langle \hat{z}^2_{{}_{ \underline{\texttt{0}}}}(\tau), \hat{z}^2_{{}_{ \underline{\texttt{0}}}}(\tau') \rangle^{\{0\}}_F = \langle \hat{z}^3_{{}_{ \underline{\texttt{0}}}}(\tau), \hat{z}^3_{{}_{ \underline{ \texttt{0}}}}(\tau') \rangle^{\{0\}}_F =
\frac{\hbar s}{\pi\bar{m}} \times  
\nonumber\\
&&\frac{1}{2s^2\alpha^2{\cal S}_\alpha}\bigg\{ 
-2  
+ {\cal S}_\alpha \alpha (\tau'-\tau)   
-{\cal S}_\alpha \left( 1-
{\cal S}_\alpha^2
\right) \pi  e^{{\cal S}_\alpha \alpha 
(\tau'-\tau)} \cot\left( 
{\cal S}_\alpha \pi\right)
\nonumber\\
&&- 2 {\cal S}_\alpha \ln \Big[ 1 - e^{\alpha (\tau'-\tau)} \Big] + 
\left( 1 - {\cal S}_\alpha^2
\right) \Big[F_{0|-}(\tau -\tau') -F_{0|+}(\tau -\tau') \Big] \nonumber\\
&& -2 e^{-{\cal S}_\alpha \alpha 
\eta} + 
{\cal S}_\alpha \alpha (\tau^{}_0 - \tau'_0) -2 {\cal S}_\alpha \ln \Big[1 - e^{\alpha (\tau^{}_0 - \tau'_0)}\Big] + 
\left( 1 -{\cal S}_\alpha^2 
\right) \Big[2 e^{-{\cal S}_\alpha \alpha 
\eta} -2 e^{-{\cal S}_\alpha \alpha 
\eta'} \nonumber\\
&& \hspace{2cm} -e^{-{\cal S}_\alpha \alpha 
\eta'}\left(e^{-{\cal S}_\alpha \alpha 
\eta} - 2 \right) F_{0|-}(\tau'_0-\tau^{}_0) 
+ e^{-{\cal S}_\alpha \alpha 
\eta}\left(e^{-{\cal S}_\alpha \alpha 
\eta'} - 2\right)F_{0|+}(\tau'_0-\tau^{}_0) 
\Big] \nonumber\\
&& - \frac{2 {\cal S}_\alpha  e^{\alpha (\tau^{}_0 - \tau'_0)}}{(e^{\alpha (\tau^{}_0-\tau'_0)} -1)^2} \bigg[ 1+ e^{ -{\cal S}_\alpha\alpha 
(\eta+\eta')} 
\nonumber\\ && \hspace{2cm} 
-e^{-{\cal S}_\alpha \alpha 
\eta} \left(1 + {\cal S}_\alpha 
\left[e^{\alpha (\tau^{}_0-\tau'_0)}-1\right]\right) - e^{-{\cal S}_\alpha \alpha 
\eta'}\left(1-{\cal S}_\alpha 
\left[e^{\alpha (\tau^{}_0-\tau'_0)}-1\right]\right)  \bigg]\nonumber\\
&& +2 e^{-{\cal S}_\alpha \alpha 
\eta} +2 e^{-{\cal S}_\alpha \alpha 
\eta'} + {\cal S}_\alpha \alpha (\eta+\eta') \nonumber\\
&& + {\cal S}_\alpha \left(1-
{\cal S}_\alpha^2\right)\pi 
e^{-{\cal S}_\alpha \alpha 
(\eta  +\tau_0-\tau'_0)} 
\left(2 -e^{-{\cal S}_\alpha \alpha 
\eta'}\right)\cot\left(
{\cal S}_\alpha\pi\right) 
  \nonumber\\
&& +  2 {\cal S}_\alpha^2 
\left[\frac{e^{-{\cal S}_\alpha\alpha
\eta'}-1}{e^{\alpha (\tau - \tau'_0)}-1} + \frac{ e^{-{\cal S}_\alpha \alpha 
\eta}-1 }{e^{\alpha (\tau'-\tau^{}_0)}-1} \right]
+2 {\cal S}_\alpha \left( \ln \Big[1 - e^{-\alpha (\tau -\tau'_0)}\Big]+ \ln \Big[ 1 - e^{-\alpha (\tau'-\tau^{}_0)}\Big] \right)  \nonumber\\
&&+\left(1 -{\cal S}_\alpha^2 
\right) \Big[(e^{-{\cal S}_\alpha\alpha 
\eta'} -2) (F_{0|-}(\tau-\tau'_0)-1)
+ e^{-{\cal S}_\alpha\alpha 
\eta'}  (F_{0|+}(\tau-\tau'_0)-1)  \nonumber\\
&& \hspace{2.05cm}
+(e^{-{\cal S}_\alpha\alpha
\eta}-2 ) (F_{0|-}(\tau'-\tau^{}_0)-1) 
+ e^{-{\cal S}_\alpha\alpha 
\eta}  (F_{0|+}(\tau'-\tau^{}_0)-1) \Big] \bigg\} \label{zTzTFcloseform}
\end{eqnarray}
with $\eta \equiv \tau-\tau^{}_0$, $\eta' \equiv \tau'-\tau'_0$, ${\cal S}_\alpha\equiv s\alpha/\varsigma = s\alpha \big( 1 + O(s^2\alpha^2) \big)$, and $F_{n|\pm}(T) \equiv {}_2F_1 \left( 1, n\pm \frac{s\alpha}{\varsigma}, n+1\pm\frac{s\alpha}{\varsigma}, e^{-\alpha T} \right)$ for $T>0$, while $\langle \hat{z}_{}^i(\tau), \hat{z}_{}^j(\tau') \rangle^{}_F \propto \eta_{}^{ij}$. We choose $\epsilon^{}_1 = \tau-\tau'$ and $\epsilon^{}_0 = \tau'_0-\tau^{}_0$, such that $\tau > \tau' \ge \tau'_0 > \tau^{}_0$.

In obtaining the early-time and long-time limits of the radiated power contributed by the transverse deviations, since $s\alpha/\varsigma \ll 1$ with our parameter values, we have used the formulas
\begin{eqnarray}
{}_2 F_1 (1,-1+b; b;z) &=& -z/b +  1 + z [1 + \ln (1-z)] + 
    b z[-\ln(1-z) + {\rm Li}^{}_2(z) ] \nonumber\\ &&
    -b^2 z [ {\rm Li}^{}_2(z) + {\rm Li}^{}_3(z)] +O(b^3), \label{2F1lim1}\\
{}_2 F_1 (1, 1+b; 2+b;, z) &=& -\ln(1-z)/z -b\big[ \ln(1-z)+{\rm Li}^{}_2(z) \big]/z 
    \nonumber\\ && -b^2\big[ {\rm Li}^{}_2(z)- {\rm Li}^{}_3(z) \big]/z + O(b^3), \\
{}_2 F_1 (1, b; 1+b; z) &=& 1 - b \ln(1-z)-b^2 {\rm Li}^{}_2(z) +b^3 {\rm Li}^{}_3(z) +O(b^4),\label{2F1lim3}
\end{eqnarray}
as $b \ll 1$. Here ${\rm Li}^{}_n(z)$ is the polylogarithm function. The above formulas are also useful in checking the regularity of (\ref{zTzTFcloseform}) in the $s\to 0$ and $\alpha\to 0$ limits.


\end{document}